\documentclass[graybox]{svmult}

\usepackage{mathptmx}       
\usepackage{helvet}         
\usepackage{courier}        
\usepackage{type1cm}        
\usepackage{makeidx}         
\usepackage{graphicx}        
\usepackage{multicol}        
\usepackage[bottom]{footmisc}

\makeindex             

\begin{document}

\title*{MHD Turbulence, Turbulent Dynamo and Applications}
\titlerunning{MHD Turbulence}

\author{Andrey Beresnyak and Alex Lazarian}

\institute{Andrey Beresnyak \at Ruhr-Universit\"at Bochum, 44780 Bochum, Germany, \email{andrey.at.astro@gmail.com}
\and A. Lazarian \at University of Wisconsin-Madison, WI, USA \email{lazarian@astro.wisc.edu}}

\maketitle

\abstract*{MHD Turbulence is common in many space physics and astrophysics
environments. We first discuss the properties of incompressible MHD turbulence.
 A well-conductive fluid amplifies initial magnetic fields in a process
called small-scale dynamo. Below equipartition scale for kinetic and magnetic
energies the spectrum is steep (Kolmogorov -5/3) and is represented by critically
balanced strong MHD turbulence. In this paper we report the basic reasoning behind
universal nonlinear small-scale dynamo and the inertial range of MHD turbulence.
We measured the efficiency of the small-scale dynamo $C_E=0.05$, Kolmogorov constant
$C_K=4.2$ and anisotropy constant $C_A=0.63$ for MHD turbulence in high-resolution
direct numerical simulations. We also discuss so-called imbalanced or cross-helical MHD
turbulence which is relevant for in many objects, most prominently in the solar wind. 
We show that properties of incompressible MHD turbulence are similar to the properties of
Alfv\'enic part of MHD cascade in compressible turbulence. The other parts of the cascade
evolve according to their own dynamics. The slow modes are being cascaded by Alfv\'enic modes,
while fast modes create an independent cascade. We show that different ways of decomposing
compressible MHD turbulence into Alfv\'en, slow and fast modes provide consistent results
and are useful in understanding not only turbulent cascade, but its interaction with fast particles.}

\section{Introduction}
Historically, most of the turbulence studies were concerned with non-conductive
fluids, described by the Navier-Stokes equations. This is because most fluids
present on Earth are non-conductive. In the context of a larger Cosmos, this
situation is not a rule but rather an exception. Indeed, space is filled with
ionizing radiation and only the protection of our atmosphere, which is very
dense by astronomical standards, allows us to have a big volumes of insulating
fluids, such as the atmosphere and the oceans. In contrast, most of the ordinary
matter in the Universe is ionized, i.e. in a state of plasma. The description of
ionized, well-conductive fluids must include the Lorentz force and the
induction equation for the magnetic field. As it turned out, turbulent
conductive fluids tend to quickly generate their own magnetic fields in the
process known as dynamo. On the other hand, the presence of the dynamically
important magnetic field could be considered an observational fact. In spiral
galaxies magnetic field has a regular component, usually along the arms and a
random turbulent component of the same order. The value of the magnetic field,
around 5 $\mu G$, roughly suggests equipartition between magnetic and kinetic
forces. 

Observations of magnetized turbulence in the interstellar medium, galaxy clusters and the
solar wind have confirmed that turbulence is indeed ubiquitous in 
astrophysical flows and has been detected in almost all astrophysical and space
environments, see, e.g., \cite{goldstein1995,armstrong1995,chepurnov2010}.
The Reynolds numbers of astrophysical turbulence are, typically, very high, owing
to astrophysical scales which are enormous compared to dissipative scales.
Recent years have been marked by new understanding of the key role that
turbulence plays in a number of astrophysical processes \cite{CLV03,elmegreen2004}.
Most notably, turbulence has drastically changed the paradigms of interstellar
medium and molecular cloud evolution \cite{stone1998,ostriker2001,vazquez2007},
see also review \cite{McKee2007}. While small scale, kinetic turbulence has been probed by a variety of
approaches such as gyrokinetics, Hall MHD and electron MHD \cite{howes2006,schekochihin2007,cho2004},
in this review we concentrate mostly on the fluid-scale MHD turbulence which is the most
important for star formation and interaction with cosmic rays. 

The theoretical understanding of magnetized turbulence can be roughly subdivided in two big domains:
MHD dynamo and the inertial range of MHD turbulence. In the first part of this review we will explain
small-scale dynamo, which is fast, universal mechanism to wind up magnetic fields. The second half is 
devoted to the properties of the inertial range cascades of incompressible and compressible MHD turbulence. 
In particular we discuss both the spectrum of fundamental MHD modes and the intermittency properties of
turbulence. The third part is devoted to applications of our knowledge of MHD cascades, which includes
turbulent reconnection, cosmic ray propagation \cite{BYL11} and damping of instabilities \cite{BL08b}.

\def\L{{\Lambda}}
\def\l{{\lambda}}

\section{Astrophysical Dynamo}
\label{sec:1}
One of the central processes of MHD dynamics is how conductive fluid generates
its own magnetic field, a process known broadly as ``dynamo''.
Turbulent dynamo has been subdivided into ``large-scale/mean-field dynamo''
and ``small-scale/fluctuation dynamo'' depending on whether magnetic
fields are amplified on scales larger or smaller than outer scale of turbulence.

Although several ``no-dynamo'' theorems have been proved for flows with symmetries, a generic
turbulent flow, which possesses no exact symmetry, was expected to amplify
magnetic field by stretching, due to the particle separation in a turbulent flow.
For the large-scale dynamo, a ``twist-stretch-fold'' mechanism was
introduced \cite{vainshtein1972}. Turbulent flow possessing perfect
statistical isotropy can not generate large-scale field, so the observed large-scale fields,
such as in the disk galaxies, are generated when statistical symmetries of turbulence are
broken by large-scale asymmetries of the system, such as stratification, rotation
and shear, see, e.g., \cite{vishniac2001,kapyla2009}. Large-scale dynamo is often
investigated using so-called mean field theory, see, e.g.\cite{krause1980}, where
the magnetic and velocity field are decomposed into mean and fluctuating parts
and the equations for the mean field are closed using statistical or volume
averaging over the fluctuating turbulent part. 

The studies of large-scale dynamo are very rich and diverse due to the variety of conditions
in astrophysical flows in different objects, one of the most ambitious goals is to explain the
solar cycle, see, e.g., \cite{brandenburg2005}. In this review we decided to concentrate
on the small-scale dynamo as it is fast and generic and almost always generate
magnetic fluctuations with energy of the order of the kinetic energy (so-called equipartition).
Magnetic fluctuations could be subsequently ordered by slower large-scale dynamo and produce large-scale
magnetic fields. Some objects, such as galaxy clusters, are dominated by small-scale dynamo, however.

Most studied was so-called kinematic regime of small-scale dynamo, which ignores
the backreaction of the magnetic field \cite{kazantsev1968,kraichnan1967,kulsrud1992}.
However, from these models it was not clear whether magnetic energy will continue to grow
after the end of kinematic regime. In astrophysical objects with very large $Re$
it becomes inapplicable at very short timescales.
Also magnetic spectrum of kinematic dynamo, possessing positive spectral index, typically 3/2,
is incompatible with observations in galaxy clusters \cite{Laing2008}. These observations
clearly indicate steep spectrum with negative power index at small scales.
In fact, from theoretical viewpoint, kinematic dynamo is inapplicable in most astrophysical
environment, because the Alfv\'en speed is typically many orders of magnitude higher
than the Kolmogorov velocity.

\begin{figure}[t]
\begin{center}
\includegraphics[width=0.8\columnwidth]{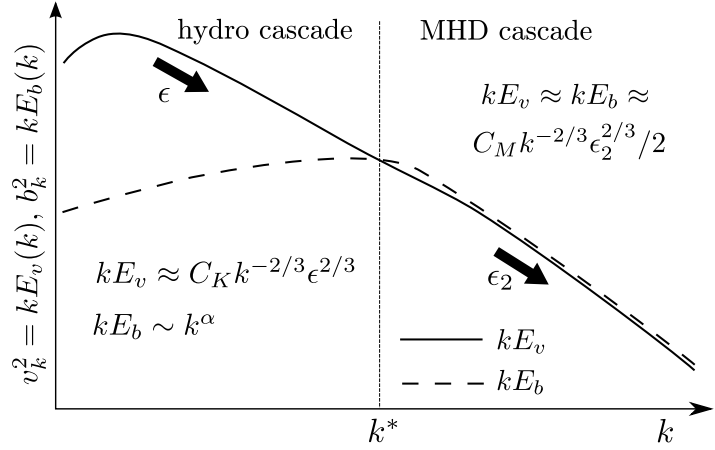}
\end{center}
\caption{A cartoon of kinetic and magnetic spectra in small-scale dynamo, at a particular moment of
time when equipartition wavenumber is $k^*$.}
\label{spect_dyn}
\end{figure}

The understanding of nonlinear small-scale dynamo was developing slowly and was influenced by analytical
kinematic studies. For a long time a popular belief was that after becoming nonlinear
the small-scale dynamo will saturate in one way or another. If we assume that the magnetic
energy indeed saturates as soon as the dynamo become nonlinear. The saturation level in this case will be $\rho v_\eta^2/2$,
where $v_\eta$ is a Kolmogorov velocity scale. This is a factor of $Re^{-1/2}$ smaller than the kinetic energy density and is
completely unimportant in high-Re astrophysical environments.
The price tag to discover what will happen in the nonlinear regime
was, therefore, fairly high. The early work by Schl\"uter and Bierman \cite{schluter1950} suggested
that the dynamo will continue to grow and will saturate on each subsequent scale by its dynamical time.
The true revival of small-scale dynamo happened relatively recently due
to availability of direct numerical simulations. First simulations were concerned with saturated state
of small scale dynamo and produced steep spectrum at small scales and significant outer-scale
fields, and the saturated state seems to be only weakly dependent on $Re$ and $Pr$
as long as $Re$ is large, see, e.g, \cite{haugen2004}. Furthermore it was suggested in 
\cite{schekochihin2007,CVB09,ryu2008,BJL09,B12a} that there is a linear growth stage.
In subsequent sections we will follow the argumentation of \cite{B12a}, who provided sufficient
analytical and numerical argumentation behind the universality of the nonlinear small-scale dynamo.

\subsection{Universal Nonlinear Small-Scale Dynamo}
We assume that the spectra of magnetic and kinetic energies
at a particular moment of time are similar to what is presented on Fig.~\ref{spect_dyn}.
Magnetic and kinetic spectra cross at some ``equipartition'' scale $1/k^*$,
below which both spectra are steep due to MHD cascade \cite{GS95,B11}. This assumption is suggested
by both numerical evidence \cite{BL09b,CVB09} and observations of magnetic fields in clusters
of galaxies \cite{Laing2008}. Also, if we start with assuming $Pr=1$ and magnetic energy very small and
follow standard kinematic dynamo calculations, e.g. \cite{kulsrud1992} the magnetic energy will
grow exponentially till the magnetic spectrum intersect kinetic spectrum at the viscous scales.
This will roughly correspond to the beginning of the nonlinear regime with equipartition scale 
equal to the dissipation scale.

At larger scales magnetic spectrum is shallow, $k^\alpha,\, \alpha>0$,
while kinetic spectrum is steep due to the hydro cascade. Most
of the magnetic energy is concentrated at scale $1/k^*$. We designate $C_K$ and $C_M$ as Kolmogorov constants
of hydro and MHD respectively. The hydrodynamic cascade rate is $\epsilon$ and the MHD cascade rate as $\epsilon_2$.
Due to the conservation of energy in the inertial range, magnetic energy will grow at a rate $\epsilon-\epsilon_2$.
We will designate $C_E=(\epsilon-\epsilon_2)/\epsilon$ as an ``efficiency of the small-scale dynamo'' and
will argue that this is a true constant, since: a) turbulent dynamics is local in scale in
the inertial range; b) neither ideal MHD nor Euler equations contain any scale explicitly.
Magnetic energy, therefore, grows linearly with time if $\epsilon=const$. 
The equipartition scale $1/k^*$ will grow with time as $t^{3/2}$ \cite{BJL09}. This
is equivalent to saying that small-scale dynamo saturates at several dynamical times
at scale $1/k^*$ and proceeds to a twice larger scale \cite{schekochihin2007}. If magnetic
energy grows approximately till equipartition \cite{haugen2004,CVB09}, the whole process
will take around several dynamical timescales of the system, or more quantitatively, $(C_K^{3/2}/C_E)(L/v_L)$.

\subsection{Locality of the Small-Scale Dynamo}
We will use ``smooth filtering'' approach with
dyadic-wide filter in k-space \cite{aluie2010}. We designate a filtered vector quantity
as ${\bf a}^{[k]}$ where $k$ is a center of a dyadic Fourier filter in the range of wave numbers
$[k/2,2k]$. The actual logarithmic width of this filter is irrelevant to further
argumentation, as long as it is not very small. We will assume that the vector field ${\bf a}$ is
H\"older-continuous, i.e., $|{\bf a(x)-a(y)}|<|\bf x-y|^h$ with exponent $0<h<1$ and designate $a_k=\langle|{\bf a}^{[k]}|^3\rangle^{1/3}$
(angle brackets are averages over ensemble),
which is expected to scale as $a_k\sim k^{\sigma_3}$, e.g., $k^{-1/3}$ for velocity in Kolmogorov
turbulence. The energy cascade rate is $\epsilon=C_K^{-3/2}kv_k^3$, where 
we defined Kolmogorov constant $C_K$ by third order, rather than second order quantities.
We will keep this designation, assuming that traditional Kolmogorov constant could be used instead.
We use spectral shell energy transfer functions such as
$T_{vv}(p,k)=-\langle{\bf v}^{[k]}({\bf v}\cdot{\bf \nabla}){\bf v}^{[p]}\rangle, \,
T_{w^+w^+}(p,k)=-\langle{\bf w^+}^{[k]}({\bf w^-}\cdot{\bf \nabla}){\bf w^+}^{[p]}\rangle$
 \cite{alexakis2005}, applicable to incompressible ideal MHD equations,
where $w^\pm$ are Els\"asser variables and $v$, $b$ and $w^\pm$ are measured in the same Alfv\'enic units.
Using central frequency $k$ and studying ``infrared`` (IR) transfers
from $p\ll k$, and ''ultraviolet'' (UV) transfers, from $q\gg k$,
we will provide absolute bounds on $|T|$, in units of energy transfer rate as in \cite{aluie2010,eyink2005},
and {\it relative} volume-averaged bounds which are divided by the actual energy rate and are dimensionless.
We will consider three main
$k$ intervals presented on Fig.~\ref{spect_dyn}: $k\ll k^*$ (``hydro cascade''), $k\sim k^*$ (``dynamo'')
and $k \gg k^*$ (``MHD cascade'').

\subsection{MHD cascade, $k \gg k^*$}
The only energy cascades here are Els\"asser cascades and, by the design of
our problem, $w^+$ and $w^-$ have the same statistics, so we will drop $\pm$.
For an exchange with $p\ll k$ band, for $|T_{ww}|$, using H\"older inequality and wavenumber
conservation we get an upper bound of $pw_pw_k^2$ and for $q \gg k$ band it is $kw_q^2w_k$,
these bounds are asymptotically small. For the full list of transfers and
limits refer to Table~\ref{transfers}.
The relative bound should be taken with respect to
$C_M^{-3/2}kw_k^3$, where $C_M$ is a Kolmogorov constant for MHD,
from which we get that most of the energy transfer with the $[k]$ band should
come from $[kC_M^{-9/4},kC_M^{9/4}]$ band, see \cite{B11}.
The global transfers between kinetic and magnetic energy must average out in this regime,
nevertheless, the pointwise IR and UV transfers can be bounded by $pb_pv_kb_k$ and $kb_q^2v_k$
and are small \cite{eyink2005}.

\begin{table}[t]
\begin{center}
\caption{Transfers and Upper Limits}
  \begin{tabular*}{0.99\columnwidth}{@{\extracolsep{\fill}}r c l c c}
    \hline\hline

Transfers & & & $p\ll k$ & $q\gg k$ \\

   \hline

$T_{vv}(p,k)$&=&$-\langle{\bf v}^{[k]}({\bf v}\cdot{\bf \nabla}){\bf v}^{[p]}\rangle$
 & $pv_pv_k^2$ & $kv_kv_q^2$  \\

$T_{bb}(p,k)$&=&$-\langle{\bf b}^{[k]}({\bf v}\cdot{\bf \nabla}){\bf b}^{[p]}\rangle$
 &  $pb_pv_k b_k$ & $kb_kv_qb_q$ \\

$T_{vb}(p,k)$&=&$\langle{\bf b}^{[k]}({\bf b}\cdot{\bf \nabla}){\bf v}^{[p]}\rangle$
 &  $pv_pb_k^2$ & $kb_kv_qb_q$  \\

$T_{bv}(p,k)$&=&$\langle{\bf v}^{[k]}({\bf b}\cdot{\bf \nabla}){\bf b}^{[p]}\rangle$
 &  $pb_pv_k b_k$ & $kv_kb_q^2$  \\

$T_{w^+w^+}(p,k)$&=&$-\langle{\bf w^+}^{[k]}({\bf w^-}\cdot{\bf \nabla}){\bf w^+}^{[p]}\rangle$
 &  $pw_pw_k^2$ & $kw_kw_q^2$  \\

   \hline

\end{tabular*}
  \label{transfers}
\end{center}
\end{table}

\subsection{Hydro cascade, $k\ll k^*$}
Despite having some magnetic energy
at these scales, most of the energy transfer is dominated by velocity field.
Indeed, $|T_{vv}|$ is bounded by $pv_pv_k^2$ for $p \ll k$ and by $kv_q^2v_k$ for $q \gg k$. Compared to these,
$|T_{bv}|$ transfers are negligible: $pb_pv_kb_k$ and $kb_q^2v_k$. For magnetic energy in
$p\ll k$ case we have $|T_{vb}|$ and $|T_{bb}|$ transfers bounded by $pv_pb_k^2$, $pb_pv_kb_k$ and for
$q\gg k$ case $|T_{vb}|$ and $|T_{bb}|$ are bounded by $kb_kv_qb_q$.
Out of these three expressions the first two go to zero, while the third goes
to zero if $\alpha-2/3<0$ or have a maximum at $q=k^*$ if $\alpha-2/3>0$.
This means that for the transfer to magnetic energy we have IR locality, but not necessarily
UV locality. Note that magnetic energy for $k\ll k^*$ is small
compared to the total, which is dominated by $k=k^*$.
We will assume that $\alpha-2/3>0$ and that the spectrum of $b_k$ for $k<k^*$ is formed
by nonlocal $|T_{vb}|$ and $|T_{bb}|$ transfers from $k^*$, namely magnetic structures at $k$ are formed
by stretching of magnetic field at $k^*$ by velocity field at $k$. Magnetic spectrum
before $k^*$ is, therefore, nonlocal and might not be a power-law, but our further argumentation will
only require that $b_k<v_k$ for $k<k^*$.

\subsection{Dynamo cascade $k=k^*$}
In this transitional regime our estimates
of Els\"asser UV transfer and kinetic IR transfer from two previous sections will
hold. We are interested how these two are coupled together and produce
observed magnetic energy growth.
IR $p\ll k^*$ $|T_{vb}|$ and $|T_{bb}|$ transfers will be bounded by $pv_pb_{k^*}^2$ and
$pb_pv_{k^*}b_{k^*}$,
which go to zero, so there is a good IR locality. Ultraviolet transfers
will be bounded by $k^*b_{k^*}b_qv_q$. This quantity also goes to
zero as $q$ increases, so there is an UV locality for this regime as well.
Let us come up with bounds of relative locality. Indeed, the actual growth of
magnetic energy was defined as $\epsilon_B=\epsilon-\epsilon_2=C_EC_K^{-3/2}kv_k^3$.
So, $p\ll k^*$ IR bound is $k^*C_E^{3/2}C_K^{-9/4}$
and UV bound is $k^*C_E^{-3/2}C_M^{9/4}$. We conclude that most of the interaction
which result in magnetic energy growth must reside in the wavevector interval of 
$k^*[C_E^{3/2}C_K^{-9/4},C_E^{-3/2}C_M^{9/4}]$. Numerically, if we substitute
$C_K=1.6$, $C_M=4.2$, $C_E=0.05$ we get the interval of $k^*[0.004,2000]$.
So, despite being asymptotically local, small-scale dynamo can be fairly nonlocal in practice.

\begin{table}[t]
\begin{center}
\caption{Three-dimensional MHD Dynamo Simulations}
  \begin{tabular*}{0.99\columnwidth}{@{\extracolsep{\fill}}c c c c c c c}
    \hline\hline
Run  & n & $N^3$ & Dissipation & $\langle\epsilon\rangle$ &  Re & $C_E$ \\

   \hline

M1-6 & 6 & $256^3$   &  $-7.6\cdot10^{-4}k^2$  & 0.091 & 1000 & $0.031\pm0.002$ \\

M7-9 & 3 &  $512^3$  &  $-3.0\cdot10^{-4}k^2$  & 0.091 & 2600 & $0.034\pm0.004$  \\

M10-12 & 3 & $1024^3$  &  $-1.2\cdot10^{-4}k^2$ & 0.091 & 6600  & $0.041\pm0.005$ \\

M13 & 1 & $1024^3$ & $-1.6\cdot10^{-9}k^4$   & 0.182 & -- & $0.05\pm0.005$  \\

M14 & 1 & $1536^3$ & $-1.5\cdot10^{-15}k^6$  & 0.24  & -- & $0.05\pm0.005$ \\

   \hline

\end{tabular*}
  \label{experiments1}
\end{center}
\end{table}

Summarizing, the kinetic cascade at large scales and the MHD cascade at small scales
are dominated by local interactions. The transition between the kinetic
cascade and the MHD cascade is also dominated by local interactions, and since ideal MHD equations
do not contain any scale explicitly, the efficiency of small-scale dynamo $C_E$ is a true universal
constant.
Note that $C_E$ relates energy fluxes, not energies, so this claim is unaffected by
the presence of intermittency. Magnetic spectrum at $k\ll k^*$ is dominated by nonlocal
triads that reprocess magnetic energy from $k= k^*$ but, since this
part of the spectrum contains negligible magnetic energy, our
universality claim is unaffected by this nonlocality.
\begin{figure}[t]
\begin{center}
\includegraphics[width=0.7\columnwidth]{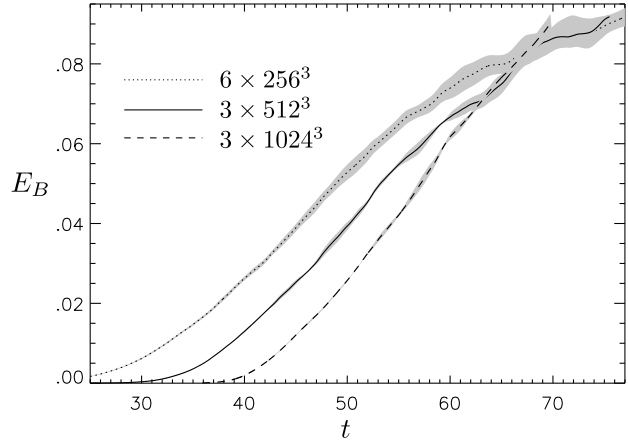}
\end{center}
\caption{Magnetic energy growth vs. time in code units, observed in simulations
M1-6 ($\tau_\eta=0.091$ in code units), M7-9 ($\tau_\eta=0.057$) and M10-12 ($\tau_\eta=0.036$). We used sample averages which greatly
reduced fluctuations and allowed us to measure $C_E$ with sufficient precision.}
\label{dyn1}
\end{figure}

\subsection{Numerical Results}
We performed numerical simulations of statistically homogeneous isotropic
small-scale dynamo by solving MHD equations with stochastic non-helical driving and explicit
dissipation with $Pr_m=1$.
The details of the code and driving are described in detail in our earlier publications  \cite{BL09a,BL09b} and 
Table~\ref{experiments1} shows simulation parameters. We started each simulation from previously well-evolved driven
hydro simulation by seeding low level white noise magnetic field. We ran
several statistically independent simulations in each group and obtained growth rates and errors
from sample averages. In all simulations, except M14, the energy injection rate was controlled.
Fig~\ref{dyn1} shows sample-averaged time evolution of magnetic energy. Growth is initially exponential
and smoothly transition into the linear stage. Note, that scatter is initially small, but grows
with time, which is
consistent with the picture of magnetic field growing at progressively larger scales and
having progressively less independent realizations in a single datacube.

\subsection{On the Efficiency of Small-Scale Dynamo}
Our $C_E$ is much smaller than unity.
One would expect a quantity of order unity because this is
a universal number, determined only by strong interaction on equipartition scale.
If we refer to the ideal incompressible MHD equations, written in terms of Els\"asser variables,
$\partial_t{\bf w^\pm}+\hat S ({\bf w^\mp}\cdot\nabla){\bf w^\pm}=0$, the dynamo could be understood
as decorrelation of ${\bf w^\pm}$ which are originally equal to each other
in the hydrodynamic cascade. In our case this decorrelation is happening at the equipartition scale $1/k^*$.
Being time-dependent, it propagates upscale, while ordinarily energy cascade goes downscale.
The small value of $C_E$ might be due to this. As opposed to picture with multiple reversals
and dissipation due to microscopic diffusivity, typical for kinematic case, in our picture we appeal
to {\it turbulent diffusion} which helps to create large-scale field. Both stretching and diffusion
depend on turbulence at the same designated scale $1/k^*$, so
in the asymptotic regime of large Re one of these processes must dominate.
As $C_E$ is small, stretching and diffusion are close to canceling each other.

\section{MHD Turbulence in the Inertial Range} 

Inertial range of turbulence was introduced by \cite{kolm41} as a range of spatial
scales where driving and dissipation are unimportant and perturbations exist due
to energy transfer from one scale to another.
In the inertial range of MHD turbulence perturbations of both velocity and magnetic
field will be much smaller than the local Alfv\'enic velocity $v_A=B/\sqrt{4\pi\rho}$, due to the turbulence
spectrum being steeper than $k^{-1}$, therefore local mean magnetic
field will strongly affect dynamics in this range \cite{iroshnikov,kraichnan}.
Furthermore, the large-scale dynamo we mentioned in Section~\ref{sec:1} will operate
in such objects as spiral galaxies and provide the mean field for the MHD turbulence
in the interstellar medium (ISM).

As in the case of hydrodynamics, the study of MHD turbulence began with
weakly compressible and incompressible cases which are directly applicable to
many environments, such as stellar interiors, ICM and hot phases of the ISM.
Later it was realized that many features of incompressible MHD turbulence
are still present even in supersonic dynamics, due to the dominant effect of
Alfv\'enic shearing \cite{cho2003c,BLC05}. It had been pointed out by \cite{GS95} that
strong mean field incompressible turbulence is split into
the cascade of Alfv\'enic mode, described by Reduced MHD or RMHD \cite{kadomtsev1974,strauss1976} and the passive cascade
of slow (pseudo-Alfv\'en) mode. In the strong mean field case it was sufficient
to study only the Alfv\'enic dynamics, as it will determine all statistical
properties of turbulence, such as spectrum or anisotropy. This decoupling
was also observed in numerics. Luckily, being the limit
of very strong mean field, RMHD has a two-parametric symmetry, which we will
discuss further in Section~\ref{basic}, which, under certain conditions,
makes universal cascade with power-law energy spectrum possible.

Interaction of Alfv\'enic perturbations propagating in a strong mean field is unusual
due to a peculiar dispersion relation of Alfv\'enic mode, $\omega=k_\| v_A$, where $k_\|$
is a wavevector parallel to the mean magnetic field. This results in a tendency of MHD
turbulence to create ``perpendicular cascade'', where the flux of energy is preferentially
directed perpendicular to the magnetic field. This tendency enhances the nonlinearity
of the interaction, described by $\xi=\delta v k_\perp/v_A k_\|$, which is the ratio of the mean-field
term to the nonlinear term, and results in development
of essentially strong turbulence. As turbulence
becomes marginally strong, $\xi\sim 1$, the cascading timescales
become close to the dynamical timescales $\tau_{\rm casc}\sim\tau_{\rm dyn}=1/wk_\perp$ and
the perturbation frequency $\omega$ has a lower bound due to an
uncertainty relation $\tau_{\rm casc}\omega>1$ \cite{GS95}. This makes turbulence
being ``stuck'' in the $\xi\sim 1$ regime, which is known as ``critical balance''.
There is another lower bound on $\omega$, due to the {\it directional} uncertainty
of the ${\bf v}_A$, which was discovered in \cite{BL08}. In the case of balanced
MHD turbulence, which we consider in the next few sections, this two bounds
coincide. We consider more general imbalanced case in Section~\ref{imbal}.

Goldreich-Sridhar (1995, \cite{GS95} henceforth GS95) model is predicting a $k^{-5/3}$ energy spectrum with anisotropy\footnote{The anisotropy should
be understood in terms of local magnetic field direction, i.e. the magnetic field direction at the given scale. The
original treatment, e.g. the closure relations employed, in the Goldreich-Sridhar paper uses the global frame of reference
which was noticed later in \cite{Lazarian1999} and used in the numerical works that validated the theory \cite{cho2000,maron2001,CLV02a}.}
described as $k_\|\sim k_\perp^{2/3}$. Numerical studies \cite{cho2000,maron2001,CLV02a} confirmed
steep spectrum and scale-dependent anisotropy, but \cite{maron2001,muller2005} claimed
a shallower than $-5/3$ spectral slope in the strong mean field case, which was close to $-3/2$.
This motivated adjustments to the GS95 model
\cite{galtier2005,boldyrev2005,gogoberidze2007}. A model with so called ``dynamic
alignment'' \cite{boldyrev2005,boldyrev2006} became popular after the
scale-dependent alignment was discovered in numerical simulations \cite{BL06}. This model
is based on the idea that the alignment between velocity and magnetic perturbations
decreases the strength of the interaction scale-dependently, and claims that the alignment
goes as $k^{-1/4}$. This would, as they argue, modify the spectral slope
of MHD turbulence from the $-5/3$ Kolmogorov slope to the observed $-3/2$ slope.
It also claims \cite{boldyrev2006} that there is a self-consistent turbulent mechanism that produces
such an alignment. Below we examine both the alignment and the spectrum.

\subsection{Basic Equations}
\label{basic}
Ideal MHD equations describe the dynamics of ideally conducting inviscid fluid with magnetic field
and can be written in Heaviside and $c=1$ units as
\begin{eqnarray}
\partial_t \rho+{\bf \nabla\cdot}(\rho{\bf v})=0,\\
\rho(\partial_t+{\bf v}\cdot\nabla){\bf v}=-\nabla P+{\bf j\times B},\\
{\bf \nabla\cdot B}=0,\\
\partial_t{\bf B}=\nabla\times({\bf v \times B}),\label{mhd0}
\end{eqnarray}
with current ${\bf j}={\bf \nabla \times B}$ and vorticity ${\bf \omega}={\bf \nabla \times v}$.
This should be supplanted with energy equation and a prescription for pressure $P$.
The incompressible limit assumes that the pressure is so high that the density is constant
and velocity is purely solenoidal (${\bf \nabla \cdot v}=0$). This does not necessarily refer to the ratio of outer scale
kinetic pressure to molecular pressure, but could be interpreted as scale-dependent condition.
Indeed, if we go to the frame of the fluid, local
perturbations of velocity will diminish with scale and will be much smaller than the speed of sound.
In this situation it will be possible to decompose velocity into low-amplitude
sonic waves and essentially incompressible component of ${\bf v}$, as long as we are not in the vicinity
of a shock. The incompressible component, bound by ${\bf \nabla\cdot v}=0$,
will be described by much simpler equations: 
\begin{eqnarray}
\partial_t{\bf v}=\hat S(-{\bf \omega \times v}+{\bf j \times b}),\\
\partial_t{\bf b}=\nabla\times({\bf v \times b}),\label{mhd1}
\end{eqnarray}
where we renormalized magnetic field to velocity units ${\bf b=B}/\rho^{1/2}$ (the absence of $4\pi$
is due to Heaviside units) and used solenoidal projection operator
$\hat S=(1-\nabla\Delta^{-1}\nabla)$ to get rid of pressure.
Finally, in terms of Els\"asser variables ${\bf w^\pm=v\pm b}$ this could
be rewritten as
\begin{equation}
\partial_t{\bf w^\pm}+\hat S ({\bf w^\mp}\cdot\nabla){\bf w^\pm}=0.\label{mhd2}
\end{equation}
This equation resembles incompressible Euler's equation. Indeed, hydrodynamics
is just a limit of $b=0$ in which ${\bf w}^+={\bf w}^-$. This resemblance, however, is misleading,
as the local mean magnetic field could not be excluded by the choice of reference
frame and, as we noted earlier, will strongly affect dynamics on all scales.
We can explicitly introduce local mean field as ${\bf v_A}$, assuming that it is constant,
so that $\delta{\bf w^\pm=w\pm v}_A$:
\begin{equation}
\partial_t{\bf \delta w^\pm}\mp({\bf v_A}\cdot\nabla){\delta \bf w^\pm}
+\hat S ({\delta \bf w^\mp}\cdot\nabla){\delta \bf w^\pm}=0\label{mhd3}.
\end{equation}
In the linear regime of small $\delta w$'s they represent perturbations, propagating
along and against the direction of the magnetic field, with nonlinear term describing
their interaction. As we noted earlier, due to the resonance condition of Alfv\'enic
perturbations they tend to create more perpendicular structure,
making MHD turbulence progressively more anisotropic. This was empirically known
from tokamak experiments and was used in so-called reduced MHD approximation, which neglected
parallel gradients in the nonlinear term \cite{kadomtsev1974,strauss1976}.
Indeed, if we denote $\|$ and $\perp$ as directions parallel and perpendicular to ${\bf v}_A$,
the mean field term $(v_A \nabla_\|) \delta w^\pm$ is much larger than
$(\delta w^\mp_\| \nabla_\|)\delta w^\pm$ and the latter could be ignored in the inertial
range where $\delta w^\pm \ll v_A$. This will result in Equation~\ref{mhd3} being split into
\begin{equation}
\partial_t{\bf \delta w^\pm_\|}\mp({\bf v_A}\cdot\nabla_\|){\delta \bf w^\pm_\|}
+\hat S ({\delta \bf w^\mp_\perp}\cdot\nabla_\perp){\delta \bf w^\pm_\|}=0,\label{mhd4}
\end{equation}
\begin{equation}
\partial_t{\bf \delta w^\pm_\perp}\mp({\bf v_A}\cdot\nabla_\|){\delta \bf w^\pm_\perp}
+\hat S ({\delta \bf w^\mp_\perp}\cdot\nabla_\perp){\delta \bf w^\pm_\perp}=0,\label{mhd5}
\end{equation}
which, physically represent a limit of very strong mean field where ${\bf \delta w^\pm_\|}$
is a slow (pseudo-Alfv\'en) mode and ${\bf \delta w^\pm_\perp}$ is the Alfv\'en mode and
Equation~\ref{mhd4} describes a passive dynamics of slow mode which is sheared by the
Alfv\'en mode, while Equation~\ref{mhd5} describes essentially nonlinear dynamics of the Alfv\'en mode
and is known as reduced MHD. For our purposes, to figure out asymptotic
behavior in the inertial range, it is sufficient to study Alfv\'enic dynamics and slow
mode can be always added later, because it will have the same statistics. 

It turns out that reduced MHD is often applicable beyond incompressible MHD limit, in a highly
collisionless environments, such as tokamaks or the solar wind. This is due to the fact that
Alfv\'en mode is transverse and does not require pressure support. Indeed, Alfv\'enic perturbations
rely on magnetic tension as a restoring force and it is sufficient that charged particles
be tied to magnetic field lines to provide inertia \cite{Schekochihin2009}.

A remarkable property of RMHD is that it has a precise two-parametric symmetry:
${\bf w} \to {\bf w}A,\ \lambda \to \lambda B,\ t \to t B/A,\ \Lambda \to \Lambda B/A$.
Here $\lambda$ is a perpendicular scale, $\Lambda$ is a parallel scale, $A$
and $B$ are arbitrary parameters of the transformation. 
This is similar to the symmetry in Euler equation (${\bf B=0}$ limit of MHD), except for
a different prescription for parallel scale $\Lambda$ which now scales as time.
It is due to this precise symmetry and the absence
of any designated scale, that we can hypothesize universal regime,
similar to hydrodynamic cascade of \cite{kolm41}. In nature, the universal regime for MHD can be
achieved with $\delta w^\pm \ll v_A$. In numerical
simulations, we can directly solve RMHD equations, which have
precise symmetry already built in. From practical viewpoint, the statistics
from the full MHD simulation with $\delta w^\pm \sim 0.1 v_A$ is virtually
indistinguishable from RMHD statistics and even $\delta w^\pm \sim v_A$ is still fairly similar to the
strong mean field case \cite{BL09a}.

\begin{table}[t]
\begin{center}
\caption{Three-Dimensional RMHD Balanced Simulations}
  \begin{tabular*}{1.00\columnwidth}{@{\extracolsep{\fill}}c c c c c c}
    \hline\hline
Run  & $n_x\cdot n_y \cdot n_z$ & Dissipation & $\langle\epsilon\rangle$ &  $L/\eta$ \\

   \hline

R1 & $256\cdot 768^2$ & $-6.82\cdot10^{-14}k^6$   & 0.073 &   200 \\
R2 & $512\cdot 1536^2$ & $-1.51\cdot10^{-15}k^6$  & 0.073  &  400 \\
R3 & $1024\cdot 3072^2$ & $-3.33\cdot10^{-17}k^6$ & 0.073 &  800\\
\hline
R4 & $768^3$ & $-6.82\cdot10^{-14}k^6$            & 0.073 &   200 \\
R5 & $1536^3$ & $-1.51\cdot10^{-15}k^6$            & 0.073 &   400 \\
\hline
R6 & $384\cdot 1024^2$ & $-1.70\cdot10^{-4}k^2$  & 0.081  & 280 \\
R7 & $768\cdot 2048^2$ & $-6.73\cdot10^{-5}k^2$ & 0.081 & 560 \\
\hline
R8 & $768^3$ & $-1.26\cdot10^{-4}k^2$            & 0.073 & 350 \\
R9 & $1536^3$ & $-5.00\cdot10^{-5}k^2$            & 0.073 & 700 \\

   \hline
\end{tabular*}
  \label{experiments}
\end{center}
\end{table}

\subsection{Basic Scalings in the Balanced Case}
As was shown in a rigorous perturbation study of weak MHD turbulence, it has a tendency of becoming {\it stronger}
on smaller scales \cite{galtier2000}. Indeed, if $k_\|$ is constant and $k_\perp$ is increasing, $\xi=\delta wk_\perp/v_A k_\|$
will increase, due to $\delta w \sim k_\perp^{-1/2}$ in this regime. This will naturally lead to strong turbulence, where
$\xi$ will stuck around unity due to two competing processes: 1) increasing interaction by perpendicular cascade
and 2) decrease of interaction due to the uncertainty relation $\tau_{\rm casc}\omega>1$, where $\tau_{\rm casc}$
is a cascading timescale. Therefore, MHD turbulence will be always marginally strong in the inertial range, which
means that cascading timescale is associated with dynamical
timescale $\tau_{\rm casc}\sim\tau_{\rm dyn}=1/\delta w k_\perp$ \cite{GS95}. 
In this case, assuming that energy transfer is local
in scale and, therefore, depend only
on perturbations amplitude on each scale, we can write Kolmogorov-type phenomenology as
\begin{equation}
\epsilon^+=\frac{(\delta w^+_\lambda)^2\delta w^-_\lambda}{\lambda},
\ \ \ \ \epsilon^-=\frac{(\delta w^-_\lambda)^2\delta w^+_\lambda}{\lambda},
\label{fluxes}
\end{equation}
where $\epsilon^\pm$ is an energy flux of each of the Els\"asser variables and
$\delta w^\pm_\lambda$ is a characteristic perturbation amplitude on a scale
$\lambda$. Such an amplitude can be obtained by Fourier filtering with a
dyadic filter in k-space, see, e.g., \cite{B12a}.

Since we consider so-called balanced case with both $w$'s having the same statistical
properties and energy fluxes, one of these equations is sufficient. This will result
in a $\delta w \sim \lambda^{1/3}$, where $\lambda$ is a perpendicular scale,
or, in terms of energy spectrum $E(k)$,
\begin{equation}
E(k)=C_K \epsilon^{2/3} k^{-5/3},\label{spec53}
\end{equation}
where $C_K$ is known as Kolmogorov constant. We will be interested in Kolmogorov constant
for MHD turbulence. This scaling is supposed to work until dissipation effects kick in.
In our further numerical argumentation dissipation scale will play a big role, but
not from a physical, but rather from a formal point of view. We will introduce an idealized
scalar dissipation term in a RHS of Eq.~\ref{mhd2} as $-\nu_n(-\nabla^2)^{n/2}{\bf w^\pm}$,
where $n$ is an order of viscosity and $n=2$ correspond to normal Newtonian viscosity,
while for $n>2$ it is called hyperviscosity. The dissipation scale for this GS95
model is the same as the one for Kolmogorov model, i.e.
$\eta=(\nu_n^3/\epsilon)^{1/(3n-2)}$. This is a unique combination of $\nu_n$ and $\epsilon$
that has units of length. Note that Reynolds number, estimated as $v L/\nu_2$, where $L$ is an
outer scale of turbulence, is around $(L/\eta)^{4/3}$.

Furthermore, the perturbations of $w$ will be strongly anisotropic
and this anisotropy can be calculated from the critical balance condition $\xi \approx 1$, so that
$k_\| \sim k_\perp^{2/3}$. Interestingly enough this could be obtained directly from units and
the symmetry of RMHD equations from above. Indeed, in the RMHD limit,
$k_\|$ or $1/\Lambda$ must be in a product with $v_A$, since only the product enters the original
RMHD equations. We already assumed above that turbulence is local and each scale of turbulence
has no knowledge of other scales, but only the local dissipation rate $\epsilon$. In this case
the only dimensionally correct combination for the parallel scale $\Lambda$, corresponding
to perpendicular scale $\lambda$ is 
\begin{equation}
\Lambda=C_A v_A \lambda^{2/3} \epsilon^{-1/3}\label{anis},
\end{equation}
where we introduced a dimensionless ``anisotropy constant'' $C_A$.
Equations ~\ref{spec53} and ~\ref{anis} roughly describe the spectrum
and anisotropy of MHD turbulence.
Note, that GS95's $-5/3$ is a basic scaling that should be
corrected for intermittency. This correction is negative due to
structure function power-law exponents being a concave function of their order \cite{frisch1995}
and is expected to be small in three-dimensional case. This correction for hydrodynamic
turbulence is around $-0.03$. Such a small deviation should be irrelevant in the context
of debate between $-5/3$ and $-3/2$, which differ by about $0.17$.

A modification of the GS95 model was proposed by Boldyrev (2005, \cite{boldyrev2005}, 2006\cite{boldyrev2006}, henceforth B06)  who suggested
that the original GS95 scalings can be modified by a scale dependent factor that decreases the strength
of the interaction, so that RHS of the Equation~\ref{fluxes} is effectively multiplied
by a factor of $(l/L)^{1/4}$, where $L$ is an
outer scale.
In this case the spectrum will be expressed
as $E(k)=C_{K2} \epsilon^{2/3} k^{-3/2}L^{1/6}$. Note that this spectrum
is the only dimensionally correct spectrum with $k^{-3/2}$ scaling, which
does not contain dissipation scale $\eta$. The absence of $L/\eta$, is due to so-called zeroth
law of turbulence which states that the amplitude at the outer scale should not depend on the viscosity.
This law follows from the locality of energy transfer has been know empirically to hold very well.
The dissipation scale of B06 model is different from that of the GS95
model and can be expressed as
$\eta'=(\nu_n^3/\epsilon)^{1/(3n-1.5)}L^{0.5/(3n-1.5)}$.  
\begin{figure}[t]
\includegraphics[width=0.49\textwidth]{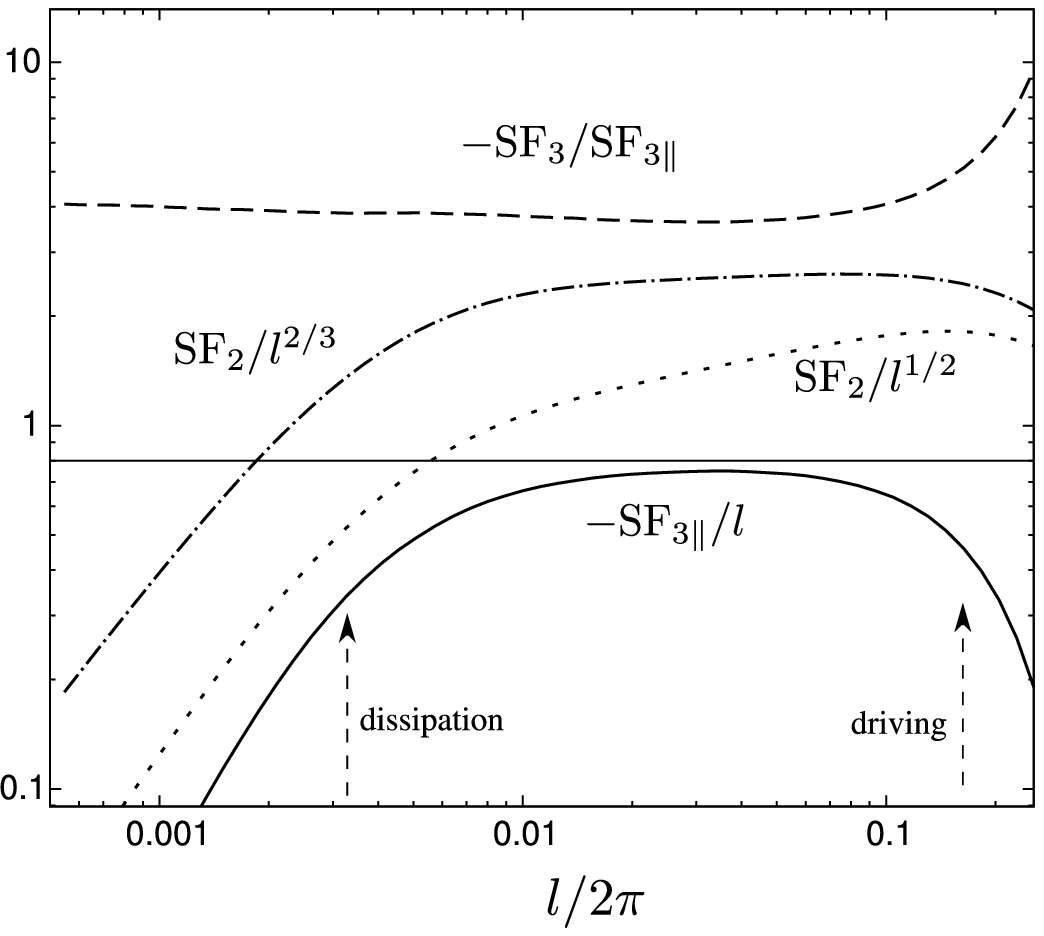}
\includegraphics[width=0.49\textwidth]{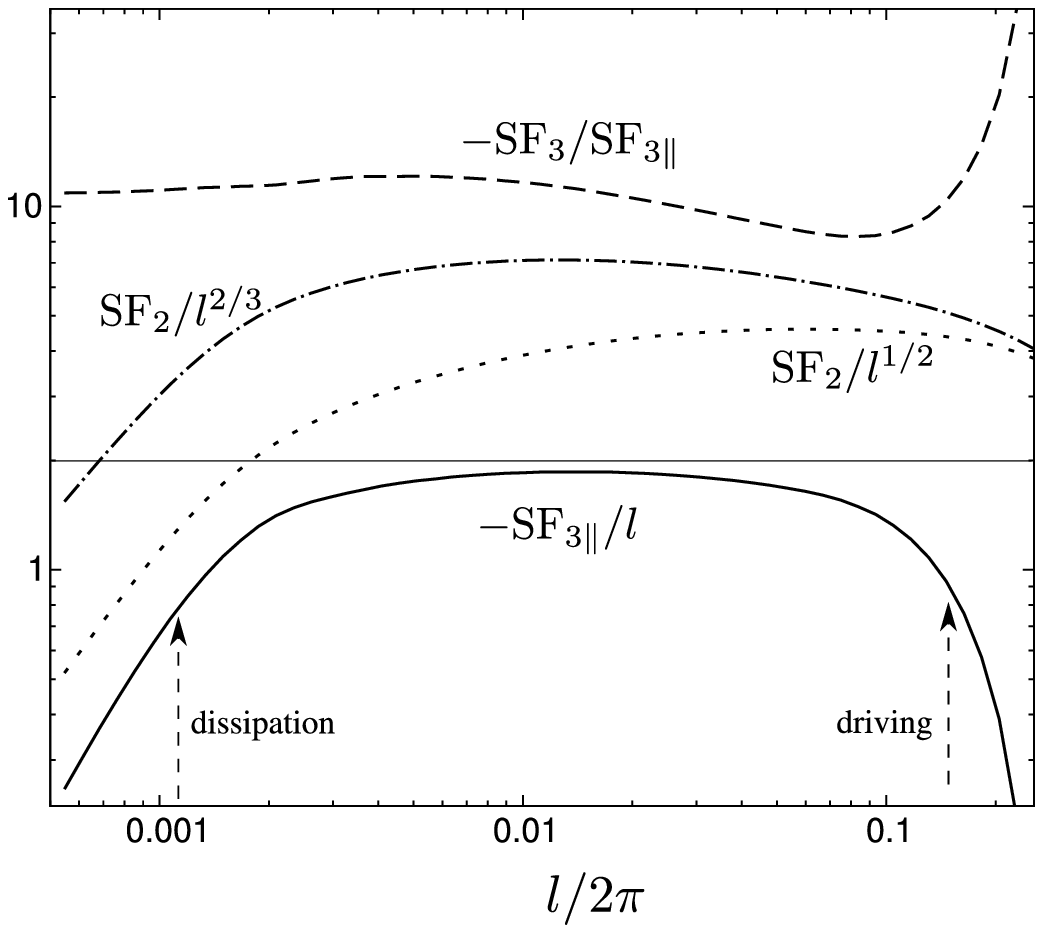}
\caption{Different structure functions vs the distance $l$,
measured in hydrodynamic (left) and MHD (right) simulations.
Solid lines show $-SF_{3\|}/l\epsilon$. The influence of driving and dissipation
is minimized in the point where $-SF_{3\|}/l\epsilon$ is closer to its theoretical value.
The dashed line indicates the ratio of the third order SF, defined in the text
to the parallel third order SF. This ratio is a test for turbulence self-similarity,
as long as this ratio is constant, the turbulence is well self-similar.
Finally, dotted and dash-dotted lines indicate the same second order structure functions,
compensated by $l^{1/2}$ and $l^{2/3}$ correspondingly, in arbitrary units.
Here $l^{2/3}$ is the Richardson-Kolmogorov scaling and $l^{1/2}$ is
the scaling that appears in Kraichnan DIA model for hydrodynamics,
Iroshnikov-Kraichnan model for MHD and B06 model.
}
\label{sfs}
\end{figure}

\subsection{Structure Functions and Spectra}
Structure and correlation functions (SF and CF) has been traditionally used in turbulence research for a long time. 
In theory these are quantities statistically averaged over ensemble, while in numerics the averaging is usually
over time and volume using homogeneity and stationarity. The typical quantity people
use in isotropic hydrodynamic turbulence is an isotropic second order structure function of velocity:
\begin{equation}
SF^2(l)=\langle(v({\bf r}-{\bf l})-v({\bf r}))^2\rangle_{\bf r}.
\end{equation}
This is a difference in velocity between two points separated by vector ${\bf l}$, squared
and averaged over the volume, i.e. the vector ${\bf r}$. This quantity could be represented
by the sum of the ``longitudinal`` and ''transverse'' components with velocity decomposed
into a direction perpendicular and parallel to ${\bf l}$. The longitudinal
structure function is important in experimental research of hydrodynamic turbulence, since this
is the primary quantity measured by the heated wire technique.  

MHD turbulence is not isotropic, therefore, there is a wider variety of structure functions
that one can possibly measure. However, in the RMHD limit there is a particular structure function
which plays the similar role as the isotropic SF in hydrodynamics, the perpendicular SF 
\begin{equation}
SF^2_\perp (l)=\langle(w^\pm({\bf r}-l{\bf n})-w^\pm({\bf r})  )^2\rangle_{\bf r},
\end{equation}
where ${\bf n}$ is a vector perpendicular to the magnetic field.
Power spectra, on the other hand, are produced by obtaining a Fourier transform ${\bf \hat v(k)}$ of original quantity ${\bf v(r)}$
and taking the product $\frac{1}{2} {\bf v_i(k)v_i^*(k)}$, where $*$  is a complex conjugate. Relations between spectra and structure functions
are well-known, see, e.g. \cite{monin1975}.

A number of exact relations for structure functions are known both for hydro and MHD, see, e.g., \cite{biskamp2003}.
The famous Kolmogorov $-4/5$ law relates a parallel signed structure function for velocity in the inertial range
with the dissipation rate:
\begin{equation}
SF_{3\|h}(l)=\langle(\delta v_{l\|})^3 \rangle=-\frac{4}{5}\epsilon l.
\end{equation}
Another exact relation, similar to the Yaglom's -4/3 law for incompressible hydro exists for axially symmetric
MHD turbulence:
\begin{equation}
SF_{3\|}(l)=\langle\delta w^{\mp}_{l\|} (\delta w^{\pm}_{l})^2 \rangle=-2\epsilon l,\label{PP}
\end{equation}
where $l$ is taken perpendicular to the axis of statistical symmetry -- the direction of the mean magnetic field
${\bf B}$ \cite{politano1998}. One can measure SFs above and argue about influence of dissipation and driving in
each particular simulation. Fig.~\ref{sfs} shows several structure functions, compensated by
various powers of $l$ and the ratio of parallel third order structure function and full third
order SF, $SF_3=\langle|v({\bf r}-{\bf l})-v({\bf r})|^3\rangle$.

\begin{figure}[t]
\begin{center}
\includegraphics[width=0.7\textwidth]{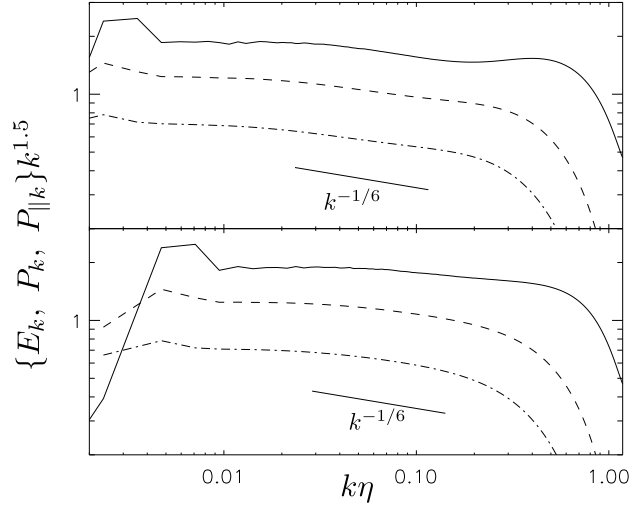}
\end{center}
\caption{Three types of spectra from a numerical simulations R1, R4. $E_k$ -- solid, $P_k$ -- dashed,
$P_{\|k}$ -- dash-dotted. In a simulation with limited resolution all three spectra have different shapes.
}
\label{spec1d3d}
\end{figure}

Normally, the inertial range in a simulation is defined as a range
of scales where $-SF_{3\|}/l$ is closest to its theoretical value, i.e. the
influence of energy injection from driving and energy dissipation from viscous
term is minimized. Another test for the inertial range is the test
for turbulence self-similarity, in particular one can take the above ratio
of the unsigned and signed third order SFs. This ratio must be constant
as long as turbulence is self-similar. Fig.~\ref{sfs} shows that hydrodynamic
turbulence is rather self-similar and the scaling of the second-order
structure function in the inertial range is around $l^{0.7}$,
i.e. close to the Kolmogorov scaling. In the MHD simulation the self-similarity is 
broken and although one can argue that the scaling is closer to the $l^{2/3}$
in the point where $-SF_{3\|}/l\epsilon$ is closest to its theoretical value of $2$,
claiming a certain scaling based on these data would be an overstatement. In the next section
we will describe a rigorous method to claim a certain scaling based on numerical
convergence in a series of simulations.

Power spectra are the measures, complimentary to second order structure functions.
In particular, so-called one-dimensional power spectrum $P_k$ is a Fourier transform
of the $SF_2$. This function is popular in the satellite measurements of the solar wind
turbulence, where a particular quantity, ${\bf v}$ or ${\bf B}$ is measured
as a function of time. It is then interpreted as an instant measurement along a line
in a turbulent realization (so-called Taylor hypothesis). The power spectra from
many samples like this are averaged to obtain $P_k$ for either velocity or magnetic
field. Another experimental measure is the so-called parallel power spectrum $P_{\|k}$. It
is obtained in the measurements of hydrodynamic turbulence by heated wire technique.
A scalar quantity is measured in this technique, which is the velocity perturbation
parallel to the average flow velocity. Similarly this is interpreted as a
measurement in space by using Taylor frozen flow hypothesis. Finally, there is
a power spectrum favored by numerics, which is a three-dimensional spectrum $E(k)$.
This spectrum is obtained from full three-dimensional power spectrum $\frac12{\bf v(k)\cdot v^*(k)}$ by
integrating over the solid angle in ${\bf k}$ space, so that $E(k)$ is only a function of
scalar $k$. In statistically isotropic hydro and MHD turbulence the integration is in spherical
shells, while in RMHD, the parallel wavenumber is infinitely small compared to other
wavenumbers, so the integration is, effectively, along all $k_\|$ and the circle in ${\bf k_\perp}$
space, i.e. the isotropic spectrum is equivalent to the perpendicular spectrum.
Three spectra  $P(k)$, $P_{\|}(k)$ and $E(k)$ of the solenoidal vector field
are related by the following expressions, see, e.g., \cite{monin1975}:
\begin{equation}
P(k)=\int^\infty_k E(k_1) \frac{dk_1}{k_1},
\end{equation}
\begin{equation}
P_{\|}(k)=\int^\infty_k E(k_1)\left(1-\frac{k^2}{k_1^2} \right)\frac{dk_1}{k_1},
\end{equation}
Fig.~\ref{spec1d3d} shows three types of spectra from the simulation. The primary spectrum was $E_k$
and the two other spectra were calculated by the above expressions.
All three spectra have different shapes. If one would want to claim a particular scaling by
qualitatively estimating the scaling from numerical spectrum, the estimate will depend on the
type of the spectrum and the chosen range of $k$ used for fitting the scaling. Based on
Fig.~\ref{spec1d3d} one can claim any spectral slope between $-5/3$ and $-3/2$. This further
reiterates the need of rigorous quantitative measurement based on numerical convergence,
presented in the next section.

\subsection{The Numerical Scaling Argument}
As was noted before, turbulence with very long range of scales
is common in astrophysics. Numerics, however, is not only unable to reproduce
such range, but actually struggles to obtain any good ``inertial range''.
In this situation a rigorous quantitative arguments
have to be invented to investigate asymptotic scalings.

\begin{figure}
\includegraphics[width=1.0\columnwidth]{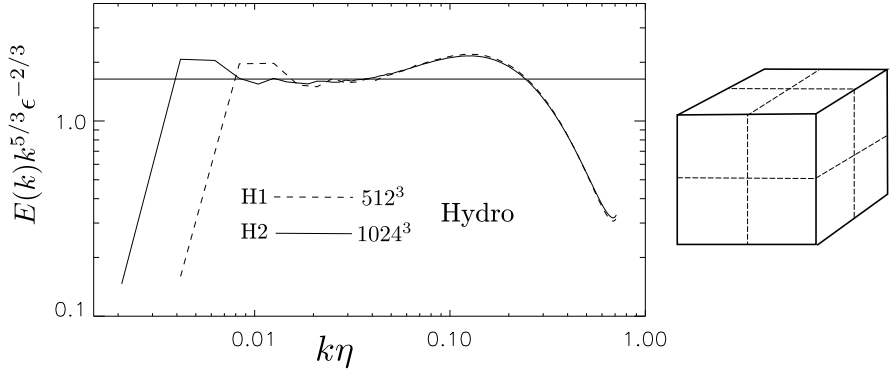}
\caption{The spectra from hydrodynamic simulations illustrate the numerical scaling argument.
 The large cube on the right can be split into smaller cubes with the same small scale statistics.
  Therefore as long as turbulence is scale-local and the effects of large scales could be neglected,
  the smaller simulation demonstrate the same statistics, as evident from convergence of dimensionless
  spectra on the left.}
\label{hydro}
\end{figure}

Suppose we performed several simulations with different Reynolds
numbers. If we believe that turbulence is universal,
and the scale separation between forcing scale and dissipation scale
is large enough, the properties of small scales should not depend on
how turbulence was driven and on the scale separation itself. This is because
neither MHD nor hydrodynamic equations explicitly contain any scale,
so simulation with a smaller dissipation scale could be considered,
due to symmetry from above, as a simulation with the same dissipation
scale, but larger driving scale. E.g., the small scale statistics
in a $1024^3$ simulation should look the same as small-scale statistics
in $512^3$, if the physical size of the elementary cell is the same
and the dissipation scale is the same.
Naturally, this scaling argument
in numerics require that the geometry of the elementary cells are
the same and the actual numerical scheme used to solve the equations
is the same. Also, numerical equations should not contain
any scale explicitly, but this is normally satisfied. What
scaling argument does not require is a high precision on the
dissipation scale or a particular form of dissipation, whether
explicit or numerical. This is because we need
that the statistics on small scales is similar in two simulations,
which is the case when numerics is the same on dissipation scale
and the influence of the outer scale is small by assumption
of turbulence locality.

In practice the scaling argument or a resolution study is
done in a following way: the averaged spectra in two
simulations are expressed in dimensionless units corresponding
to the expected scaling, for example a $E(k)k^{5/3}\epsilon^{-2/3}$
is used for hydrodynamics, and plotted versus dimensionless
wavenumber $k\eta$, where dissipation scale $\eta$ correspond
to the same model, e.g. $\eta=(\nu^3/\epsilon)^{1/4}$ is used
for scalar second order viscosity $\nu$ and Kolmogorov
phenomenology. Plotted this way the two spectra should collapse
onto the same curve on the viscous scales, see, e.g., Fig.~\ref{hydro}.
This method has been used in hydrodynamics
since long time ago, see, e.g., \cite{yeung1997,gotoh2002,kaneda2003}.
Although for hydrodynamics good convergence on the
dissipation scale has been observed starting with rather
moderate resolutions, which signifies that hydrodynamic
cascade has good, narrow locality, the larger
the resolution, the better the convergence should be.
Note that in \cite{kaneda2003}, which had very high resolution
even the intermittency correction
to the spectrum has been captured. So, the optimal
strategy for MHD would be to perform the largest resolution
simulations possible and do a resolution study with
particular models in mind.

\subsection{Numerical Experiments}
We will briefly explain the numerical setup and methods
used in \cite{BL09a,BL09b,BL10,B11,B12b}. For further detail the
reader is referred to these publications.
We used pseudospectral dealiased code
to solve RMHD equations. Same code was used earlier for RMHD, incompressible
MHD and incompressible hydrodynamic simulations. The RHS of Eq.~\ref{mhd5}
was complemented by an
explicit dissipation term $-\nu_n(-\nabla^2)^{n/2}{\bf w^\pm}$ and forcing
term ${\bf f}$. Diffusive terms with $n=2$ are referred to as normal viscosity and
with $n>2$ are referred to as hyperviscosity.
Table~\ref{experiments} shows the parameters of the balanced simulations. The
Kolmogorov scale is defined as $\eta=(\nu_n^3/\epsilon)^{1/(3n-2)}$,
the integral scale $L=3\pi/4E\int_0^\infty k^{-1}E(k)\,dk$ (which was
approximately 0.79 for R1-3). Dimensionless ratio $L/\eta$ could serve as a
``length of the spectrum'', although spectrum is actually significantly shorter
for n=2 viscosity and somewhat shorter for n=6 hyperviscosity.

\begin{figure}[t]
\includegraphics[width=1.0\textwidth]{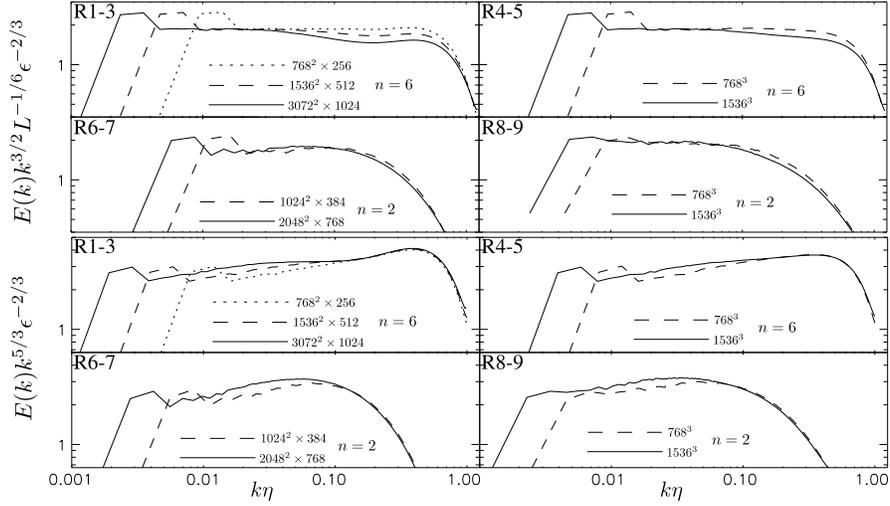}
\caption{Numerical convergence of spectra in all simulations. Two upper
rows are used to study convergence assuming B06 model and two
bottom rows -- assuming GS95 model. Note that definition
of dissipation scale $\eta$ depends on the model, this difference is tiny
in hyperviscous simulations R1-5, but significant in viscous simulations R6-9.
Numerical convergence require that spectra will be similar on small scales, including
the dissipation scale, see, e.g. \cite{gotoh2002}. As we see from the
plots, numerical convergence is absent for B06 model. For GS95
model the convergence is reached only at the dissipation scale. Higher-resolution
simulations are required to demonstrate convergence in the inertial range.}
\label{converg}
\end{figure}

Since we would like to use this review to illustrate the resolution
study argument we used a variety of resolution, dissipation and
driving schemes. There are four schemes, presented in Table~\ref{experiments1},
and used in simulations R1-3, R4-5, R6-7 and R8-9. In some of
the simulations the resolution in the direction parallel
to the mean magnetic field, $n_x$, was
reduced by a factor compared to perpendicular resolution.
This was deemed possible due to an empirically known lack of
energy in the parallel direction in $k$-space and has been
used before \cite{muller2005}. The R4-5 and R8-9
groups of simulations were fully resolved in parallel direction.
One would expect that roughly the same resolution
will be required in parallel and perpendicular direction
\cite{BL09a}. In all simulation groups time step was strictly
inversely proportional to the resolution, so that we can utilize
the scaling argument.

Driving had a constant energy injection rate for all simulations
except R6-7, which had fully stochastic driving. All simulations
except R8-9 had Els\"asser driving, while R8-9 had velocity driving.
All simulations were well-resolved and R6-7 were over resolved
by a factor of 1.6 in scale (a factor of 2 in Re).
The anisotropy of driving was that of a box, while injection rate
was chosen so that the amplitude was around unity on outer scale,
this roughly corresponds to critical balance on outer scale.
Indeed, as we will show in subsequent section, since anisotropy
constant is smaller than unity, our driving with $\lambda\sim \Lambda\sim 1$
and $\delta w\sim 1$ on outer scale is somewhat over-critical,
so $\Lambda$ decreases after driving scale to satisfy uncertainty
relation (see Fig.~\ref{anisotropy}). This is good for maintaining critical balance over
wide range of scales as it eliminates possibility for weak turbulence.

In presenting four groups of simulations, with different
geometries of elementary cell, different dissipation terms
and different driving, our intention is to show
that the scaling argument works irrespective of numerical effects,
but rather relies on scale separation and the assumption of
universal scaling. Simulations R1-3 are the same
as those presented in \cite{B11}.

\subsection{Resolution Study for Balanced Spectra}
\label{resolsub}

Fig.~\ref{converg} presents a resolution study all simulations. The upper
rows assume B06 scaling, while the bottom rows assume GS95 scaling.
Reasonable convergence on small scales was achieved only for GS95 scaling.
The normalized amplitude at the dissipation scale for two upper rows of plots
systematically goes down with resolution, suggesting that $-3/2$ is
not an asymptotic scaling. The flat part of the normalized spectrum on R1-3
plots was fit to obtain Kolmogorov constant of $C_{KA}=3.27\pm 0.07$
which was reported in \cite{B11}. The total Kolmogorov constant
for both Alfv\'en and slow mode in the above paper
was estimated as $C_K=4.2\pm0.2$ for the case of isotropically
driven turbulence with zero mean field, where the energy ratio of slow and Alfv\'en mode
$C_s$ is between 1 and 1.3. This larger
value $C_K=C_{KA}(1+C_s)^{1/3}$ is due to slow mode being passively
advected and not contributing to nonlinearity. The measurement of $C_{KA}$ had relied
on an assumption that the region around $k \eta\approx 0.07$ represent asymptotic
regime. Recently, we performed simulations with resolution up to $4096^3$, which also
confirmed the $-5/3$ spectrum \cite{B14a}. Furthermore, it appears from these 
simulations that the residual energy, $E_B-E_v$ have the same spectral slope
as the total energy, i.e. there is a constant fraction of residual energy in
the inertial range. This fraction was measured in \cite{B14a} to be around $0.15$.
Previously, the most popular model \cite{muller2005} suggested that the spectrum
of the residual energy follows $k^{-2}$ scaling, which is problematic both conceptually
and theoretically. We confirmed that the residual energy is a fraction of the total energy
in the inertial range and made explanations suggesting different scalings
for magnetic and kinetic energies unnecessary.  

\begin{figure}[t]
\includegraphics[width=0.44\columnwidth]{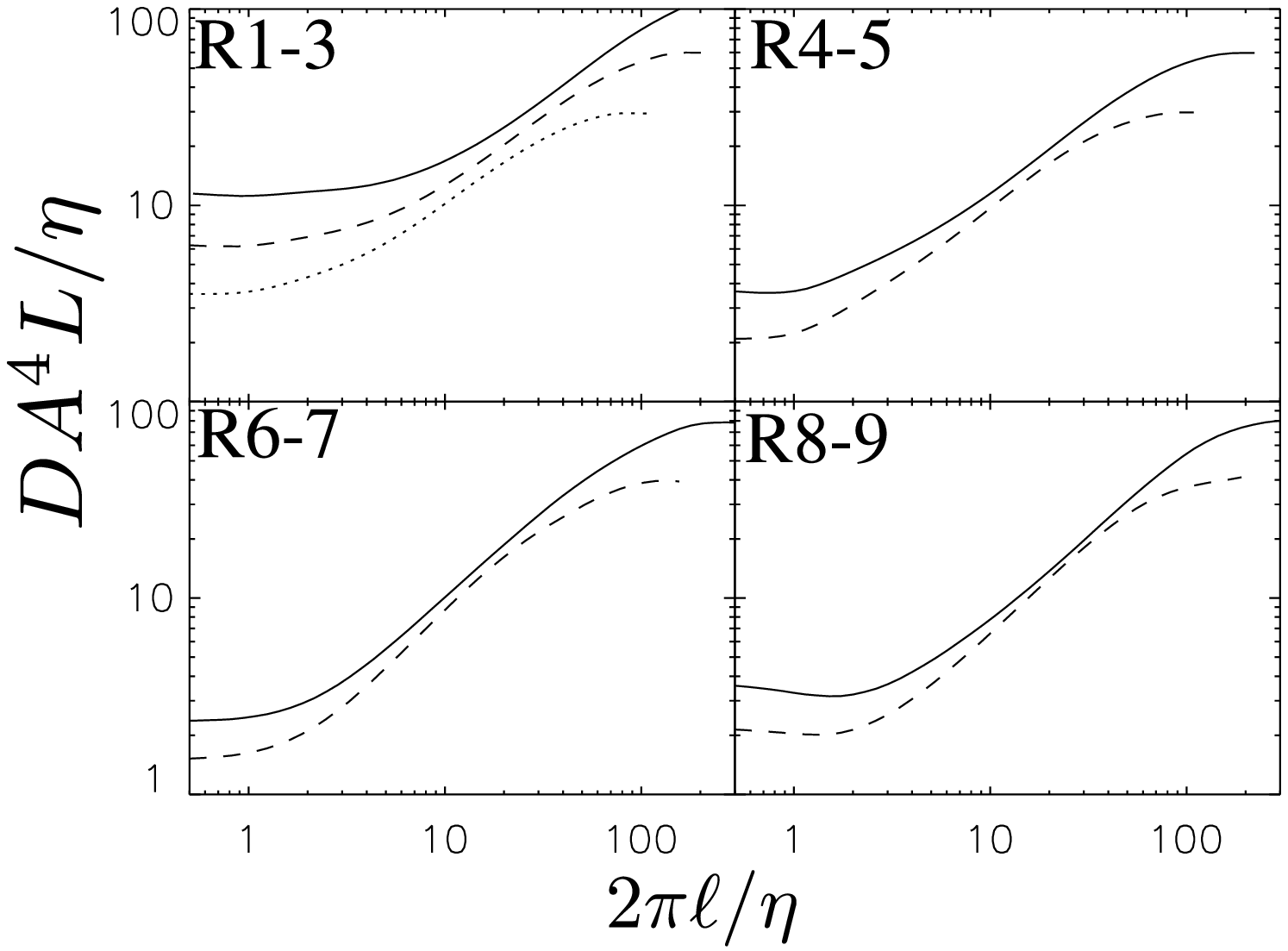}
\includegraphics[width=0.56\columnwidth]{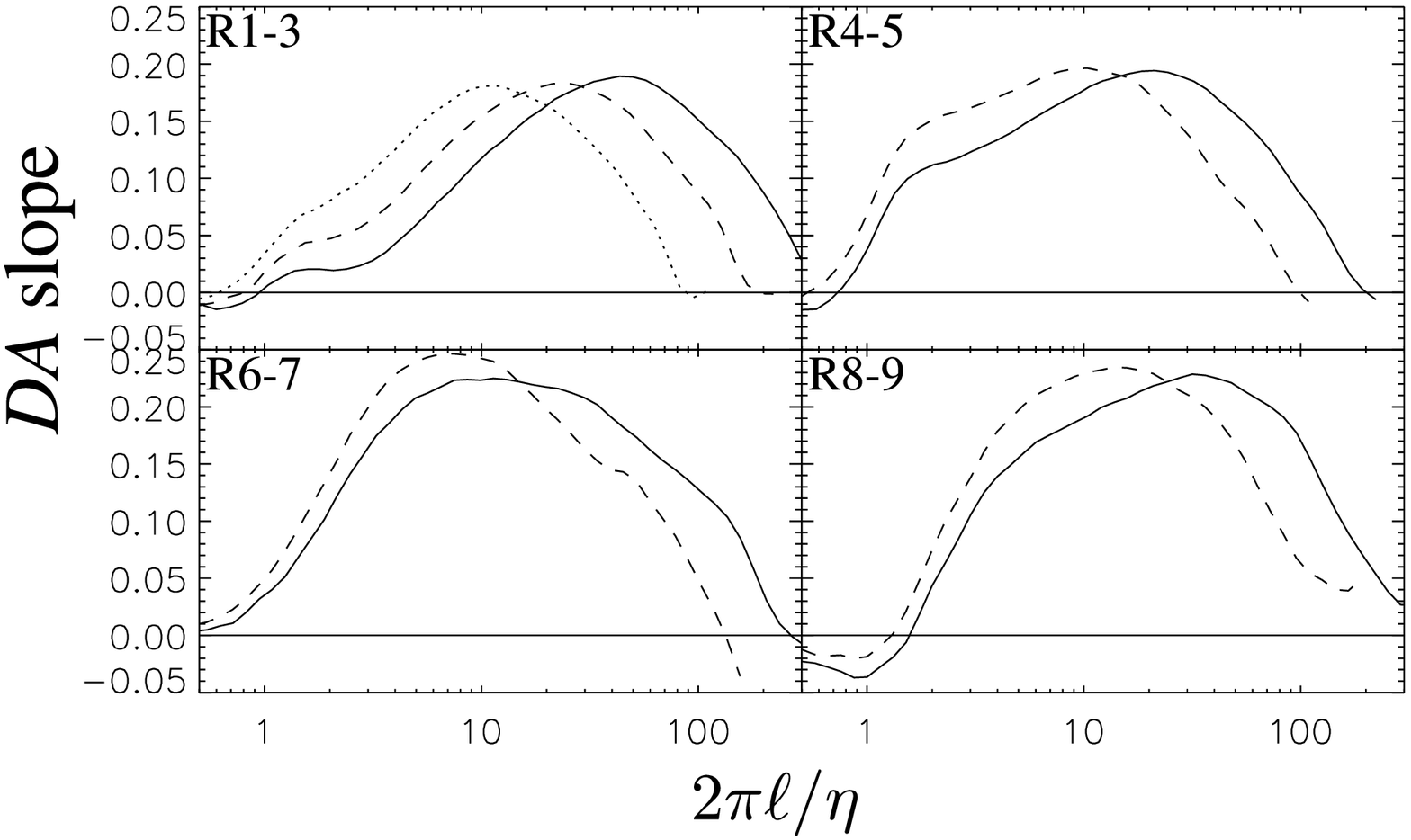}
\caption{Left: resolution study for ``dynamic alignment``, assuming B06 scaling.
  Both axis are dimensionless, solid is higher resolution and dashed is lower resolution.
  Convergence is absent for all simulations. This suggests that $l^{0.25}$
  is not a universal scaling for alignment. Right: $DA$ slope, defined as $l/{DA} \partial {DA}/\partial l$,
  solid is higher resolution and dashed is lower resolution. Dynamic alignment slope does not converge and has a tendency
  of becoming smaller in higher-resolution simulations. 
  This may indicate that the asymptotic alignment slope is zero,
  which will correspond to the GS95 model.}
\label{align_res}
\end{figure}

\subsection{Dynamic Alignment -- Theories vs Measurements}
\label{align}

Recent simulations, as we discussed earlier, support GS95 model and
therefore it can be considered correct in the zeroth approximation. However, we are
far from believing that we understand all the effects of MHD turbulence. For instance,
it is not clear how different alignment effects that we considered in \S \ref{resolsub} 
may affect the basic properties of MHD turbulence at the limited range of scales
when they exhibit scale-dependent properties.

An attempt to construct a model for such a behavior taking into account $DA$ was done in
\cite{boldyrev2005}. There it was proposed that ${\bf w}^+$ and ${\bf w}^-$ eddies
are systematically aligned and therefore, GS95 model should be amended
and the inertial range scaling should be modified. 
As we discussed earlier, this suggestion is not supported by either resolution studies
or studies of the alignment/polarization effects that we performed. For instance, 
the original alignment idea was investigated numerically in
in \cite{BL06} and no significant alignment was found for the averaged
angle between ${\bf w}^+$ and ${\bf w}^-$,
$AA=\langle|\delta {\bf w}^+_\lambda\times \delta {\bf w}^-_\lambda|/|\delta
{\bf w}^+_\lambda||\delta {\bf w}^-_\lambda|\rangle$, but when this angle
was weighted with the amplitude $PI=\langle|\delta {\bf
  w}^+_\lambda\times \delta {\bf w}^-_\lambda|\rangle/\langle|\delta {\bf
  w}^+_\lambda||\delta {\bf w}^-_\lambda|\rangle$, some alignment was found.
Later \cite{boldyrev2006} proposed the alignment between ${\bf v}$
and ${\bf b}$ and \cite{mason2006} suggested a particular
amplitude-weighted measure, $DA=\langle|\delta {\bf
  v}_\lambda\times \delta {\bf b}_\lambda|\rangle/\langle|\delta {\bf
  v}_\lambda||\delta {\bf b}_\lambda|\rangle$. We note that $DA$ is similar
to PI but contain two effects: alignment and local imbalance. The
latter could be measured with
$IM=\langle |\delta (w^+_\lambda)^2- \delta (w^-_\lambda)^2|\rangle /\langle \delta
(w^+_\lambda)^2+ \delta (w^-_\lambda)^2\rangle$, \cite{BL09b}.

In this section we check the assertion of \cite{boldyrev2005,boldyrev2006}
that alignment depends on scale as $\lambda^{1/4}$, by using $DA$ which is, by
some reason, favored by aforementioned group. We did a resolution study of $DA$,
assuming suggested scaling, which is presented on Fig.~\ref{align_res}.
Convergence was absent in all simulations.
It appears that the claims of \cite{boldyrev2006} were not substantiated by a proper resolution study.
In general, a result from a single isolated simulation could be easily contaminated by the effects
of outer scale, since it is not known a-priori how local MHD turbulence is
and what resolution is sufficient to get rid of such effects. On the contrary,
the resolution study offers a systematic approach to this problem.

Fig.~\ref{align_res} also shows ``dynamic alignment'' slope for all simulations. Although
there similar to the previous plot there is no convergence, it is interesting
to note that alignment slope decreases with resolution. This suggests
that most likely the asymptotic state for the alignment slope is zero,
i.e. alignment is scale-independent and GS95 model is
recovered. Also, alignment from simulations R1-5 seems to indicate
that the maximum of the alignment slope is tied to the outer scale,
therefore alignment is a transitional effect.

\begin{figure}[t]
\begin{center}
\includegraphics[width=0.7\columnwidth]{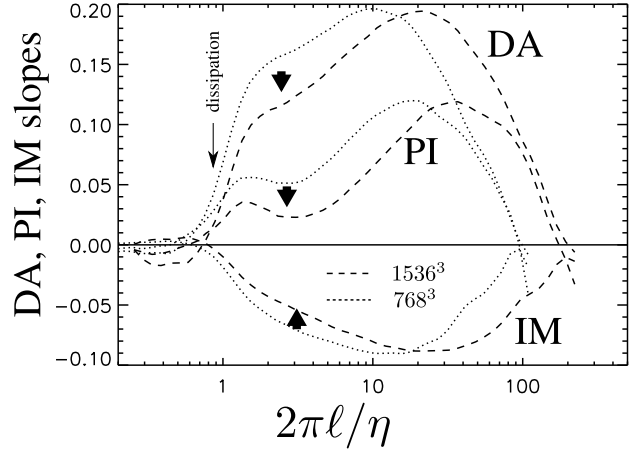}
\end{center}
\caption{Slopes of several alignment measures vs scale in R4-5 (for definitions see the text).
  Each measure follows its own scaling, however there are indications that they are all
  tied to the outer scale, due to the maximum of alignment being a fraction
  of the outer scale, which is an indication that their scale-dependency is of transient nature.}
\label{align_sl_all}
\end{figure}

In our earlier studies \cite{BL06,BL09b} we measured several types
of alignment and found no evidence that all alignment measures
follow the same scaling, see, e.g., Fig.~\ref{align_sl_all}.
As one alignment measure, PI, has been already known to be well scale-dependent \cite{BL06} prior
to $DA$, it appears that a particular measure of the alignment in \cite{mason2006} was
hand-picked for being most scale-dependent and no thorough explanation
was given why it was preferred. 

We are not aware of any convincing physical argumentation explaining why
alignment should be a power-law of scale. \cite{boldyrev2006}
argues that alignment will tend to increase, but
will be bounded by field wandering, i.e. the alignment on each scale will be
created independently of other scales and will be proportional to the relative
perturbation amplitude $\delta B/B$. But this violates two-parametric
symmetry of RMHD equations mentioned above, which suggests that
field wandering can not destroy alignment or imbalance. Indeed, a perfectly
aligned state, e.g., with $\delta {\bf w}^-=0$ is a precise solution of
MHD equations and it is not destroyed by its own field wandering.
The alignment measured in simulations of strong MHD
turbulence with different values of $\delta B_L/B_0$ showed very little or
no dependence on this parameter \cite{BL09b}. 

Some alignment measures are scale-dependent over about
one order of magnitude in scale. The origin of this scale-dependency
was not yet clearly identified. However, the most plausible explanation
is the combinations of two facts: a) MHD turbulence is less local than
hydro turbulence \cite{BL09b,BL10,B11} and b) the driving used in MHD turbulence does not
particularly well reproduce the statistical properties of the inertial range.
Thus transition to asymptotic statistics of the inertial range takes larger scale
separation than in the hydrodynamic case.

The contribution to energy flux from different $k$
wavebands is important to understand, since most cascade models
assume locality, or rather to say the very term ''cascade'' assumes
locality. An analytical upper bound on locality suggests
that the width of the energy transfer window can scale as $C_K^{9/4}$
\cite{B12a}, however, in practice turbulence
can be more local. The observation of \cite{BL09b} that MHD simulations
normally lack bottleneck effect, even with high-order dissipation,
while hydrodynamic simulations always have bottleneck, which is
especially dramatic with high-order dissipation, is consistent with
above conjecture on locality, since bottleneck effect relies
on locality of energy transfer.
As locality constraint depends on the efficiency
of the energy transfer, so that the efficient energy transfer must be local,
while inefficient one could be nonlocal \cite{BL10,B11,B12a}.
As we observe larger $C_K$ in MHD turbulence compared to hydrodynamic
turbulence, the former could be less local than the latter, which
is consistent with our earlier findings.

\subsection{Dynamic Alignment -- Relation to Spectra}
The papers \cite{boldyrev2006, mason2006} and subsequent papers assert that the particular
measure of alignment, $DA$ in our notation, is weakening interaction scale-dependently, so that
the energy spectral slope is modified. In particular, the above papers claim that if $DA\sim \lambda^{\alpha}$,
then the spectrum $E(k)\sim k^{-5/3+2/3\alpha}$ which, in the case of $\alpha=1/4$ will result in $E(k)\sim k^{-3/2}$.
Numerics does not show flat spectra if one compensates $E(k)$ slope with $2/3$ of $DA$ slope, however.
Let us critically examine the claim $E(k)\sim k^{-5/3+2/3\alpha}$ from the theoretical viewpoint.
The exact relation describing energy flux through scales is given by Eq.~\ref{PP}. Let us analyze
this statistical average for a ``+'' component at a particular value of $l$:
$\langle\delta w^{-}_{l\|} (\delta w^{+}_{l})^2 \rangle$.
Indeed, it appears that the anti-correlation of $\delta w^{-}_{l}$ and $\delta w^{+}_{l}$ could result
in a reduction of the above statistical average, as \cite{boldyrev2006} seems to allege. There are
three arguments against this, however. 

Firstly, the $DA$ does not describe such an anti-correlation, and something
different, such as $IM$ should be taken instead. So, the assumption of the interaction weakening rely on the
claim that alignment measures scale similarly. As we see from Fig.~\ref{align_sl_all}, the slopes of $DA$ and
$IM$ are quite different and if $DA$ reaches the maximum slope of $0.2$, the $IM$ only reaches the the maximum
slope of $0.09$ and this value does not increase with resolution. This is far from $0.25$, required in
B06 model. The numerical analysis of \cite{mason2006} and subsequent papers, however,
dealt exclusively with $DA$ and the earlier publication \cite{BL06} that reported several different
alignment measures, which scaled differently, was ignored and the strong claim
of interaction weakening was made nevertheless. However, with present numerics reported so far, even assuming
an anti-correlation argument, one can not deduce that the interaction is weakened by a factor of $l^{1/4}$.

Secondly, the $DA$ is based on a second order measure, while $\langle\delta w^{-}_{l\|} (\delta w^{+}_{l})^2 \rangle$
is third-order. We also know that $AA$ which is based on zeroth order ($\sin$ of the alignment angle) is very weakly
scale-dependent \cite{BL06}, we can extrapolate to ''third-order alignment`` having $\sim \lambda^{3 \alpha/2}$
dependence and the spectral slope will be $E(k)\sim k^{-5/3+\alpha}$. This is actually more numerically consistent
with the data than $E(k)\sim k^{-5/3+2/3\alpha}$, because $\alpha$ is typically below 0.2 and the spectral slope
is often flatter than $-3/2$ close to the driving scale.

Thirdly, and most importantly, there is no rigorous argumentation that could suggest that
the discussed anti-correlation necessarily reduces the above statistical average. Indeed, the $\delta w^{-}_{l\|}$
is a signed quantity, and so is the whole expression under the statistical average. Therefore, the value of
the statistical average is not necessarily related to the RMS value of the expression, but rather depend
on the {\it skewness} of the PDF of the expression. This is most obviously indicated by the Fig.~\ref{sfs}
where the ratio of unsigned to signed statistical average is about 10. In fact, this ratio could be arbitrarily
large, e.g. in weak MHD turbulence, where taking larger $B_0$ will result in decreased energy rates,
the above PDF becoming closer to Gaussian and its skewness going to zero. It is only
the GS95 similarity hypothesis for the case of strong MHD turbulence, which is similar
to Kolmogorov hypothesis, that asserts that the skewness is independent on scale, allows us to derive
the $k^{-5/3}$ spectrum. When one wants to explore a different similarity relations,
as \cite{boldyrev2005, boldyrev2006} did, it is necessary to argue in favor of the scale-independent skewness again.
In MHD turbulence, which has fluctuations of the imbalance ratios, it is not clear what self-similarity prescription
should be adopted. In a more detailed treatment of the imbalanced turbulence below we argue that is it
very likely that the skewness of $\delta w^{-}_{l\|} (\delta w^{+}_{l})^2$ and $\delta w^{+}_{l\|} (\delta w^{-}_{l})^2$
could be very different in the imbalanced case, due to the fact that the stronger component is cascaded weakly,
i.e. $\langle\delta w^{-}_{l\|} (\delta w^{+}_{l})^2\rangle$ is not the constant fraction of
$\langle|\delta w^{-}_{l}| (\delta w^{+}_{l})^2\rangle$, but could be much smaller.

To summarize, the assertion that the interaction is weakened by the $DA$ factor is at best heuristic
and could be seriously questioned by both numerical data and theoretical argumentation.
Apart from this, we reiterate the arguments of previous sections that the numerical evidence
strongly suggests that $DA$ and other  alignment measures become constant in the inertial
range and that the asymptotic inertial-range scaling for MHD turbulence is closer to $-5/3$.

\begin{figure}[t]
\begin{center}
\includegraphics[width=0.8\columnwidth]{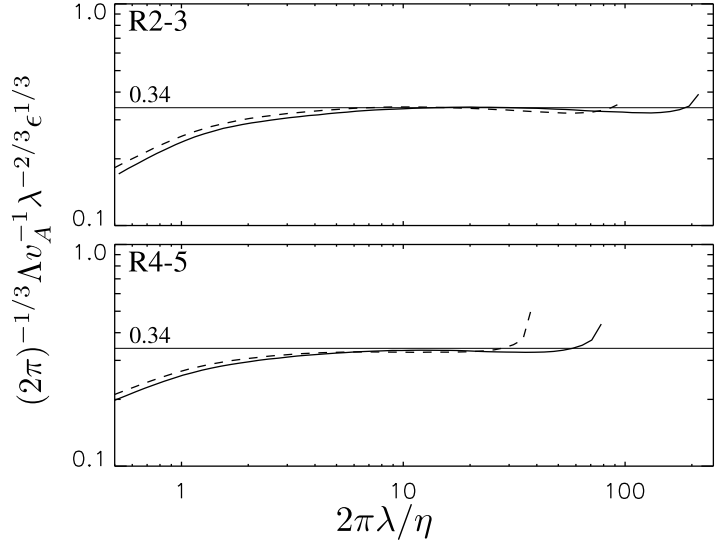}
\end{center}
\caption{The scaling study for anisotropy shows moderately good
convergence to a universal anisotropy $\Lambda=C_A v_A \lambda^{2/3} \epsilon^{-1/3}$ with
anisotropy constant $C_A$ of around 0.63.}
\label{anisotropy}
\end{figure}

\subsection{Anisotropy -- Balanced case}
In Section 3 we suggested that anisotropy
should be universal in the inertial range
and expressed as $\Lambda=C_A v_A \lambda^{2/3} \epsilon^{-1/3}$,
where $C_A$ is an anisotropy constant to be determined from the numerical
experiment or observation. Note, that both Alfv\'enic and slow modes
should have the same anisotropy. This is because they have the same
ratio of propagation to nonlinear timescales. Fig.~\ref{anisotropy} shows
anisotropy for the two best resolved groups R1-3 and R4-5. We used
a model independent method of minimum parallel structure function,
described in detail in \cite{BL09a}. Alternative definitions
of local mean field give comparable results, as long as they are reasonable.
From R1-3 we obtain $C_A=0.63$. Note, that the conventional definition of
critical balance involve the amplitude, rather than $(\epsilon\lambda)^{1/3}$,
so the constant in this classical formulation will be $C_A C_K^{1/2}\approx 1.1$,
which is closer to unity. Together with energy spectrum this is a full
description of universal axisymmetric two-dimensional spectrum of MHD
turbulence in the inertial range.

\subsection{Basic Properties of Balanced MHD turbulence}

In this review we argue that the properties of Alfv\'en and slow components
of MHD turbulence in the inertial range will be determined only by the Alfv\'en speed
$v_A$, dissipation rate $\epsilon$ and the scale of interest $\lambda$. The energy spectrum
and anisotropy of Alfv\'en mode will be expressed as
\begin{equation}
E(k)=C_K \epsilon^{2/3} k^{-5/3},
\end{equation}
\begin{equation}
\Lambda/\lambda=C_A v_A (\lambda \epsilon)^{-1/3},
\end{equation}
with $C_K=3.3$ and $C_A=0.63$. If the slow mode is present, its anisotropy will
be the same, and it will contribute to both energy and dissipation rate. Assuming
the ratio of slow to Alfv\'en energies between 1 and 1.3, the latter was observed in
statistically isotropic high resolution MHD simulation with no mean field, we can
use $C_K=4.2$ for the total energy spectrum \cite{B11}.

Anisotropy of MHD turbulence is an important property that affects such processes
as interaction with cosmic rays, see, e.g., \cite{yan2002}. Since cosmic ray
pressure in our Galaxy is of the same order as dynamic pressure, their importance
should not be underestimated. Another process affected is the three-dimensional
turbulent reconnection, see, e.g., \cite{Lazarian1999}.

Previous measurements of the energy slope relied on the
highest-resolution simulation and fitted the slope in the fixed $k$-range
close to the driving scale, typically between $k=5$ and $k=20$. We argue that such
a fit is unphysical unless a numerical convergence has been demonstrated.
We can plot the spectrum vs dimensionless $k\eta$ and if we clearly
see a converged dissipation range and a bottleneck range, we can assume that
larger scales, in terms of $k \eta$ represent inertial range.
In fitting fixed $k$-range at low $k$ we will never
get rid of the influence of the driving scale. In fitting a fixed $k \eta$
range, the effects of the driving will diminish with increasing resolution.

Since we still have trouble transitioning into the inertial range in 
large mean field simulations, for now it is impossible to 
demonstrate inertial range in statistically isotropic simulations
similar to once presented in \cite{muller2005}. This is because
we do not expect a universal power-law scaling in transAlfv\'enic
regime, due to the absence of appropriate symmetries and the
transitioning to subAlfv\'enic regime, where such scaling is possible,
will require some extra scale separation.
These two transitions require numerical resolution that is
even higher than the highest resolution presented in this paper and for now
seem computationally impossible.

Full compressible MHD equations contain extra degrees of freedom, which, in a weakly compressible case, entails the
additional cascade of the fast MHD mode, possibly of weak nature. Supersonic simulations with
moderate Mach numbers \cite{cho2003c} show that Alfv\'enic cascade is pretty resilient and is not much affected
by compressible motions. The models of the "universal" supersonic turbulence covering supersonic
large scales and effectively subsonic small scales are based mainly on simulations with limited resolution
and unlikely to hold true. This is further reinforced by the results of this paper which demonstrated that even
a much simpler case of sub-Alfv\'enic turbulence require fairly high resolutions to obtain an asymptotic scaling.

\section{Imbalanced MHD turbulence}
\label{imbal}
While hydrodynamic turbulence have only one energy cascade, the
incompressible MHD turbulence has two, due to the exact conservation of
the Els\"asser (oppositely going wave packets') ``energies''. This can
be also formulated as the conservation of total energy and
cross-helicity\footnote{The latter, $\int {\bf v}\cdot {\bf B}\, d^3x $
is a quantity conserved in the absence of dissipation.}.
The situation of zero total cross-helicity, which we considered in previous sections
has been called ``balanced'' turbulence as the amount of oppositely moving
wavepackets balance each other, the alternative being ``imbalanced''
turbulence.  Most of the above studies concentrated on the balanced
case, and, without exception, the GS95 model, which is the strong
cascading model with critical balance, can only be kept
self-consistent assuming balanced case.

The real MHD turbulence, however, is often imbalanced,
such as in situations when the mean magnetic field is present and we
have a strong localized source of perturbations. The perfect example
is the solar wind, where satellite measurements discovered
strong correlations between ${\bf v}$ and ${\bf B}$ since long time
ago. These correlations actually correspond to the imbalanced turbulence
with the dominant component propagating away from the Sun. If the mean magnetic
field of the Parker spiral is directed locally outwards the Sun then
the dominant component will be ${\bf w^-}$, otherwise it'll be ${\bf w^+}$.
\begin{figure}
\begin{center}
\includegraphics[width=0.7\columnwidth]{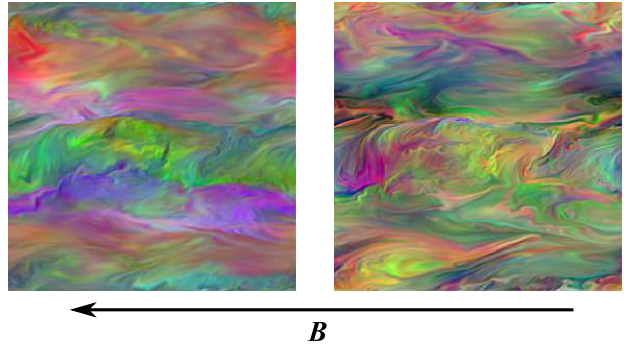}
\end{center}
\caption{The slices of ${\bf w}^+={\bf v}+{\bf B}/\sqrt{4\pi\rho}$ (left) and ${\bf w}^-={\bf v}-{\bf B}/\sqrt{4\pi\rho}$ (right) from the three-dimensional MHD
simulation with strong mean magnetic field and imbalance.}
\label{imb_slices}
\end{figure}

Certainly, we expect similar phenomena happen in the active galactic nuclei (AGN),
where the jet has a strong large mean magnetic field component and the perturbations
will propagate primarily away from the central engine, where they will be excited
by either Blandford-Znajek mechanism, for the inside jet, or by the motions of
the magnetic field footpoints, embedded into the turbulent accretion disk.
Another example is the interstellar medium (ISM) turbulence in spiral galaxies.
Indeed, in spiral galaxies, due to the action of the large-scale dynamo there is
a large-scale component of the magnetic field, spanning the radius of the disk itself.
The ISM turbulence, however, is inhomogeneous, due to the energy sources
for turbulence (supernovas and stellar winds) distributed unevenly in the disk.
This will create imbalanced turbulence, which might properties different from the
balanced one, which has implications for ISM heating, cosmic ray propagation
and many other physical processes in the ISM. 

Finally, from the theoretical viewpoint, it is impossible to fully
understand balanced turbulence by itself, if the more general imbalanced case is
not treated. This is due to the fact that turbulence is a
stochastic phenomena with all quantities fluctuating and every piece
of turbulence at any given time can have imbalance in it. In this
respect, while the mean-field Kolmogorov model can be expanded to
include fluctuations of the dissipation rate in the volume,
the mean field GS95 model can not.

Imbalanced turbulence, or ``turbulence with non-zero cross-helicity''
has been discussed long ago by a number of authors
\cite{dobrowolny1980,matthaeus1980,grappin1983,pouquet1988}.
This work testified that the non-zero cross-helicity modifies the turbulence.
Although these studies correctly reproduced separate cascades
for energy and cross-helicity, they were based on then-popular
models of MHD turbulence and later it became evident that these are problematic.
For example, the closure theory of isotropic MHD
turbulence \cite{pouquet1976}, which reproduced Iroshnikov-Kraichnan model can be
criticized on the basis that the ad-hoc term for ``relaxation of triple correlations'',
happen to be larger than real physical nonlinear interaction and makes MHD turbulence,
effectively, isotropic. Numerics, however, show that strong MHD turbulence
is locally anisotropic, as we demonstrated in previous sections.
Another class of models were based on so-called two-dimensional MHD
turbulence that, as we demonstrated in previous sections,
is unable to reproduce basic properties of the real three-dimensional
turbulence, such as strong interaction with critical balance.

\subsection{Theoretical considerations}
As we explain in the previous sections, the MHD cascade is primarily
perpendicular and as it proceeds to small scales, the applicability of
weak interaction breaks down, and Alfv\'enic turbulence becomes
strong. In this situation GS95
assumed that the frequency of the wavepacket can not be smaller than
the inverse lifetime of the wavepacket, estimated from nonlinear
interaction. In the GS95 closure model there is an explicit ad-hoc term
that allows for the increase of the wave frequency. Unlike
previous models this term is scale-dependent and is based on the
assumption of turbulence locality, i.e. that there is one characteristic
amplitude of perturbation pertaining to each scale and that this perturbation
determines the strength of the interaction and finally renormalization of
frequencies. However, as was realized as early as in the original GS95 paper
in the imbalanced case we have two characteristic amplitudes, $w^+,w^-$, and
the choice for frequency renormalization becomes unclear\footnote{We assume that
imbalanced turbulence is ``strong'' as long as the applicability
of weak Alfv\'enic turbulence breaks down. This requires that
at least one component is perturbed strongly. In the imbalanced
turbulence the amplitude of the dominant component is larger, so
that in the transition to strong regime the applicability
of weak cascading of the subdominant component breaks down first.}.
Any theory of strong imbalanced turbulence, must deal with this difficulty.

Let us first demonstrate that a straightforward generalization of GS95
for the imbalanced case does not work. If we assume that the frequency
renormalization for one wavepacket is determined by the shear rate
of the oppositely moving wavepacket, the wave with small amplitude
(say, $w^-$) may only weakly perturb large amplitude wave $w^+$ and
the frequency of cascaded $w^+$ will conserve. On the other hand,
$w^+$ may strongly perturb $w^-$ and $w^-$'s frequency will be
determined as $w^+_l/l$\footnote{Throughout this paper we assume that $w^+$ is
the larger-amplitude wave. This choice, however, is purely arbitrary and corresponds
to the choice of positive versus negative total cross-helicity.}.
This mismatch in frequencies creates an inconsistency in the paradigm of
scale-local cascade where both wavepackets must have both parallel and
perpendicular wavenumbers comparable. As the cascade proceeds to small
scales this mismatch only increases, making the cascade nonlocal and
inefficient. Such shutdown of the cascade on small scales is unacceptable,
since in the stationary case it must carry a constant energy flux for
both components. In order to deal with this fundamental difficulty,
one must assume something extra to the original GS95 critical balance. 

Currently there were several propositions how to deal with strong
anisotropic imbalanced MHD turbulence. In \cite{LGS07}, the authors
proposed that the parallel scale for both components is determined by
the shear rate of the stronger component. This model predicts the same anisotropy
for both components. In \cite{BL08} the authors
proposed a new formulation for critical balance for the stronger
component. In \cite{C08} an advection-diffusion model of cascading
was adopted, where advection was describing perpendicular cascade
and diffusion was describing the increase of frequencies. These three
models clearly state the difficulty described above and try to resolve
it with the new physical argumentation that goes beyond the original GS95 critical
balance. These three models smoothly transition to the balanced theory of GS95
in the limit of small imbalance. Several other models has been suggested,
advocating a different picture, in particular the influence of so-called
dynamic alignment. In \cite{PB09} the authors argued that the dynamic alignment
will effectively lead to the same nonlinear timescale for both components.
This has been criticized as grossly inconsistent with numerics \cite{BL09a, BL10}
and having no meaningful physical limit for large imbalances.

\subsection{Lithwick, Goldreich \& Sridhar (2007) model, LGS07}
LGS07 argue that the strong
wave $w^+$ is also cascaded strongly and its frequency is equal to
the frequency of the weak wave, i.e. the critical balance for strong wave
uses the amplitude of the strong wave itself ($w^+\L=v_A\l$).
In this case the anisotropies of the waves are identical.
The formulas for energy cascading are strong cascading formulas, i.e.
\begin{equation}
\epsilon^\mp=\frac{(w^\mp(\l))^2 w^\pm(\l)}{\l}.
\end{equation}
This lead to the prediction $w^+/w^-=\epsilon^+/\epsilon^-$.
In terms of energy spectra the model predicts
\begin{equation}
E_k^\pm=C_K(\epsilon^\pm)^{4/3}(\epsilon^\mp)^{-2/3}k^{5/3},
\end{equation}
where the Kolmogorov constant $C_K$ must be the same for the theory
to have a limit of standard balanced MHD turbulence.

\subsection{Beresnyak \& Lazarian (2008) model, BL08}
\begin{figure}[t]
\begin{center}
\includegraphics[width=0.6\textwidth]{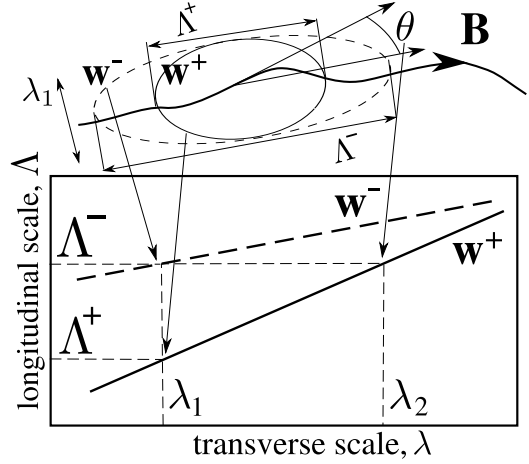}
\end{center}
\caption{Upper: a ${\bf w}^+$ wavepacket, produced
by cascading by ${\bf w}^-$ wavepacket is aligned with respect
to ${\bf w}^-$ wavepacket, but misaligned with respect
to the local mean field on scale $\lambda_1$, by the angle $\theta$.
Lower: the longitudinal scale $\L$ of the wavepackets,
as a function of their transverse scale, $\l$; $\L^+$, $\L^-$, $\l_1$, $\l_2$
are the notations used in this paper. From Beresnyak \& Lazarian model \cite{BL08}.}
\label{anis_cartoon}
\end{figure}

BL08a relaxes the assumption of local cascading for the strong component $w^+$,
while saying the $w^-$ is cascaded in a GS95-like way. In BL08a picture
the waves have different anisotropies (see Fig. \ref{anis_cartoon}) and the $w^+$ wave
actually have smaller anisotropy than $w^-$, which is opposite to what
a naive application of critical balance would predict.
The anisotropies of the waves are determined by
\begin{equation}
w^+(\l_1)\L^-(\l_1)=v_A\l_1,
\end{equation}
\begin{equation}
w^+(\l_2)\L^+(\l^*)=v_A\l_1,
\end{equation}
where $\l^*=\sqrt{\l_1\l_2}$, and the energy cascading
is determined by weak cascading of the dominant wave
and strong cascading of the subdominant wave:
\begin{equation}
\epsilon^+=\frac{(w^+(\l_2))^2 w^-(\l_1)}{\l_1}\cdot\frac{w^-(\l_1) \L^-(\l_1)}{v_A\l_1}\cdot f(\l_1/\l_2),
\end{equation}
\begin{equation}
\epsilon^-=\frac{(w^-(\l_1))^2 w^+(\l_1)}{\l_1}.
\end{equation}
One of the interesting properties of BL08a model is that, unlike LGS07
and C08, it does not produce self-similar (power-law) solutions when
turbulence is driven with the same anisotropy for $w^+$ and $w^-$
on the outer scale. BL08a, however, claim that, on sufficiently
small scales, the initial non-power-law solution will transit into
asymptotic power law solution that has
$\Lambda^-_0/\Lambda^+_0=\epsilon^+/\epsilon^-$
and $\lambda_2/\lambda_1=(\epsilon^+/\epsilon^-)^{3/2}$.
The range of scales for the transition region was not specified
by BL08a, but it was assumed that larger imbalance
will require larger transition region.

\subsection{Perez \& Boldyrev (2009) model, PB09}
Unlike the models described above PB09 employs dynamic alignment which decreases without
limit to smaller scales as $l^{1/4}$ and claims the $3/2$ spectral slope for both components.
In this respect it is similar to \cite{boldyrev2005,boldyrev2006}. It does, however,
a big step beyond these papers by claiming that alignment will effectively result
in the same nonlinear timescales for both components, which effectively lead to
$(w^+)^2/(w^-)^2=\epsilon^+/\epsilon^-$. It could be rephrased that PB09 predicts
turbulent viscosity which is equal for both components. It is not clear, however,
how this could be made consistent with the limit of large imbalances, where the weak
component will not be able to produce any sizable turbulent viscosity.

\begin{table}[t]
\caption{Three-Dimensional RMHD Imbalanced Simulations}
\begin{tabular*}{0.99\columnwidth}{@{\extracolsep{\fill}}l c c c c r}
    \hline\hline
Run  & Resolution & $f$ & Dissipation & $\epsilon^+/\epsilon^-$ & $(w^+)^2/(w^-)^2$   \\
   \hline
I1 &  $512\cdot 1024^2$ & $w^\pm$ & $-1.9\cdot10^{-4}k^2$ & 1.187 &  $1.35\pm 0.04$ \\
I2 &  $768^3$ & $w^\pm$ & $-6.8\cdot10^{-14}k^6$ & 1.187 &  $1.42 \pm 0.04$   \\
I3 &  $512\cdot 1024^2$ &$w^\pm$ & $-1.9\cdot10^{-4}k^2$ & 1.412 &  $1.88\pm 0.04$   \\
I4 &  $768^3$ & $w^\pm$ & $-6.8\cdot10^{-14}k^6$ &  1.412 & $1.98\pm 0.03$ \\
I5 &  $1024\cdot 1536^2$ & $w^\pm$ & $-1.5\cdot10^{-15}k^6$ &   2   &    $5.57\pm 0.08$  \\
I6 &  $1024\cdot 1536^2$ & $w^\pm$ & $-1.5\cdot10^{-15}k^6$ & 4.5   &  $45.2\pm1.5$   \\
   \hline

\end{tabular*}
  \label{imb_experiments}
\end{table}

\subsection{Imbalanced simulations}
Table~\ref{imb_experiments} summarizes our high-resolution experiments with imbalanced driving.
All experiments were conducted to reproduce stationary turbulence.
We started our high resolution simulations with earlier lower-resolution
runs that were evolved for a long time, typically hundreds Alfv\'enic times
and reached stationary state. The imbalanced runs were evolved for
longer times, up to 40 dynamical times, due to longer cascading timescales
for the stronger component. The energy injection rates were kept
constant in I1-6 and the fluctuating dissipation rate was within few
percent of the former.

\subsection{Nonlinear cascading and dissipation rate}
\begin{figure}[t]
\begin{center}
\includegraphics[width=0.65\columnwidth]{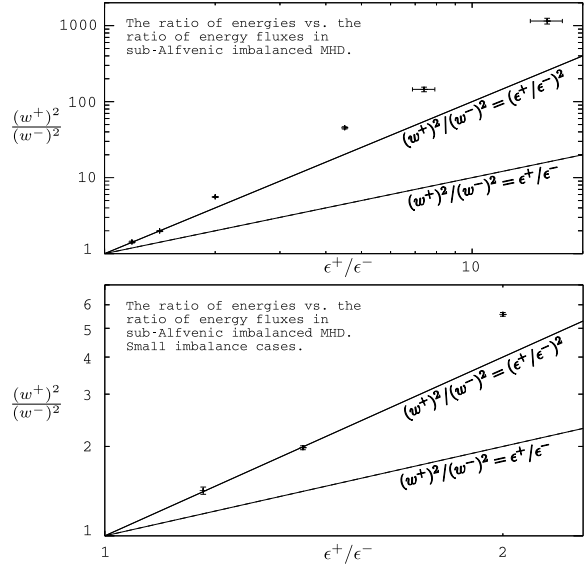}
\end{center}
\caption{Energy imbalances versus dissipation rate imbalance.
Lower panel shows a magnified portion of the
upper panel. Solid line: LGS07 prediction, dashed line: a formula from PB09,
this also is a prediction for purely viscous dissipation.
The point indicate measurements from simulations, where errorbars indicate
fluctuation in time. I1 and I3 are simulations with normal viscosity which have
slightly lower energy imbalance than I2 and I4.
This is an indication that in these simulations viscosity was affecting outer scales. 
Two high imbalance points are taken from \cite{BL09a}. 
For a fixed dissipation ratio the energy imbalance has
a tendency to only {\it increase} with resolution.}
\label{dissip}
\end{figure}

Compared to spectral slopes, dissipation rates are robust quantities that require
much smaller dynamical range and resolution to converge. Fig. \ref{dissip} shows
energy imbalance $(w^+)^2/(w^-)^2$ versus dissipation rate imbalance $\epsilon^+/\epsilon^-$ for simulations I2, I4, I5 and I6. We also use two data points from
our earlier simulations with large imbalances, A7 and A5 from BL09a.
I1 and I3 are simulations with normal viscosity similar to I2 and I4.
They show slightly less energy imbalances than I2 and I4.
We see that most data points are above the prediction of LGS07, which is consistent with BL08.
In other words, numerics strongly suggest that 
\begin{equation}
\frac{(w^+)^2}{(w^-)^2}\geq \left(\frac{\epsilon^+}{\epsilon^-}\right)^2.
\end{equation} 
Although there is a tentative correspondence between LGS07
and the data for small degrees of imbalance, the
deviations for large imbalances are significant.
The important lesson, however, that in the case of small imbalances
the cascading smoothly transition to the balanced case, i.e. the prediction of GS95 model.
This is an important verification that the exactly balanced case is not a special case, in a sense.

In the case of strong imbalance it suggests that the strong component cascading
rate is smaller than what is expected from strong cascading.
As to PB09 prediction, it is inconsistent with data
for all degrees of imbalance including those with small imbalance
and normal viscosity, i.e. I1 and I3.

\subsection{Imbalanced Spectra}
\begin{figure}[t]
\begin{center}
\includegraphics[width=\columnwidth]{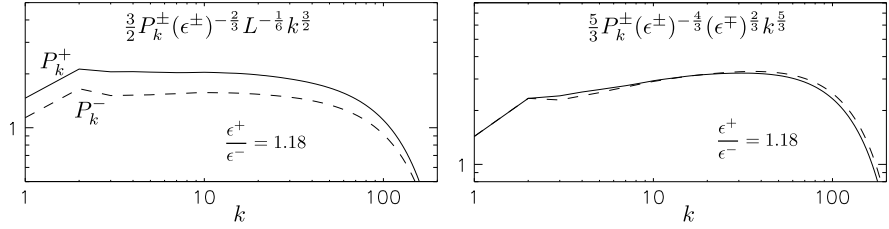}
\end{center}
\caption{Energy spectra for $w^+$ (solid) and $w^-$ (dashed) from simulation I2, compensated by
factors that correspond to PB09 (upper panel) and LGS07 (lower panel). Either theory is
confirmed for this low-imbalanced case if the spectra for $w^+$ and $w^-$ collapse
onto the same curve. We see that the collapse is much better for the LGS07 model.}
\label{imb2models}
\end{figure}

Fig.~\ref{imb2models} shows spectra from low-imbalance simulation I2,
compensated by the predictions of PB09 and LGS07.
We see that the collapse of two curves for $w^+$ and $w^-$ is much better for the LGS07 model,
however the spectral slope is much closer to $-3/2$ than to $-5/3$. The issue of spectral slope
was discussed in previous section with respect to the balanced simulations. We were arguing that
MHD cascade is less local than hydro cascade and is being influenced by driving on a larger
range of scales, more importantly the statistical properties of driving is different than
asymptotic regime of MHD cascade, which results in a transition range of scales of about one
order of magnitude. We expect the same effect to operate in the imbalanced case. Indeed, if we
neglect the part of the spectrum with k between 2 and 20, the spectrum could be considered
flat on the lower panel of Fig.~\ref{imb2models}. In this deviation of spectral slope from
$-5/3$ we do not see any significant differences between the balanced case, which was discussed
extensively in the previous sections and the low-imbalance case. 

\begin{figure}[t]
\includegraphics[width=0.99\columnwidth]{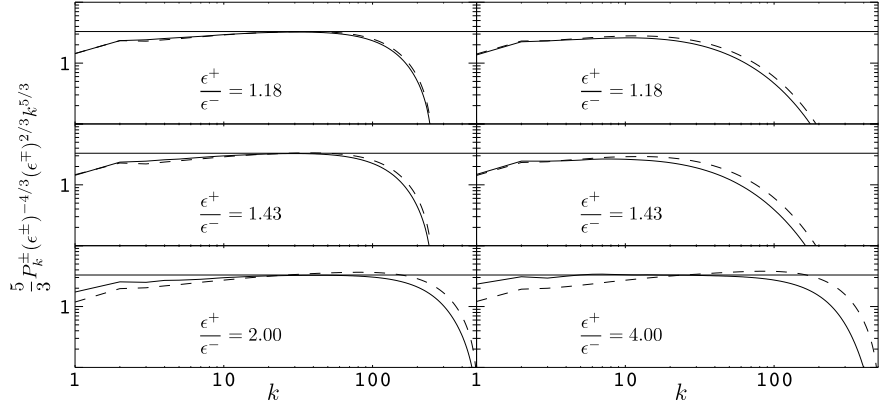}
\caption{Energy spectra for $w^+$ (solid) and $w^-$ (dashed) for simulations I1-I6,
compensated by factors that correspond to LGS07. The thin solid line corresponds to Kolmogorov
constant for Alfv\'enic turbulence $C_{KA}=3.27$. The factor $5/3$ is introduced due to
the difference between $P_k$ and $E_k$.}
\label{imb_spectra}
\end{figure}

Fig.~\ref{imb_spectra} shows spectra from all I1-6 simulations, compensated by the prediction of LGS07.
For lower imbalances the collapse is reasonably good and become progressively worse for
larger imbalances. This deviation, however, does not fully follow the prediction of the
asymptotic power-law solutions from BL08, which will predict that the solid curve will
go above $C_{KA}$ and the dashed curve -- below it. This is possibly explained by the fact
that asymptotic power law solutions were not reached in these limited resolution
experiments, this is also observed for anisotropies which we consider in the next section.

\subsection{Imbalanced Anisotropies}
We measured parallel and perpendicular structure functions in simulations I1-I6
in order to quantify anisotropies of eddies. The perpendicular structure function was defined
above. In the RMHD case which physically correspond to the case of very strong mean field
the perpendicular structure function must be calculated with respect to the global mean field.
The same is not true for the parallel structure function. Indeed, measuring
parallel SF with respect to the global field will destroy scale-dependent
anisotropy, even in the case of very strong field. If we have
$\delta B_L/B_0 \ll 1$, the field line wandering will be of the order
of $B_0/\delta B_L$, while the GS95 anisotropy on the scale $l$ will be
much higher, $\sim B_0/\delta B_l$, by a factor of $B_L/B_l$.
The direction of the mean field will deviate from the direction
of the local field by the angle which is much larger than the angle
of GS95 anisotropy. This will result in an incorrect estimation of the
parallel structure function which will be contaminated by contribution
from perpendicular direction. Therefore, one must measure anisotropy
with respect to local mean field, as was realized in \cite{cho2000,CLV02a,BL09b}.

For the parallel structure function we will use the model-independent method
suggested in \cite{BL09b} or ``minimum method'', namely
\begin{equation}
SF^2_{\|}(\L)=\min_\l \langle(w^\pm({\bf r}-\L{\bf b}_\l/b_\l)-w^\pm({\bf r})  )^2\rangle_{\bf r}.
\end{equation}
Where ${\bf b_\l}$ is the magnetic field smoothed on scale $\l$ with Gaussian kernel.
It turns out that this method gives very close results to the previously suggested methods of choosing
the local mean field, most prominently in the balanced case. We choose this method as it does not
contain any arbitrary assumptions as previous methods.  

As long as we know both parallel and perpendicular structure functions, the mapping $\L(\l)$
is obtained from the equation $SF^2_{\|}(w^\pm,\L)=SF^2_{\perp}(w^\pm,\l)$. Physically this
correspond to measurement of the parallel eddy size $\L$, whose energy is concentrated on
scales $\l$. 

\begin{figure}[t]
\begin{center}
\includegraphics[width=0.9\columnwidth]{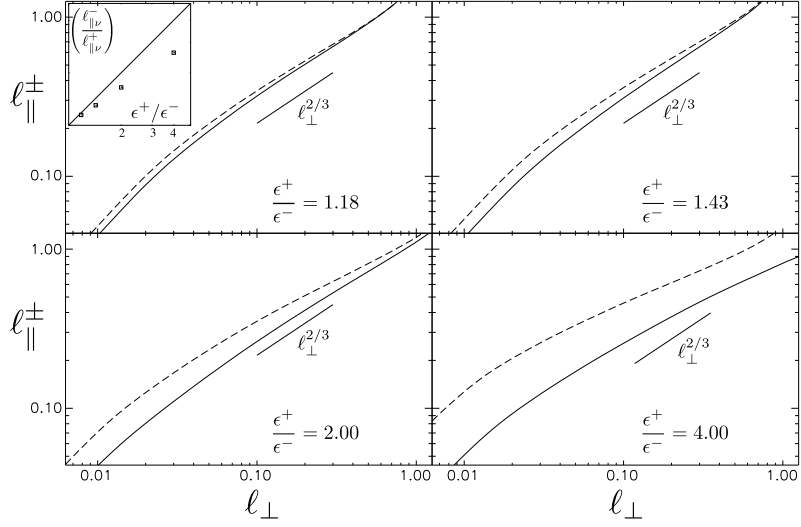}
\end{center}
\caption{Anisotropies for $w^+$ (solid) and $w^-$ (dashed), simulations I1-I6. The relation
between parallel scale $\Lambda$ and perpendicular scale $\lambda$ is obtained by second
order structure functions, as explained in the text. The small upper inset shows the ratio
of anisotropies on smallest scales vs the prediction of BL08 for the asymptotic power-law
solution, which is $\epsilon^+/\epsilon^-$.}
\label{imb_anis}
\end{figure}

Fig.~\ref{imb_anis} shows anisotropies for I1-6 simulations.
All simulations were driven by the same anisotropies on the outer scale, which is unfavorable
for obtaining the asymptotic power law solutions of BL08, which have an anisotropy ratio which
is constant through scales and equal to $\epsilon^+/\epsilon^-$. It is, however, favorable to the
LGS07 model, which predicts the same $w^+$ and $w^-$ anisotropies for all scales. Therefore, these
simulations are a sensitive test between LGS07 and BL08 models, both of which are roughly consistent
in terms of energy ratios and spectra for small imbalances. If LGS07 was true, starting with the same
anisotropies on outer scale, this should be preserved by the cascade on smaller scales, but this is not what is
observed on Fig.~\ref{imb_anis}, where anisotropies start to diverge on smaller scales.
The ratio of anisotropies is roughly consistent with BL08 asymptotic power-law
solutions for small imbalances and falls short for larger imbalances. This is explained by the fact
that it is harder to get to the asymptotic power-law solutions for larger imbalances, as was also
observed for the case of power spectra.

\section{Compressibility in MHD turbulence}

Our discussion so far was centered at the incompressible MHD turbulence. From the astrophysical point of view compressibility
is an essential property that cannot be ignored. This calls for studies to what extend our earlier description survives
in realistic set ups and what additional properties are gained by compressible MHD turbulence.

Kolmogorov turbulence is known to be applicable to compressible non-magnetized fluids and therefore one should expect
that some properties of GS95 model should persist at least for low Mach number magnetic turbulence. At the same time,
new modes are excited in MHD in the presence of compressibility. In particular, if MHD turbulence in the incompressible
limit can be decomposed into Alfv\'en and pseudo-Alfv\'en modes, in the case of compressible MHD turbulence, three modes,
namely, Alfv\'en, slow and fast are present. While the pseudo-Alfv\'en modes are a limiting case of the slow modes for compressibility
going to zero, the fast modes present a new type of motion intrinsic for compressible media\footnote{In the limiting case of 
compressibility going to zero, the fast modes are sound waves with phase speed going to infinity.}.

\subsection{Decomposition into fundamental modes}

\begin{figure*}[t] 
\begin{center}
\includegraphics[width=0.8\textwidth]{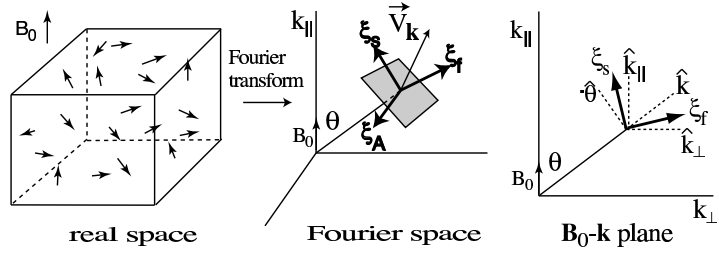}
\end{center}
 \caption{Graphical representation of the mode separation method. We separate
the Alfv\'{e}n, slow and fast modes by the projection of the velocity Fourier
component ${\bf v}_k$ on the bases ${\bf \hat\xi}_A$, ${\bf \hat\xi}_s$ and
${\bf \hat\xi}_f$, respectively. From CL03 \label{fig:separation}}
\end{figure*}

The original procedure of decomposition of MHD simulations into different modes was proposed by Cho \& Lazarian (\cite{Cho2002a, cho2003c} 
henceforth CL02, CL03, respectively). 
Unlike earlier discussions which dealt with small perturbations the aforementioned papers demonstrated the
decomposition of the transAlfv\'enic turbulence, i.e. the turbulence with substantial amplitudes. The procedure of decomposition is
performed in the Fourier space by a simple projection of the velocity Fourier
components $\hat{\bf {u}}$ on the direction of the displacement vector for each
mode (see Fig.~\ref{fig:separation}).  The directions of the displacement vectors
${\bf \hat\xi}_s$, ${\bf \hat\xi}_f$, and ${\bf \hat\xi}_A$ corresponding to the
slow mode, fast and Alfv\'{e}n modes, respectively, are defined by their unit
vectors
\begin{equation}
 {\bf \hat\xi}_s \propto (-1 + \alpha - \sqrt{D}) k_\parallel {\bf \hat{k}_\parallel} + (1 + \alpha - \sqrt{D}) k_\perp {\bf \hat{k}_\perp} \, ,
\end{equation}
\begin{equation}
 {\bf \hat\xi}_f \propto (-1 + \alpha + \sqrt{D}) k_\parallel {\bf \hat{k}_\parallel} + (1 + \alpha + \sqrt{D}) k_\perp {\bf \hat{k}_\perp} \, ,
\end{equation}
\begin{equation}
 {\bf \hat\xi}_A = - {\bf \hat\varphi} = {\bf \hat{k}}_\perp \times {\bf \hat{k}}_\parallel \, ,
\end{equation}
where ${\bf k}_\parallel$ and ${\bf k}_\perp$ are the parallel and perpendicular
to ${\bf B}_\mathrm{ext}$ components of wave vector, respectively, $D = (1 +
\alpha)^2 - 4 \alpha \cos^2 \theta$, $\alpha = a^2 / V_A^2$, $\theta$ is the
angle between ${\bf k}$ and ${\bf B}_\mathrm{ext}$, and ${\bf \hat\varphi}$ is
the azimuthal basis in the spherical polar coordinate system.  The Fourier
components of each mode can be directly used to calculate spectra.  For other
measures, such as structure functions, transforms back to the real space were
used.

The results of CL02 and CL03 revealed several important properties of MHD turbulence.
For the cases studied, they revealed that GS95
scaling is valid for {\it Alfv\'en modes}:
$$
   \mbox{ Alfv\'{e}n:~}  E^A(k)  \propto k^{-5/3}, 
                        ~~~k_{\|} \propto k_{\perp}^{2/3}. 
$$
{\it Slow modes} also follows the GS95 model for both
high $\beta$ and mildly supersonic low $\beta$ cases:
$$
   \mbox{ Slow:~~~}   E^s(k)  \propto k^{-5/3}, 
                        ~~~k_{\|} \propto k_{\perp}^{2/3}.  
$$
For the highly supersonic low $\beta$ case, the kinetic energy spectrum of 
slow modes tends to be steeper, which may be related
to the formation of shocks.\\
{\it Fast mode} spectra are compatible with
acoustic turbulence scaling relations:
$$
   \mbox{ Fast:~~~}   E^f(k)  \propto k^{-3/2}, 
                        ~\mbox{isotropic spectrum}.   
$$
The super-Alfv\'enic turbulence simulations suggested that 
the picture above was true at sufficiently small scales at which
Alfv\'en speed $V_A$ was larger than the turbulent velocity $v_l$.

\begin{figure*}[t]
  \includegraphics[width=0.99\textwidth]{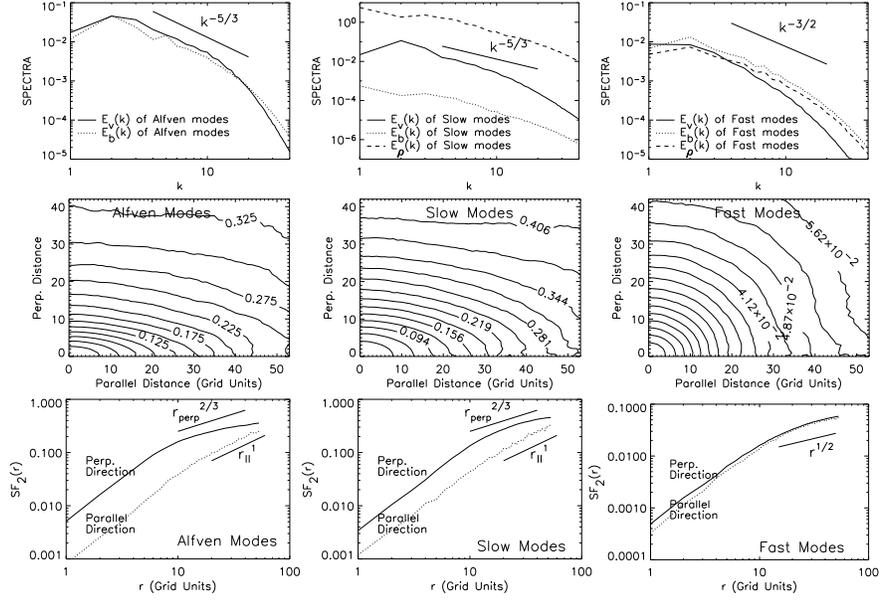}
  \caption{
      Highly supersonic low $\beta$ ($\beta\sim0.02$ and $M_s\sim$7). 
      $V_A\equiv B_0/\sqrt{4\pi\rho}=1$. $a$ (sound speed) $=0.1$. 
      $\delta V \sim 0.7$.
      Alfv\'en modes follow the GS95 scalings. Slow modes follow
      the GS95 anisotropy. But velocity spectrum of slow modes is uncertain.
      Fast modes are isotropic.
}
\label{f2ch}
\end{figure*}

Fig.~\ref{f2ch} illustrate that even in highly supersonic regime, where it was customary 
to claim that the modes were completely blended, the decomposition reveals a regular
structure of MHD modes that corresponds to the expectation of the compressible 
extension of the GS95 theory.

Surely, one can debate whether the adopted technique is
reliable. Indeed, the technique above is statistical in nature.
That is, we separate each MHD mode with respect to the {\it mean} magnetic field
${\bf B}_0$. This procedure is affected by the wandering of large scale magnetic field lines,
as well as density inhomogeneities\footnote{One way to 
      remove the effect by the wandering of field lines is to
      drive turbulence anisotropically in such a way as 
      $k_{\perp, L} \delta V \sim k_{\|,L} V_A$, where
      $k_{\perp,L}$ and $k_{\|,L}$ stand for the wavelengths 
      of the driving scale and
      $\delta V$ is the r.m.s. velocity.
      By increasing the $k_{\perp,L}/k_{\|,L}$ ratio,
      we can reduce the degree of mixing of different wave modes.}.

Nevertheless, CL03 demonstrated  that the technique gave
statistically correct results.
For instance, in low $\beta$ regime,
the velocity of a slow mode 
is nearly parallel to the {\it local} mean magnetic
field. Therefore, for low $\beta$ plasmas, we can obtain velocity statistics
for slow modes in real space as follows.
First, the direction of the {\it local} mean magnetic field was measured using the local magnetic field.
Second, the calculation of the second order structure function for slow modes was defined
by the formula 
vSF$_2({\bf r})=< |\left( {\bf v}({\bf x}+{\bf r})-{\bf v}({\bf x}) \right)
        \cdot \hat{\bf B}_l  |^2 >$, 
where $\hat{\bf B}_l$ is the unit vector along the {\it local} mean field.

\begin{figure*}[t]
  \includegraphics[width=0.3\textwidth]{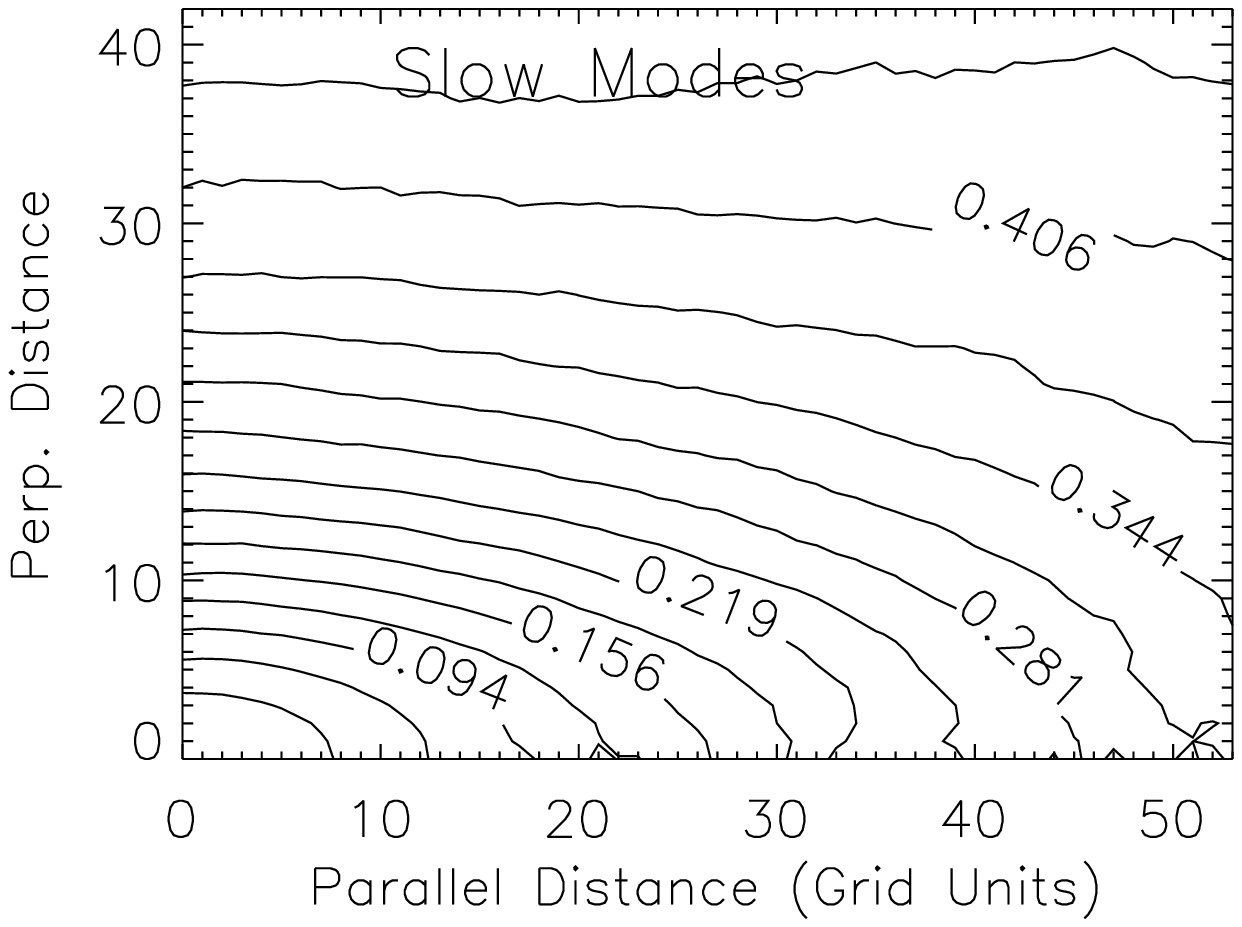}
\hfill
  \includegraphics[width=0.335\textwidth]{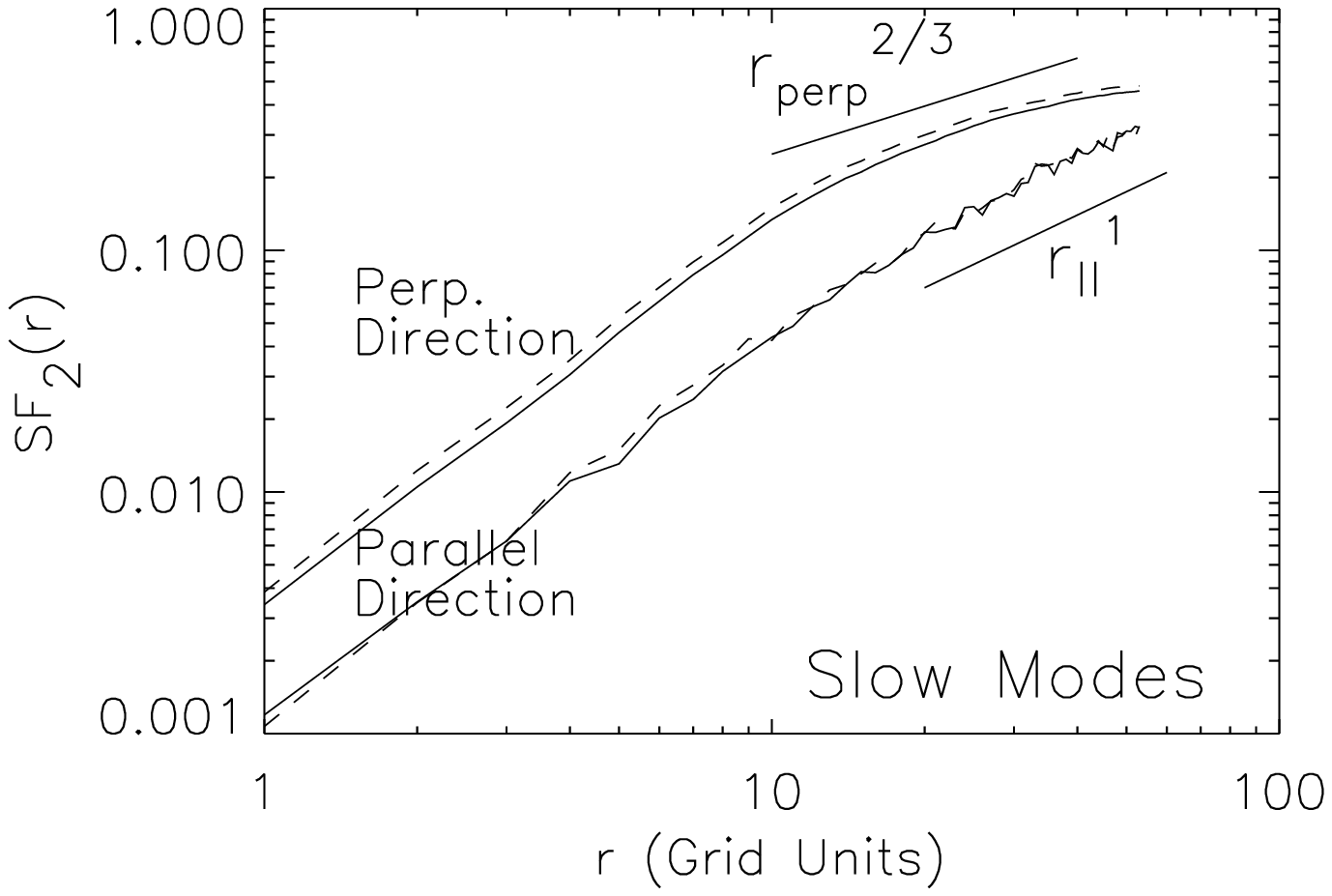}
\hfill
  \includegraphics[width=0.335\textwidth]{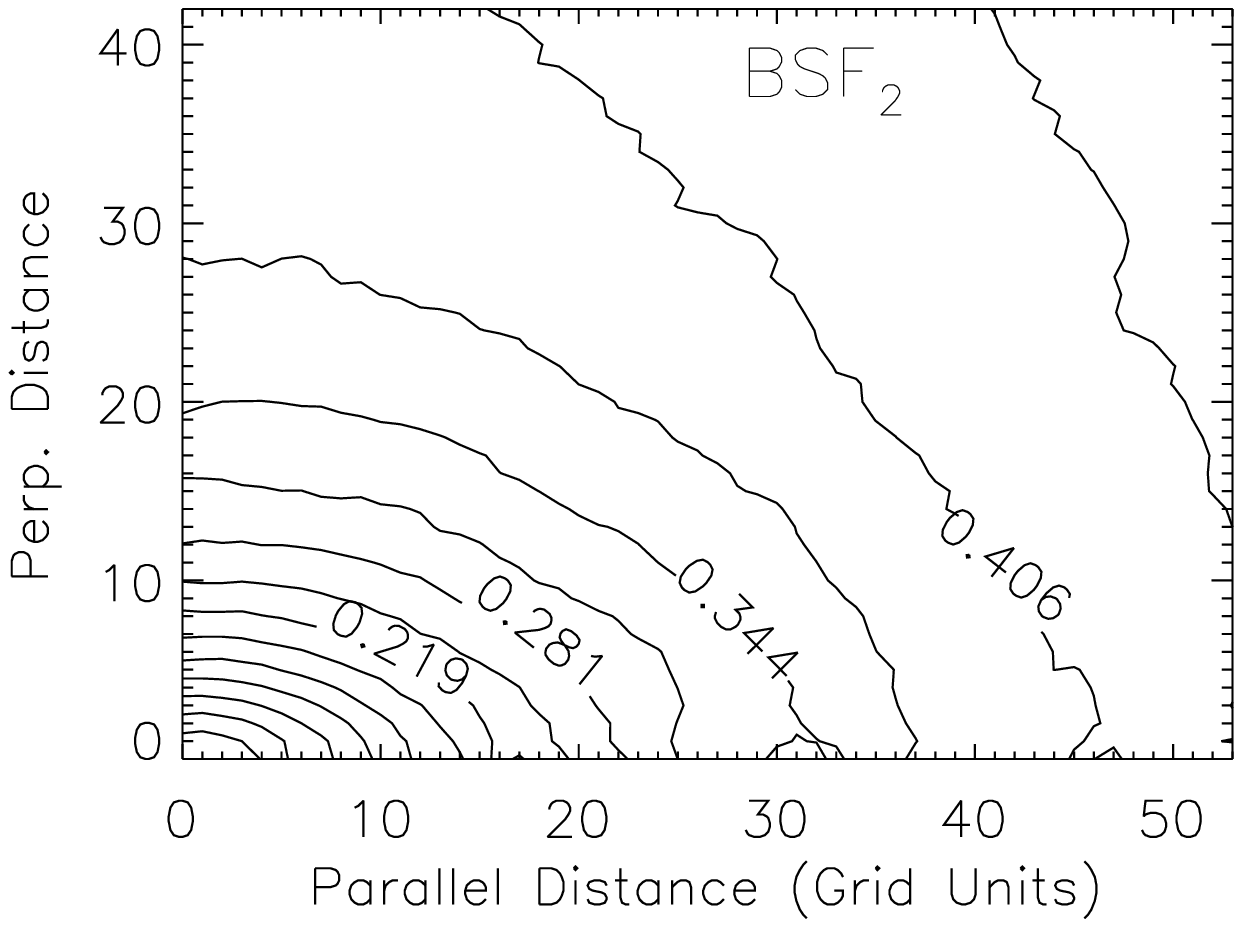}
  \caption{
      Comparison between Fourier space method and real space method.
    (a){\it left:} From real space calculation. $M_s\sim 7$.
    (b){\it middle:} Solid: Fourier space. Dashed: real space. $M_s\sim 7$.
    (c){\it right:} Similar plot for $M_s\sim$ 2.3.
}
\label{fig_comparision}
\end{figure*}

Fig. \ref{fig_comparision}(a) shows the contours obtained by the method
for the high sonic Mach number run.
In Fig. \ref{fig_comparision}(b), we compare the result obtained this way 
(dashed lines) and using CL03 technique. A  similar plot for the mildly supersonic case 
is presented in 
Fig. \ref{fig_comparision}(c).

How physical is 
this decomposition? If the coupling between the modes is strong in MHD turbulence one
cannot talk about three different energy cascades. 
Indeed, the compressible MHD turbulence is a highly non-linear phenomenon
and it has been thought that 
Alfv\'en, slow and fast modes 
are strongly coupled. Nevertheless,
one may question whether this is true.
A remarkable feature of the GS95 model is that
Alfv\'en perturbations cascade to small scales over just one wave
period, while the other non-linear interactions require more time.
Therefore one might expect
that the non-linear interactions with other types of waves
should affect Alfv\'enic cascade only marginally. 
Moreover, since the Alfv\'en waves are incompressible, the properties
of the corresponding cascade may not depend on the sonic Mach number.

The generation of compressible motions 
(i.e. {\it radial} components in Fourier space) 
{}from Alfv\'enic turbulence
is a measure of mode coupling.
How much energy in compressible motions is drained from Alfv\'enic cascade?
According to closure calculations \cite{Zank1993},
the energy in compressible modes in {\it hydrodynamic} turbulence scales
as $\sim M_s^2$ if $M_s<1$.
CL03 conjectured that this relation can be extended to MHD turbulence
if, instead of $M_s^2$, we use
$\sim (\delta V)_{A}^2/(a^2+V_A^2)$. 
(Hereinafter, we define $V_A\equiv B_0/\sqrt{4\pi\rho}$, where
$B_0$ is the mean magnetic field strength.) 
However, since the Alfv\'en modes 
are anisotropic, 
this formula may require an additional factor.
The compressible modes are generated inside the so-called
GS95 cone, which takes up $\sim (\delta V)_A/ V_A$ of
the wave vector space. The ratio of compressible to Alfv\'enic energy 
inside this cone is the ratio given above. 
If the generated fast modes become
isotropic (see below), the diffusion or, ``isotropization'' of the
fast wave energy in the wave vector space increase their energy by
a factor of $\sim V_A/(\delta V)_A$. This  results in
\begin{equation}
  \frac{\delta E_{comp}}{\delta E_{Alf}}\approx \frac{\delta V_A V_A}{V_A^2+c_s^2},
\label{eq_high2}
\end{equation}
where $\delta E_{comp}$ and $\delta E_{Alf}$ are energy
of compressible  and Alfv\'en modes, respectively.
Eq.~(\ref{eq_high2}) suggests that the drain of energy from
Alfv\'enic cascade is marginal
when the amplitudes of perturbations
are weak, i.e. $(\delta V)_A \ll  V_A$. Results of numerical
calculations
shown in CL02 support these theoretical considerations.
This justifies\footnote{A claim in the literature is that a 
strong coupling of incompressible and compressible motions is
required to explain simulations that show fast decay of MHD turbulence.
There is not true. The incompressible motions decay themselves 
in just one Alfv\'en crossing time.} our treating modes separately.

\subsection{Other ways of decomposition into fundamental modes}

Kowal \& Lazarian (2010, \cite{Kowal2010} henceforth KL10) extended the CL03 technique by introducing an additional step before the
Fourier separation, in which we decompose each component of the velocity field
into orthogonal wavelets using discrete wavelet transform:
\begin{equation}
 \textbf{U}(a, \textbf{w}_{lmn}) = a^{-N/2} \sum_{\textbf{x}_{ijk}}{\psi\left( \frac{\textbf{x}_{ijk} - \textbf{w}_{lmn}}{a} \right) \textbf{u}(\textbf{x}_{ijk}) \Delta^N \textbf{x}},
\end{equation}
where $\textbf{x}_{ijk}$ and $\textbf{w}_{lnm}$ are $N$-dimensional position and
translation vectors, respectively, $a$ is the scaling parameter,
$\textbf{u}(\textbf{x}_{ijk})$ is the velocity vector field in the real space,
$\textbf{U}(\textbf{x}_{ijk})$ is the velocity vector field in the wavelet space, and
$\psi$ is the orthogonal analyzing function called wavelet.  The sum in the
equation is taken over all position indices.  KL10 use Daubechies wavelet
as an analyzing function and fast discrete version of the wavelet transform, as a result they obtained
a finite number of wavelet
coefficients.  After the wavelet transform of the velocity the
Fourier representation of each wavelet coefficient 
was calculated and perform individual
separation into the MHD modes was performed in the Fourier space using the CL03 method and
then update the Fourier coefficients of all MHD waves iterating over all
wavelets.  In this way KL06 obtained a Fourier representation of the Alfv\'en, slow
and fast waves.  The final step is the inverse Fourier transform all all wave
components.
\begin{figure*}
  \includegraphics[width=0.3\textwidth]{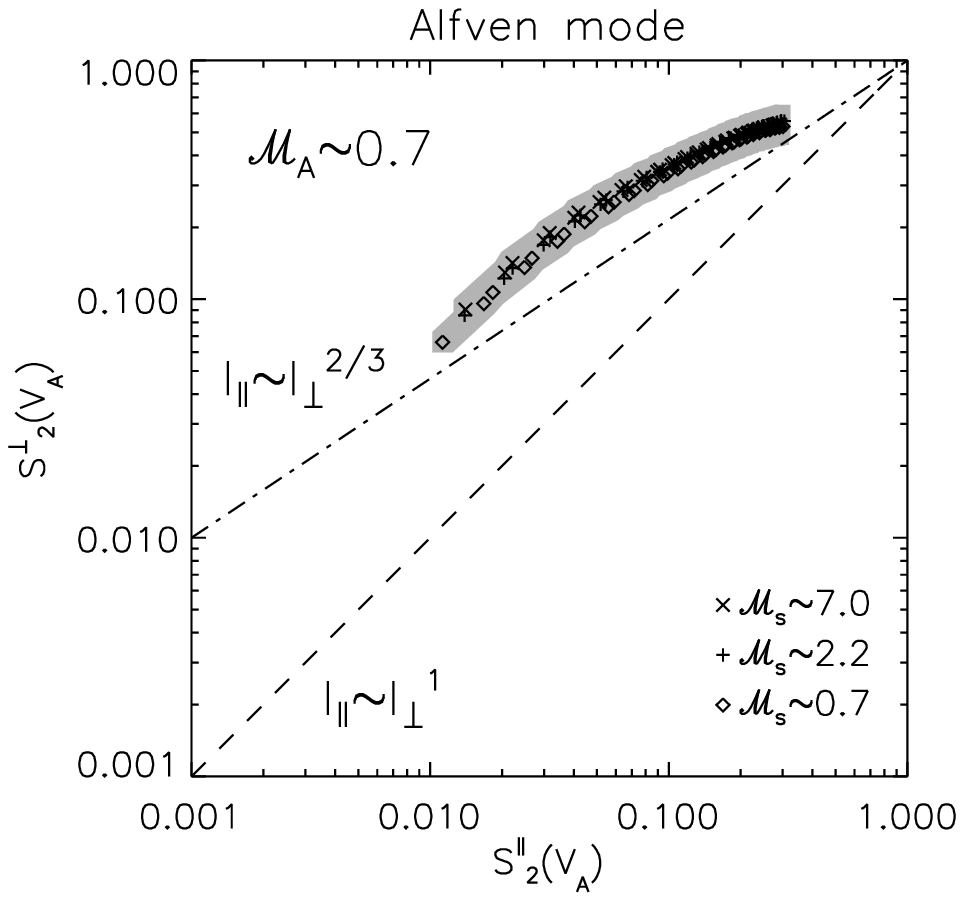}
\hfill
  \includegraphics[width=0.335\textwidth]{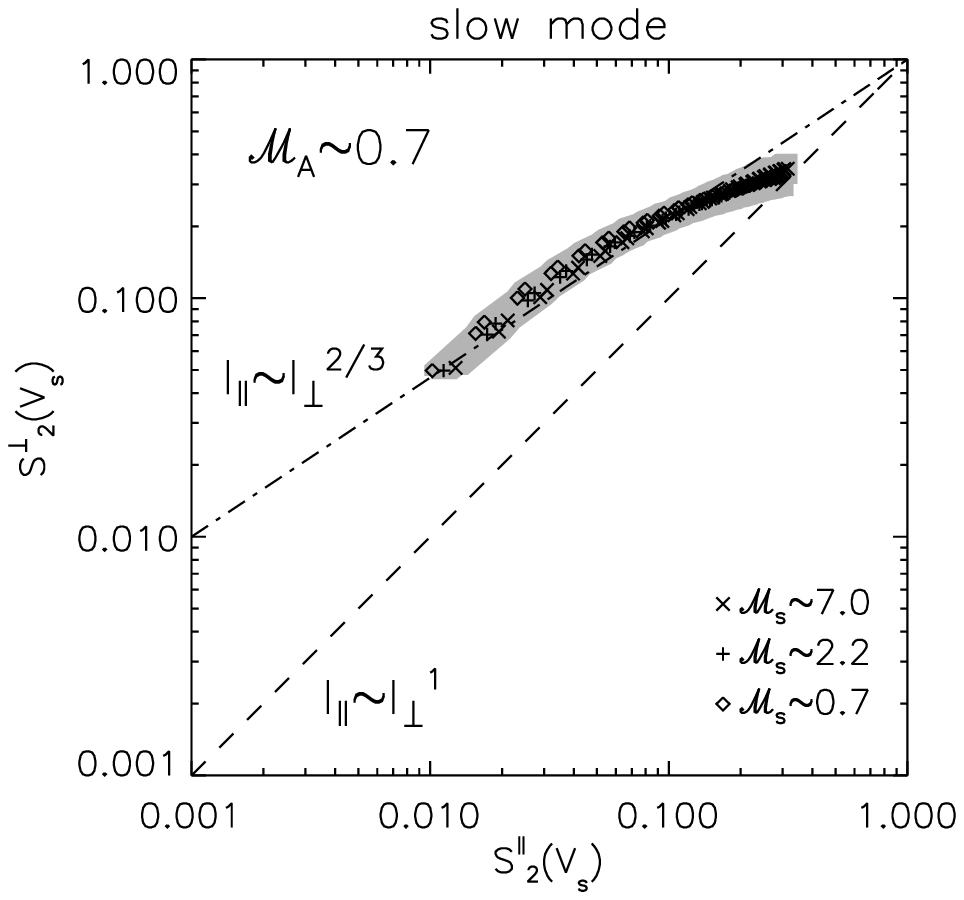}
\hfill
  \includegraphics[width=0.335\textwidth]{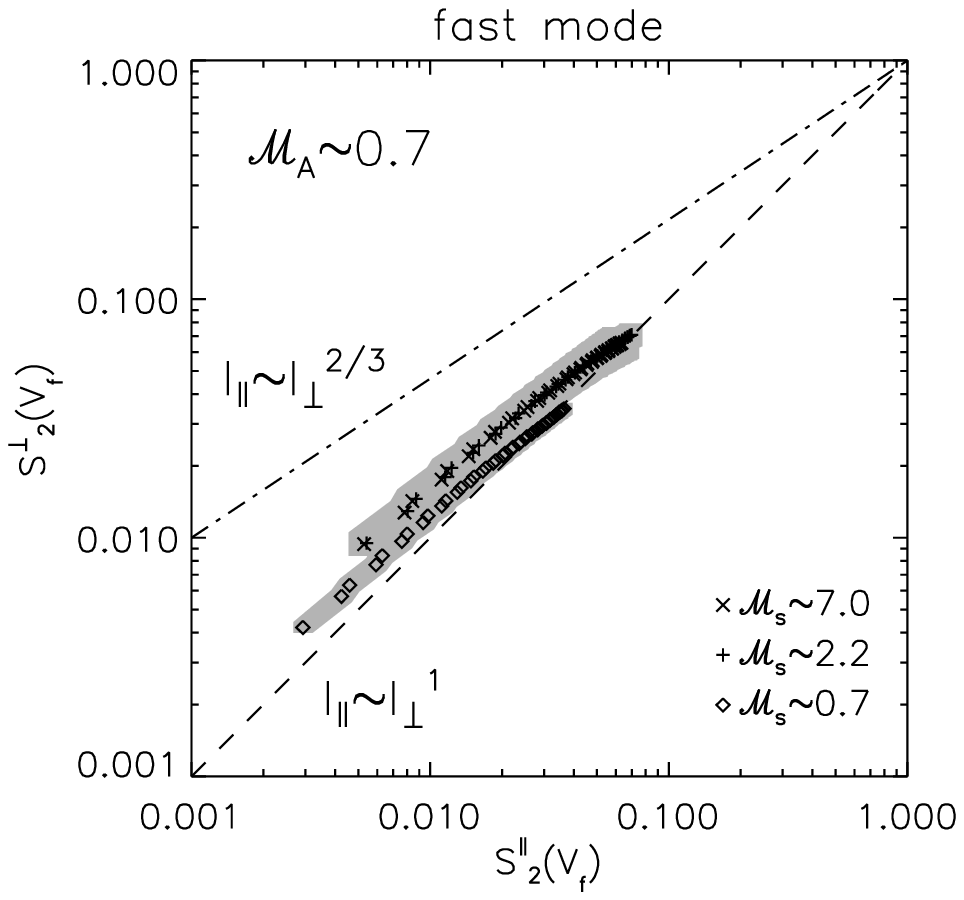}
  \caption{
  Anisotropy of the Alfv\'en, slow and fast modes. To show the anisotropy we use the 2$^\mathrm{nd}$-order total structure functions, parallel and perpendicular to the local mean magnetic field. Points correspond to the mean profiles of the structure functions averaged over several snapshots. The gray areas under points correspond to the degree of departures of the structure functions in time. From KL10.    
}
\label{f5}
\end{figure*}

This additional step allows for important extension of the CL03 method, namely,
allows for the local definition of the mean magnetic field and density used to
calculate $\alpha$ and $D$ coefficients.  Since the individual wavelets are
defined locally both in the real and Fourier spaces, the averaging of the mean
field and density is done only within the space of each wavelet.

The study in KL10 provided results consistent with the CL03 and it extended
the decomposition to new physical cases. For instance, Fig.~\ref{f5}
shows the anisotropy for subAlfv\'enic turbulence which agrees well with that
obtained in CL03.

\begin{figure*}
  \includegraphics[width=0.99\textwidth]{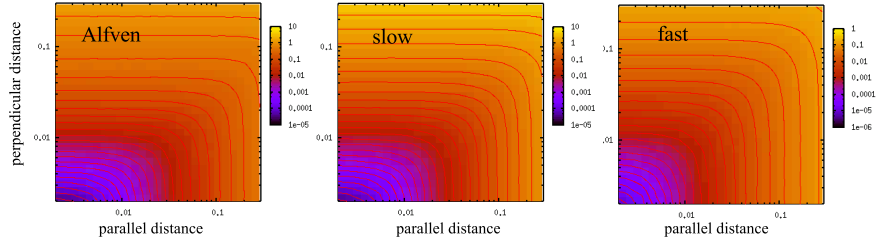}
  \caption{
  Anisotropy of the Alfv\'en, slow and fast modes as evidenced by the contours of the second order structure function. Here we used the new SF decomposition method.
 The Alfv\'en and slow mode exhibit scale-dependent anisotropy, while the fast mode is almost isotropic.}
\label{sf_decomp}
\end{figure*}

Another way to decompose into modes using structure functions has been recently proposed and tested by one of the authors (AB). 
In this method the separation vector $\vec l$ of the structure function plays the role of the wavenumber, because
there is a correspondence relation between one-dimensional structure function along the certain line and the power spectrum along the same line.
Fig.~\ref{sf_decomp} shows the contours of the structure function corresponding to each mode obtained
in datacubes from $M_s=10$ supersonic simulations used earlier in \cite{BLC05} (see also \S~\ref{density}).
The anisotropies of each mode show the same behavior as in the earlier discussed global decomposition method, see Fig.~\ref{f2ch}.
There are two advantages in using the new decomposition method. First, it is computationally efficient, as the structure functions can be
calculated by the Monte-Carlo method which samples only a fraction of data points. This way, the very high resolution simulations can be
processed in a reasonable time. The second advantage is that the structure function is a local measurement, so we can measure spectral
characteristics of the modes in a highly inhomogeneous situations. This method has been applied to the decomposition of MHD turbulence
obtained in high-resolution cosmological simulation of a galaxy cluster \cite{BXLS13}. The cluster environments has been notoriously
difficult to analyze due to the strong dependence of all quantities on the distance to the center. The new method was used to calculate the SFs
in concentric shells around the cluster center. Among other things the aforementioned paper estimated the fraction of the fast mode to around 0.25,
which is fairly high for subsonic to trans-sonic cluster environment. We hypothesized that this is due to the way the cluster turbulence is
driven -- through mergers, which are essentially compressible trans-sonic motions. 

\subsection{Decomposition into solenoidal and potential modes}

KL10 also used a different decomposition of the velocity field.
Using the Hodge generalization of the Helmholtz theorem we can split an
arbitrary vector field $\textbf{u}$ into three components:
\begin{equation}
 \textbf{u} = \textbf{u}_p + \textbf{u}_s + \textbf{u}_l ,
\end{equation}
where each component has specific properties:
\begin{itemize}
 \item[a)] Potential component ($\textbf{u}_p$) - it is curl-free component, i.e.
$\nabla \times \textbf{u}_p = 0$, so it stems from a scalar potential $\phi$:
\begin{equation}
 \textbf{u}_p = \nabla \phi.
\end{equation}
 The scalar potential $\phi$ is not unique. It is defined up to a constant. This
component describes the compressible part of the velocity field.

 \item[b)] Solenoidal component ($\textbf{u}_s$) - it is divergence-free component,
i.e. $\nabla \cdot \textbf{u}_s = 0$, so it stems from a vector potential ${\bf
{\cal A}}$:
\begin{equation}
 \textbf{u}_s = \nabla \times {\bf {\cal A}}.
\end{equation}
 The vector potential ${\bf {\cal A}}$ also is not unique. It is defined only up to
a gradient field. In the case of velocity this component describes the
incompressible part of the field.

 \item[c)] Laplace component (${\bf u}_l$) - it is both divergence-free and
curl-free. Laplace component comes from a scalar potential which satisfies the
Laplace differential equation $\Delta \phi = 0$.
\end{itemize}

Thus the decomposition can be rewritten in the form:
\begin{equation}
 \textbf{u} = \nabla \times {\bf {\cal A}} + \nabla \phi + \textbf{u}_l .
 \label{eqn:decompose}
\end{equation}

\begin{figure*}
  \includegraphics[width=0.3\textwidth]{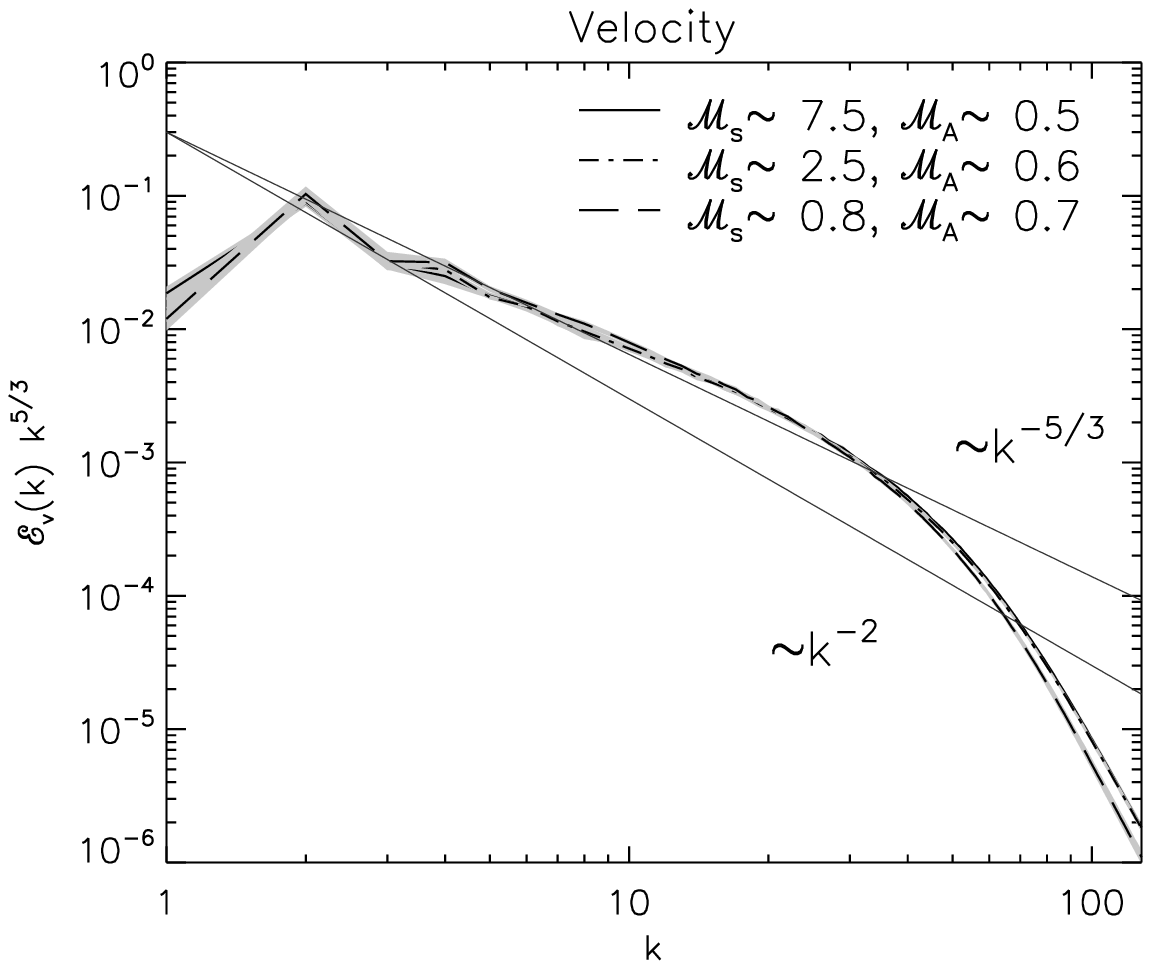}
\hfill
  \includegraphics[width=0.335\textwidth]{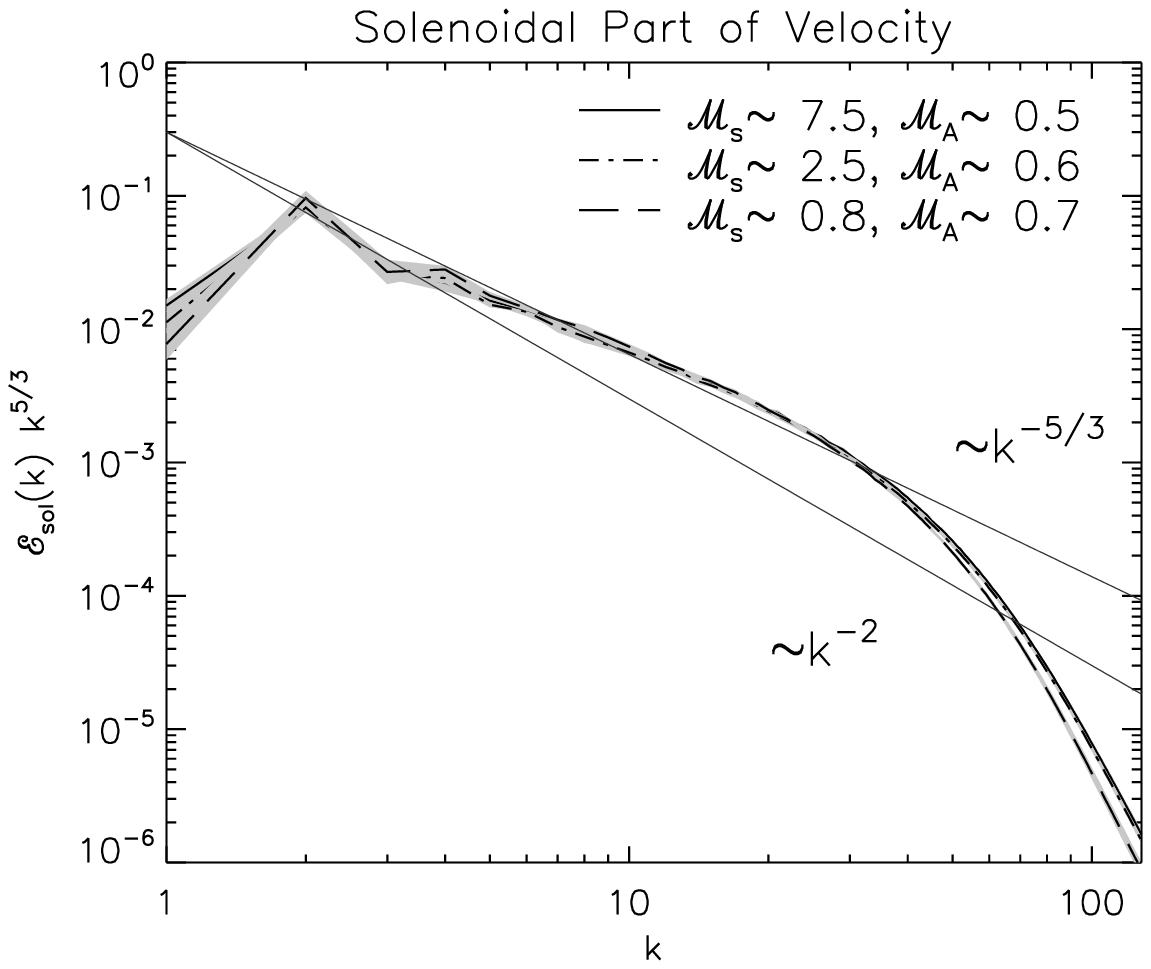}
\hfill
  \includegraphics[width=0.335\textwidth]{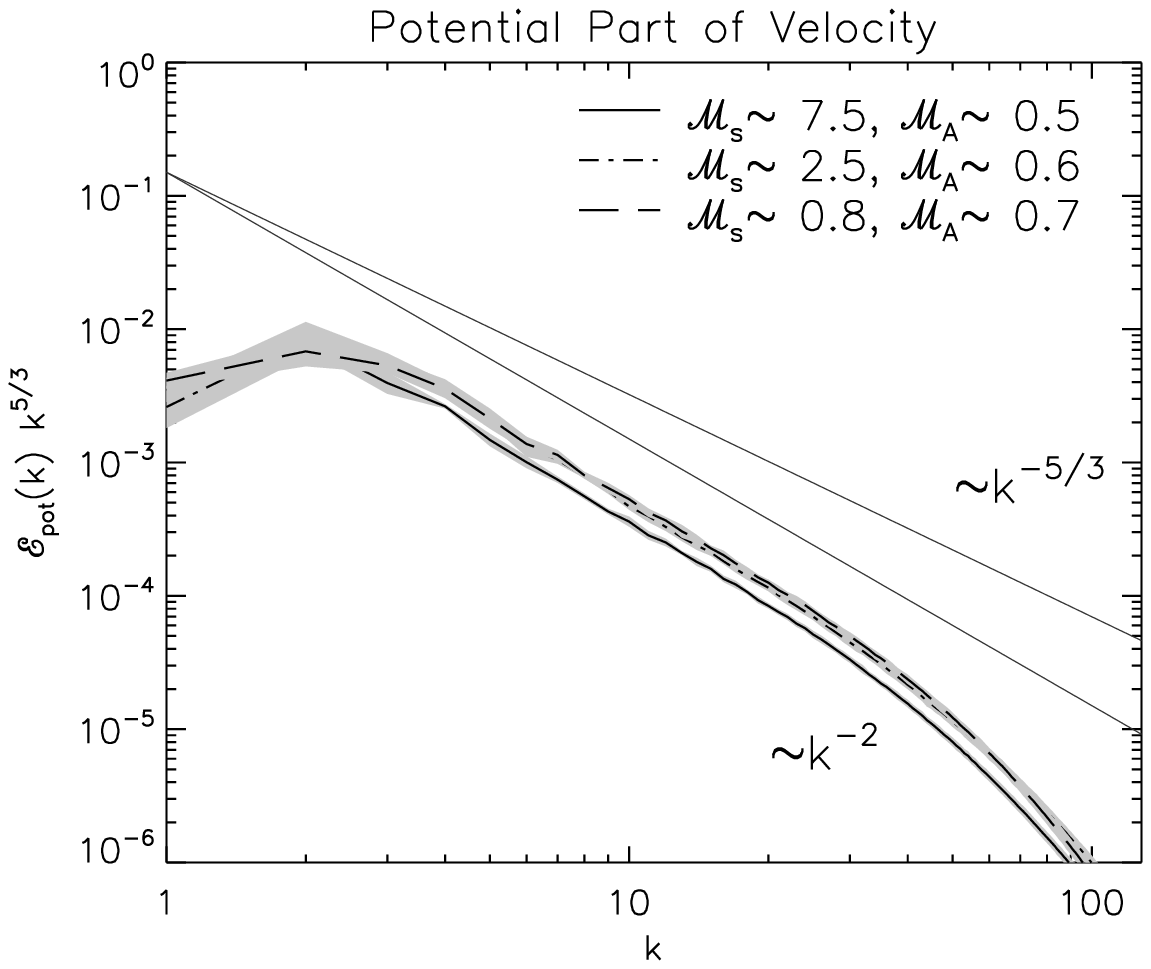}
  \caption{
  Spectra of the solenoidal and potential parts of velocity for subAlfv\'{e}nic turbulence.
  From KL10.    
}
\label{f6k}
\end{figure*}

The results of this decomposition are illustrated in Fig.~\ref{f6k}

It is clear that the compressible components of velocity correspond to shocks, while the incompressible
part is dominated by GS95-type motions.

Table \ref{tab1} illustrates how the percentage of energy changes within different components of the flow. It is clear
from the table that even for highly compressible magnetized supersonic flows most of the energy is residing in the incompressible motions.
In terms of fundamental modes the Alfv\'en modes dominate. However, the role of the fast modes increases with the increase of 
the sonic Mach number. 

\begin{table}[t]
\caption{Percentage amount of the kinetic energy contained within each velocity component.  Errors correspond to a measure of the time variation.}
\begin{tabular*}{0.99\columnwidth}{@{\extracolsep{\fill}}r cccccc}
\hline\hline
 {${\cal M}_{s}$} &
 {${\cal M}_{A}$} &
 {$V_\mathrm{incomp.}$} &
 {$V_\mathrm{comp.}$} &
 {$V_A$} &
 {$V_s$} &
 {$V_f$} \\
\hline
  $\sim 0.7$ & $\sim 0.7$ & 96.5$^{\pm0.8}$ &  3.3$^{\pm0.8}$ & 58$^{\pm4}$ & 37$^{\pm3}$ & 4.8$^{\pm0.7}$ \\
  $\sim 2.2$ & $\sim 0.7$ & 93$^{\pm2}$     &  7$^{\pm2}$     & 58$^{\pm5}$ & 33$^{\pm4}$ & 9$^{\pm2}$ \\
  $\sim 7.0$ & $\sim 0.7$ & 92$^{\pm2}$     &  7$^{\pm2}$     & 56$^{\pm4}$ & 36$^{\pm4}$ & 8.0$^{\pm0.7}$ \\
\hline
  $\sim 0.7$ & $\sim 7.4$ & 95$^{\pm2}$ & 5$^{\pm2}$  & 52$^{\pm4}$ & 42$^{\pm4}$ & 6.2$^{\pm0.8}$ \\
  $\sim 2.3$ & $\sim 7.4$ & 86$^{\pm1}$ & 14$^{\pm2}$ & 47$^{\pm3}$ & 37$^{\pm4}$ & 16$^{\pm2}$    \\
  $\sim 7.1$ & $\sim 7.1$ & 84$^{\pm2}$ & 16$^{\pm2}$ & 47$^{\pm4}$ & 33$^{\pm4}$ & 20$^{\pm2}$ \\
\hline
\end{tabular*}
\label{tab1}
\end{table}

\subsection{Density scalings}
\label{density}
\begin{figure*}[t]
  \includegraphics[width=0.295\textwidth]{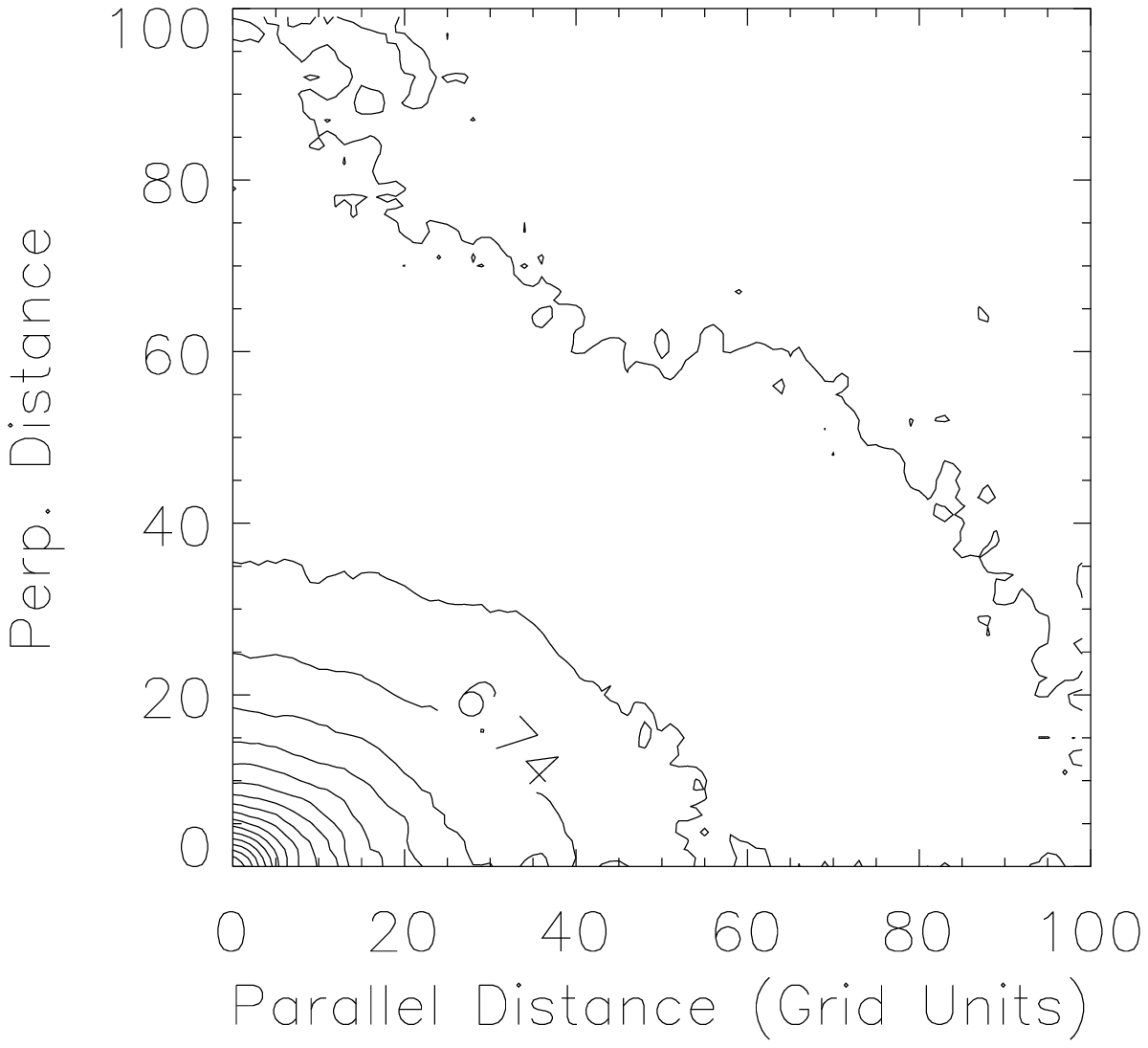}
\hfill
  \includegraphics[width=0.295\textwidth]{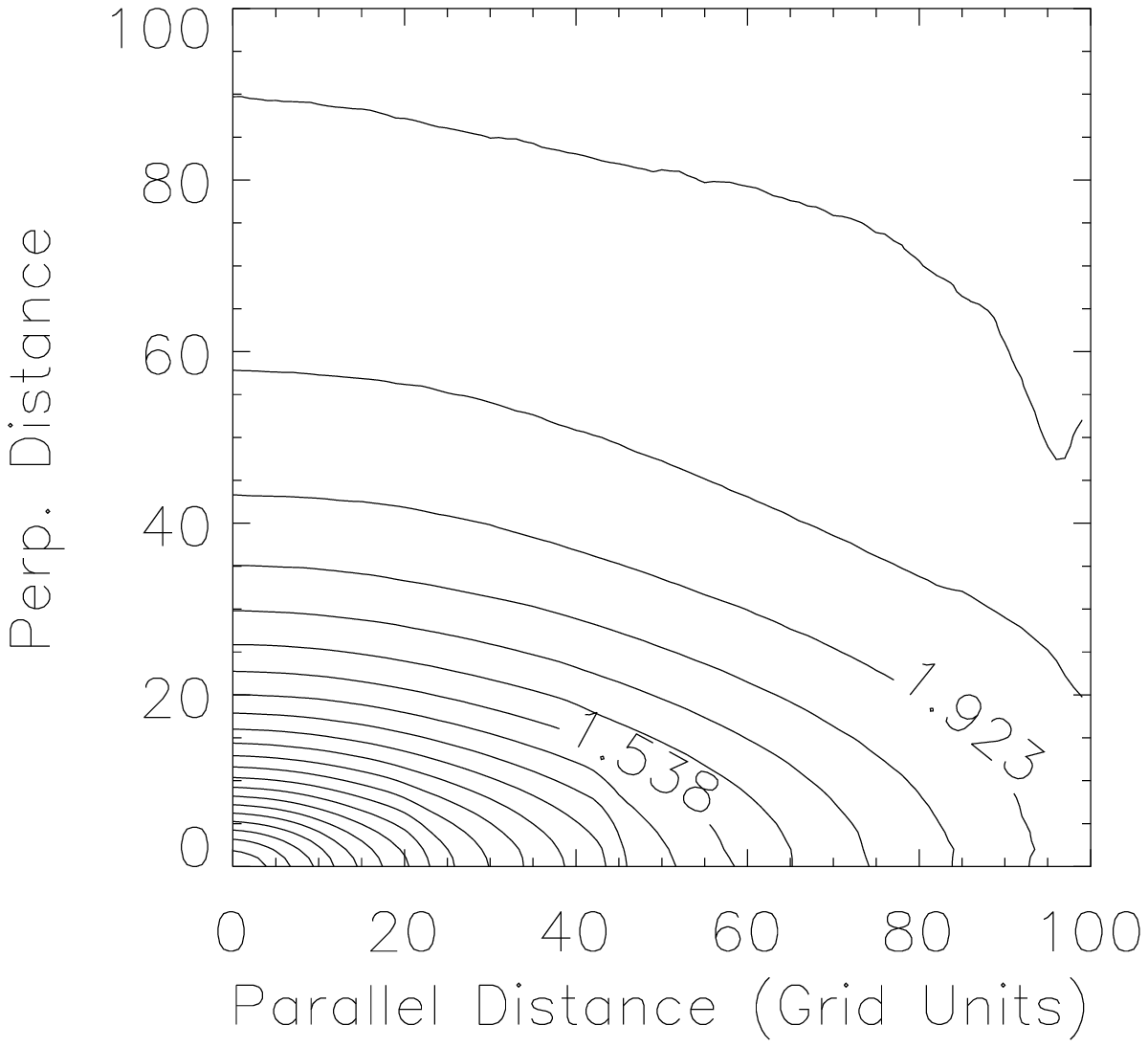}
\hfill
  \includegraphics[width=0.39\textwidth]{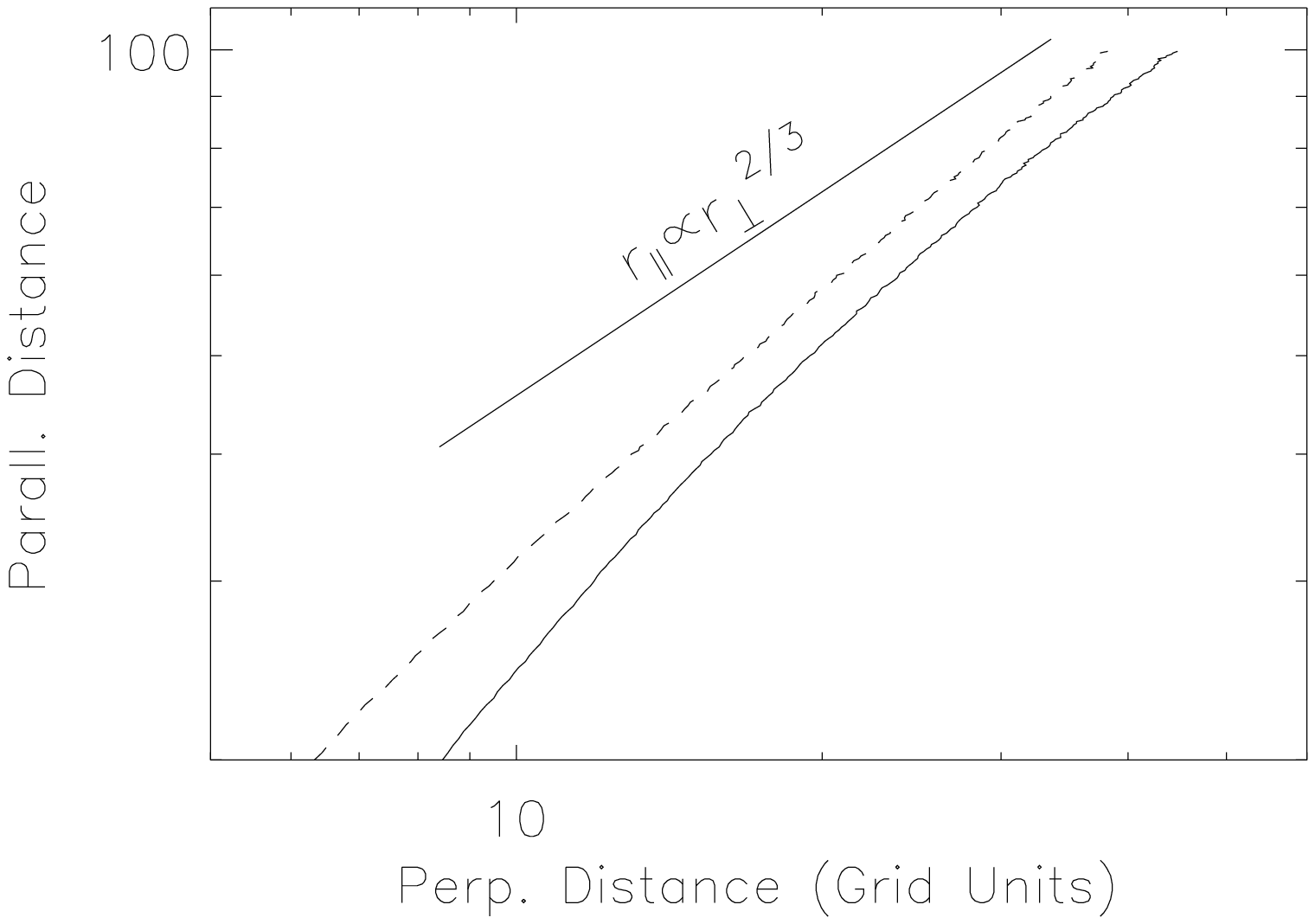}
  \caption{Contours of the structure function of density (left), log-density (center) and the anisotropy of log-density (right). From \cite{BLC05}}
\label{log_density}
\end{figure*}

The properties of density in supersonic ISM turbulence has always been of interest to astronomers due to its applications
to star formation. The density is thought be be associated primarily with the slow mode, since this is the mode that perturb
density the most in low-beta supersonic fluid. However, the structure function of density was generally observed to be
very different from the structure function of the velocity of the slow mode. In particular, while slow mode show well-pronounced
scale-dependent anisotropy, see Fig.~\ref{f2ch}, the structure function of density was almost isotropic, see
Fig.~\ref{log_density}. This mysterious difference has made applications of our knowledge of supersonic MHD turbulence
to the case of star formation difficult.

However, \cite{BLC05} proposed a simple picture which both unraveled the mystery and further
shed light on the dynamics of density in supersonic MHD. It turned out that the second-order structure function
method work appropriately only if the quantity in question has a Gaussian distribution. If we use it on density, which
distributed approximately log-normally and has high-density tail, this greatly favor high-density regions or clumps.
The apparent isotropy, therefore, is an artifact of these clumps being distributed randomly in space. Furthermore,
the flat spectrum of density comes from the same effect, namely, high-density clumps act as a delta-functions and
produce flat spectrum. When we use log-density instead of density, the spectra become steeper and the second-order
structure function shows remarkable scale-dependent anisotropy, see Fig.~\ref{log_density}.

So it turned out that while the perturbations of density and its log-normal PDF are created by random slow shocks,
the structure of density has an imprint from Alfv\'enic driving, the same imprint the structure of slow mode velocity has. 

\subsection{Viscosity-dominated regime of MHD turbulence}

In this section, we focus on the effects of viscosity.
In strong hydrodynamic turbulence energy is injected at a scale $L$,
and cascades down to smaller scales without significant viscous losses
until it reaches the viscous damping scale $l_{dv}$.
The Kolmogorov energy spectrum applies to the inertial range, i.e.
all scales between $L$ and $l_{dv}$.
This simple picture becomes more complicated when we deal with MHD turbulence 
because there are two dissipation scales - the velocity damping scale
$l_{dv}$ and the magnetic diffusion scale $l_{dm}$, where magnetic
structures are dissipated.
In fully ionized collisionless plasmas (e.g. the hottest phases of
the ISM), $l_{dv}$  is less than an order of magnitude
larger than $l_{dm}$, but both scales are very small.
However, in partially ionized plasmas (e.g. the warm or cold neutral phase
of the ISM), the two dissipation scales are very different and 
$l_{dv}\gg l_{dm}$.
In the Cold Neutral Medium (see \cite{Draine1999} for a list of the ISM phases) neutral particle transport leads to viscous
damping on a scale which is a fraction of a parsec. 
In contrast, in these same phases $l_{dm}\sim 100km$.
\begin{figure*}
  \includegraphics[width=0.49\textwidth]{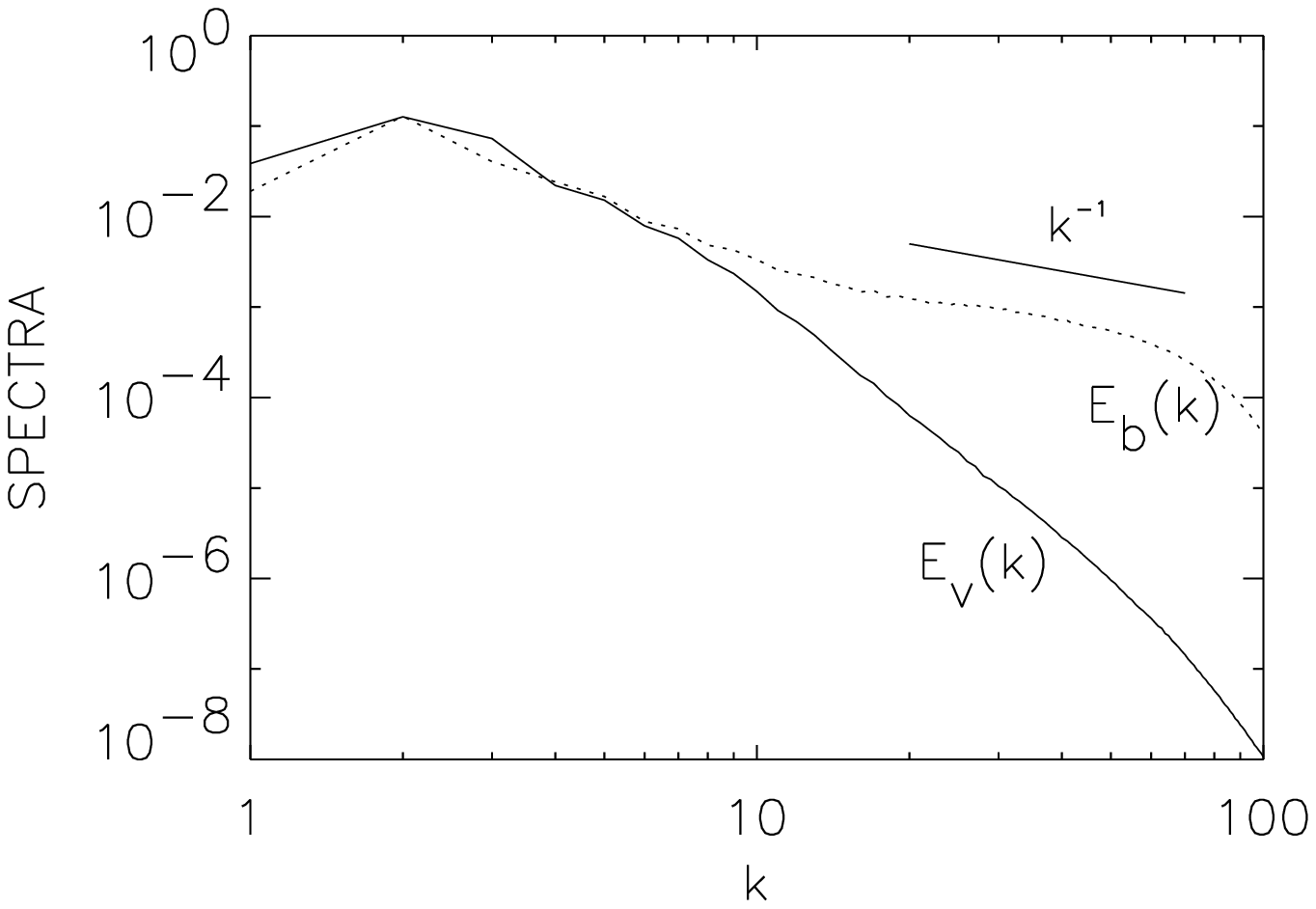}
\hfill
  \includegraphics[width=0.49\textwidth]{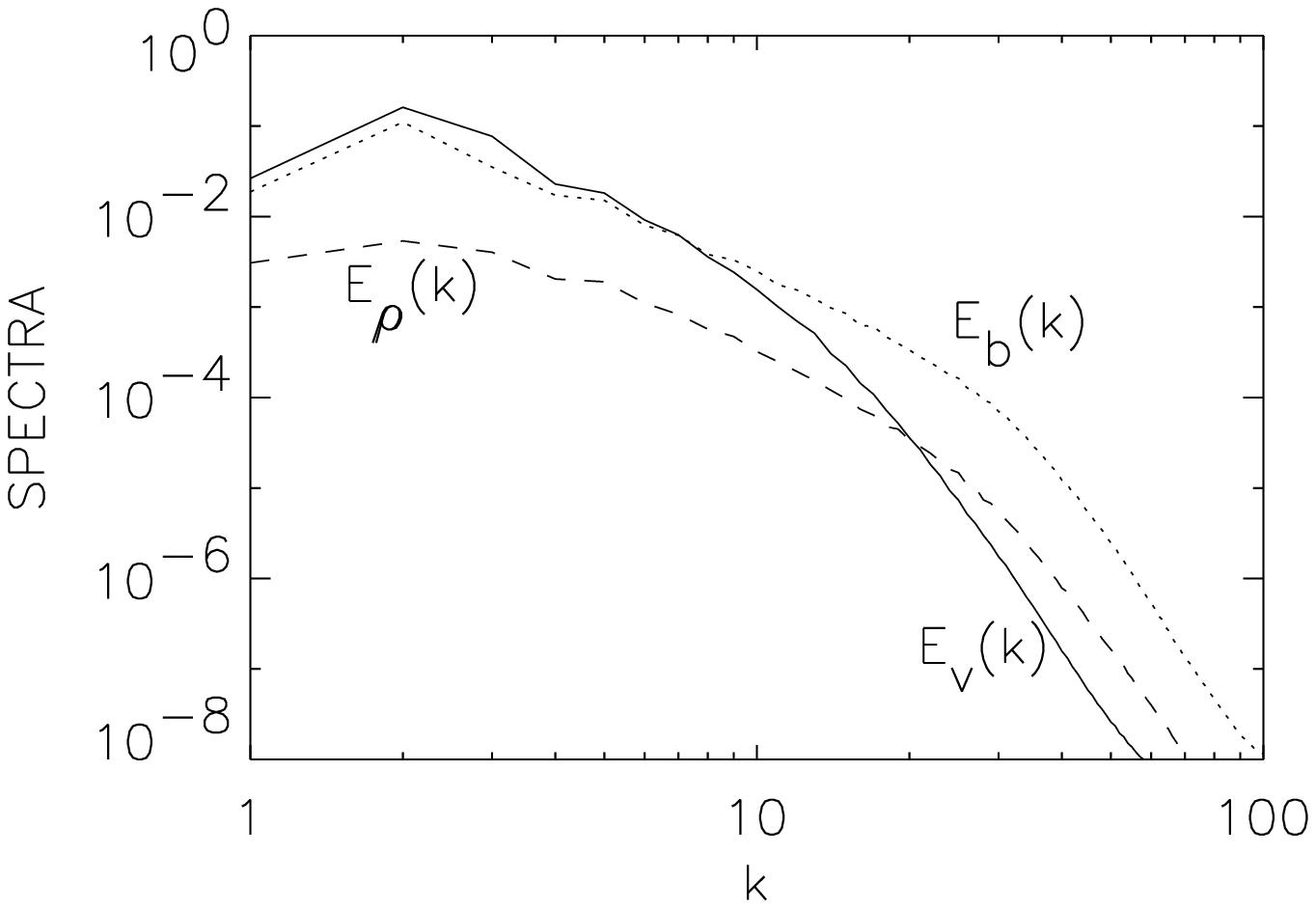}
  \caption{
      Viscous damped regime (viscosity $>$ magnetic diffusivity).
      Due to large viscosity, velocity damps after $k\sim10$.
    (a) {\it Left:} Incompressible case with $384^3$ grid points. 
        Magnetic spectra show a shallower slope ($E_b(k)\propto k^{-1}$)
        below the velocity damping scale.
        We achieve a very small magnetic diffusivity through the use
        hyper-diffusion.
       {}From CLV02b.
    (b) {\it Right:} Compressible case with $216^3$ grid points.
        Magnetic and density spectra show structures below the
        velocity damping scale at $k\sim10$.
        The structures are less obvious than the incompressible case
        because it is relatively hard to
        achieve very small magnetic diffusivity in the compressible run.
From CL03.}
\label{f7}
\end{figure*}

This has a dramatic effect on the energy cascade model in 
a partially ionized medium.
When the energy reaches the viscous damping scale $l_{dv}$,
kinetic energy will dissipate there, but the magnetic
energy will not.
In the presence of dynamically important magnetic field,
Cho, Lazarian, \& Vishniac (\cite{CLV02b}; hereafter CLV02b) reported 
a completely new regime of turbulence below the scale at which
viscosity damps kinetic motions of fluids.  
They showed that
magnetic fluctuations extend below the viscous damping scale
and form a shallow spectrum $E_b(k)\sim k^{-1}$. 
This spectrum is similar to that of the viscous-convective range of
a passive scalar in hydrodynamic turbulence.

A further numerical study of the viscosity-damped MHD
turbulence was presented in CL03 and \cite{Cho2003b}.
Figure\ref{f7} compares the results for this regime obtained for compressible
and incompressible MHD turbulence.

\begin{figure*}[t]
\includegraphics[width=0.43\textwidth]{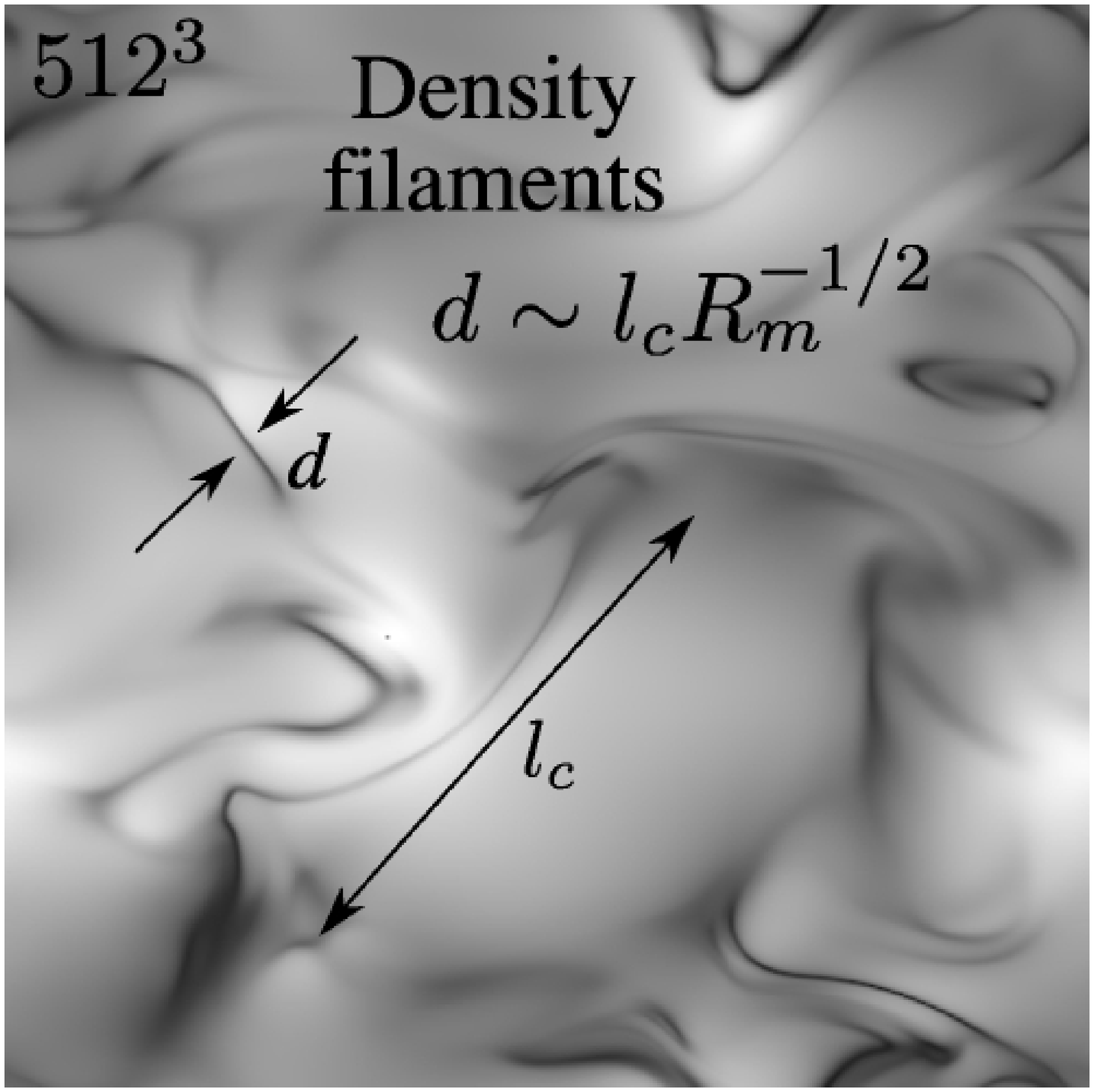}
\hfill
\includegraphics[width=0.565 \textwidth]{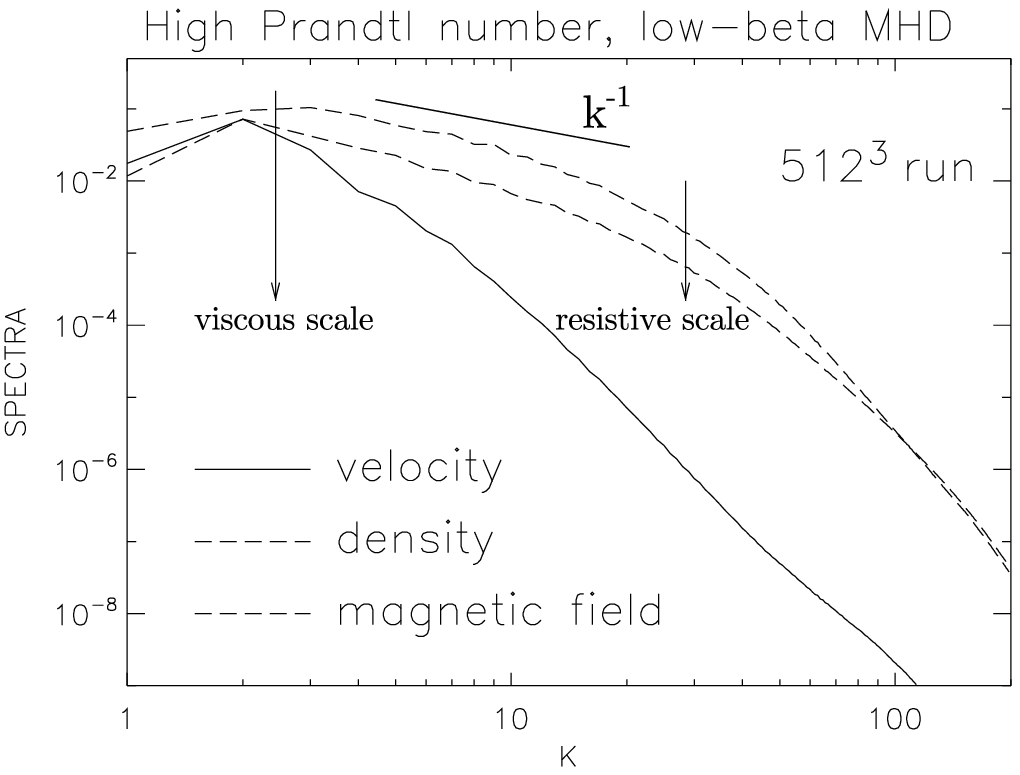}
\caption{
Simulations of supersonic, viscously damped MHD turbulence, with high viscosity emulating high drag
from ambipolar diffusion in molecular clouds.
{\it Left}: 
 Filaments of density created by magnetic
compression of the gas in this regime. Darker regions correspond to higher density. The viscous damping scale $l_c$ is much
larger than the current sheet thickness $d$.  This creates large observed density contrasts.
{\it Right}: The spectra of density, velocity and magnetic field in this case. While the density and magnetic spectra are similar,
the velocity spectrum has a cutoff due to high viscosity. 
Note that the resistive scale in this regime is not
$L/Rm$ but $L {Rm}^{-1/2}$.  }
\label{f8}
\end{figure*}

Fig.~\ref{f8} show structures and spectra in supersonic viscous MHD simulations, emulating conditions in the molecular clouds,
where high ambipolar diffusion could result in drag and damping of kinetic motions. Remarkably, the kinetic and magnetic
spectra are very similar to the incompressible and weakly compressible cases. However, the structures, observed in the datacubes
are completely different. The supersonic structures are completely dominated by the current sheets, which are also density sheets.
This is because currents sheets has low magnetic pressure and this has to be compensated by gas pressure.

The theoretical study of weakly compressible viscously damped case was performed in Lazarian, Vishniac \& Cho (2004, \cite{Lazarian2004}
henceforth LVC04). Below we present a brief summary of the theory.
Following the usual treatment of ordinary strong MHD turbulence, we define
the wavenumbers $k_{\|}$ and $k_{\perp}$ as the components of the
wavevector measured along the {\it local}
mean magnetic field and perpendicular to it, respectively.
Here the local mean magnetic field is the direction of the locally averaged
magnetic field, which depends not only on the location but also
the volume over which the average is taken.
See \cite{CV00,CLV02a} for details.

Lazarian, Vishniac, \& Cho (\cite{Lazarian2004}, henceforth LVC04) 
proposed a theoretical model for
viscosity-damped MHD turbulence.
We summarize the  model as follows.

Since there is no significant velocity fluctuation below $l_{dv}$,
the time scale for the energy cascade below $l_{dv}$
is fixed at the viscous damping scale.  Consequently
the energy cascade time scale $t_{cas}$ 
is scale-independent below $l_{dv}$ and
the requirement for a scale independent energy transfer rate $b_l^2/t_{cas}$ yields
\begin{equation}
  b_l \sim \mbox{constant, or~~~} E_b(k)\sim k^{-1},
\end{equation}
where $k E_b(k)\sim b_l^2$. 

In LVC04, we assume that the curvature of the magnetic field lines
changes slowly, if at all, in the cascade:
\begin{equation}
  k_{\|} \sim \mbox{constant}.
\end{equation}
  This is consistent
with a picture in which the cascade is driven by repeated shearing
at the same large scale.  It is also consistent with the numerical
work described in CLV02b, which yielded a constant $k_{\|}$
throughout the viscously damped nonlinear cascade.  A corollary
is that the wavevector component in the direction of the perturbed
field is also approximately constant, so that the increase in $k$ is
entirely in the third direction.

The kinetic spectrum depends on the scaling of intermittency.
In LVC04, we define  a filling factor $\phi_l$, which is the
fraction of the volume containing strong magnetic field perturbations
with a scale $l \sim k^{-1}$.  We denote the velocity and perturbed
magnetic field inside these sub-volumes with a ``$\hat{\ } $'' so
that
\begin{equation}
v_l^2=\phi_l \hat v_l^2,
\end{equation}
and
\begin{equation}
b_l^2=\phi_l \hat b_l^2.
\end{equation}
We can balance viscous and magnetic tension forces
to find
\begin{equation}
{\nu\over l^2} \hat v_l \sim \max[\hat b_lk_c,B_0k_{\|,c}] \hat b_l
\sim k_c\hat b_l^2,
\label{s3}
\end{equation}
where $k_c \sim 1/l_{dv}$
and $k_{\|,c}$ is the parallel component of the wave vector corresponding
to the perpendicular component $k_c$.
We used the GS95 scaling ($B_0k_{\|,c}\sim b_lk_c$)
   and $\hat b_l \geq b_l$ to evaluate the two terms in the square braces.
Motions on scales smaller than $l_{dv}$ will be 
continuously sheared at a rate $\tau_s^{-1}$.
These structures will reach a dynamic equilibrium if they generate a
comparable shear, that is
\begin{equation}
{\hat v_l\over l}\sim \tau_s^{-1} \sim \mbox{constant}.
\label{s1}
\end{equation}
Combining this with equation (\ref{s3}), we get
\begin{equation}
 \phi_l\sim {k_c l}
\end{equation}
and
\begin{equation}
 E_v(k) \sim k^{-4}.
\end{equation}
Note that equation (\ref{s3}) implies that
kinetic spectrum would be $ E_v(k) \sim k^{-5}$ if $\phi_l$=constant.

\subsection{Application of results to collisionless fluids}

Some astrophysical magnetized fluids are collisionless, meaning the
typical collision frequency is lower than the gyrofrequency. It is
important to understand to what extend the results obtained for MHD
can also be applied to such environments. The effective collisionality
of the medium depends on the collective effects of magnetic scattering
of ions. For instance, gyroresonance instability induced by large
scale compressions produces small scale perturbations that induce
efficient scattering of charged particles \cite{Schekochihin2006,LB06,Schekochihin2008}. Thus the free energy of turbulent
environment makes plasmas, effectively, much more collisional. Another
example of this is a collisionless shock which excite plasma waves and lead to effective
particle thermalization.

Furthermore, some subsets of MHD equations, such as reduced or Alfv\'enic MHD, which we studied in great detail in Sections~3 and 4,
are actually applicable to fully collisionless plasmas, because Alfv\'enic motions are essentially $[{\bf E\times B}]$ drift motions,
rely only on magnetic tension and do not require collisions, see, e.g., \cite{Schekochihin2009}.

A recent study in \cite{Santos2014}, using a closure for anisotropic plasma pressure, showed that for a reasonable choice of the
relaxation term the collisionless fluids behave similar to MHD. Thus we expect that both MHD turbulence scaling relations and the results
of turbulent dynamo that we discussed in above are applicable to collisionless turbulent astrophysical plasmas above the effective
collisional scale. The measurements in the solar wind indicate that the effective MHD scales could be as low as the ion skip depth or
the ion Larmor radius.

\subsection{Outlook on relativistic turbulence}

When electromagnetic energy density is much larger than the rest mass
energy density of matter, the electromagnetic fields becomes
essentially force-free, which is described with the so-called
relativistic force-free approximation.
The examples of such environments include electron-positron pulsar
magnetospheres and the inner parsec-scale AGN jets.

\begin{figure*}[t]
\begin{center}
\includegraphics[width=0.99\textwidth]{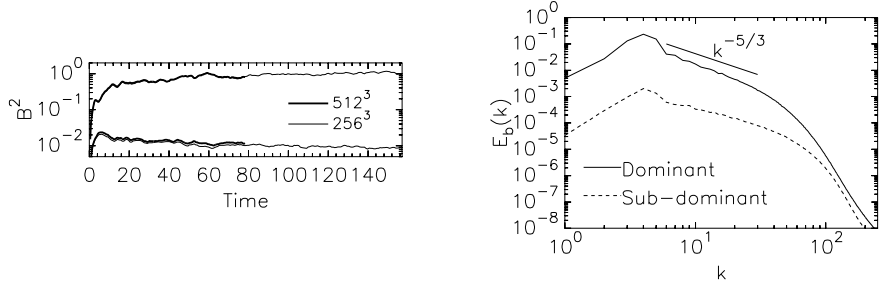}
\end{center}
 \caption{ \it Left panel: The time dependence of energy densities
for the dominant
and subdominant waves. {\it Right panel}: Energy spectra of the
dominant and subdominant waves. The spectrum of the
subdominant flux is shallower that for the dominant wave. From CL14.}
\label{relativistic}
\end{figure*}

Numerical simulations of force-free MHD turbulence by \cite{Cho2005}
reported anisotropic Goldreich-Sridhar scalings, similar to the ones
observed in Alfv\'enic turbulence, which was earlier conjectured by
\cite{Thompson1998}.
More recently a challenging numerical work of studying imbalanced
relativistic turbulence was performed in Cho \& Lazarian (2014, \cite{Cho2014},
henceforth CL14).
Fig.~\ref{relativistic} shows the energy densities and energy spectra
of the dominant and subdominant components.

The results of this study agree with the predictions of the Beresnyak-Lazarian \cite{BL08}
model for non-relativistic imbalanced turbulence that we discussed in
Section~4. In fact, CL14 concluded that the magnetic spectrum of
dominant waves is steeper than that of sub-dominant waves and the
dominant waves exhibit anisotropy which is weaker than predicted in
\cite{GS95} while the sub-dominant waves exhibit stronger than GS95
anisotropy. In addition, CL14 showed that
The energy density ratio of the dominant to subdominant waves scales in
proportion to the ratio of the energy injection rates to the power of
$n$, i.e. $(\epsilon_{+}/\epsilon{-})^n$, where $n > 2$, which is also
consistent with the Beresnyak-Lazarian \cite{BL08} predictions.

The work on imbalanced and balanced relativistic turbulence strongly
indicate that the nature of turbulence does not significantly change
with the transfer to the relativistic regime. This conclusion is
suggestive that the models based on the Goldreich-Sridhar turbulence,
e.g. the turbulent reconnection model \cite{Lazarian1999} can
be extended to relativistic phenomena (see discussion in \cite{Lyutikov2013}).
 
\section{Intermittency of MHD turbulence}

\subsection{General considerations}

So far our focus in the review was on the turbulence
self-similarity. This property, which is also called scale-invariance,
implies that
fluid turbulence can be reproduced by the magnification of some part of it.

At the dissipation scales the self-similarity is known to fail with turbulence
forming non-Gaussian dissipation structures as exemplified, e.g. in \cite{biskamp2003}.
Interestingly enough, present-day research shows that 
self-similarity is not exactly true even along the inertial range. Instead
the fluctuations tend to get increasingly sparse in time and space at 
smaller scales. This property is called {\it intermittency}. Note, that
the power-law scaling does not guarantee the scale-invariance or absence
of intermittency.

One way to do such studies is to investigate the scaling powers of longitudinal
velocity fluctuations, i.e. $(\delta V)^p$, where 
$\delta V\equiv ({\bf V}({\bf x}+{\bf
r})-{\bf V}({\bf x})) {\bf r}/r$. The infinite set of various powers of
$S^p\equiv \langle (\delta V)^p \rangle$, where $\langle ..\rangle$ denote
ensemble\footnote{In astrophysics spatial or temporal averaging is used.}
 averaging, is equivalent to the p.d.f. of the velocity increments.
For those powers one can write $S^p(r)= a_p r^\xi_p$ to fully characterize
the isotropic turbulent field in the inertial range. While the 
scaling coefficients $a_p$ are given by the values of the function
$S^p$ e.g. at  the injection scale,  the scaling
exponents $\xi_p$ are very non-trivial. It is possible to show that 
for a self-similar flow the scaling exponents are linear function of $n$,
i.e. $\xi_p\sim p$, which for Kolmogorov model $S^1\sim v_l\sim l^{1/3}$ gives
$\xi_p=p/3$. Experimental studies, however, give different results which
shows that the Kolmogorov model is an oversimplified one.

MHD turbulence, unlike hydro turbulence, deals not only with velocity 
fluctuations, but also with the magnetic ones. The intermittencies of the two 
fields can be different. In addition, MHD turbulence is anisotropic as
magnetic field affects motions parallel to the local direction
of ${\bf B}$ very different. This all makes it more challenging to
understand the properties of MHD intermittency 
more interesting. 

An interesting and yet not understood property of structure functions,
however, helps to extend the range over which $S^p$ can be studied.
Benzi et al \cite{Benzi1995} reported that for hydrodynamic turbulence
the functions $S^p(S^3)$ exhibit much broader power-law range compared
to $S^p(r)$. While for the inertial range a similarity in scaling
of the two functions stem from the Kolmogorov scaling $S^3\sim r$,
the power-law scaling of $S^p(S^3)$ protrudes well beyond the inertial
range into the dissipation range\footnote{In practical terms this means
that instead of obtaining $S^p$ as a function of $r$, one gets $S^p$ as
a function of $S^3$, which is nonlinear in a way to correct for the
distortions of $S^p$.} . This observation shows that the dissipation
``spoils'' different orders of $S$ in the same manner. Therefore there is
no particular need to use the third moment, but one can use any other
moment $S^m\sim r^m$ and obtain a good power law of the function
$S^p\sim (S^m)^{\xi_p/\xi_m}$ (see \cite{biskamp2003}).

\subsection{She-Leveque model of intermittency}

A successful model to reproduce both experimental hydro data
and numerical simulations is She-Leveque (1994, \cite{SheLeveque}) model. According to \cite{Dubrulle} 
 this model can be derived assuming that the energy from large scale
is being transferred to $f<1$ less intensive eddies and $1-f$ of more intensive
ones.   
The scaling relations suggested in \cite{SheLeveque}
related $\zeta_p$ to the scaling of the velocity $V_l\sim l^{1/g}$,
the energy cascade rate $t_l^{-1}\sim l^{-x}$, and the co-dimension of the
dissipative structures $C$:
\begin{equation}
\zeta_p={p\over g}(1-x)+C\left(1-(1-x/C)^{p/g}\right).
\label{She-Leveque}
\end{equation}
For incompressible turbulence these parameters are $g=3$, $x=2/3$, 
and $C=2$, implying that
dissipation happens over 1D structures (e.g. vortices). So far the
She-Leveque scaling has done well in reproducing the intermittency of
incompressible hydrodynamic turbulence.

\subsection{Intermittency of incompressible turbulence}

\begin{figure*}[t]
  \includegraphics[width=0.3\textwidth]{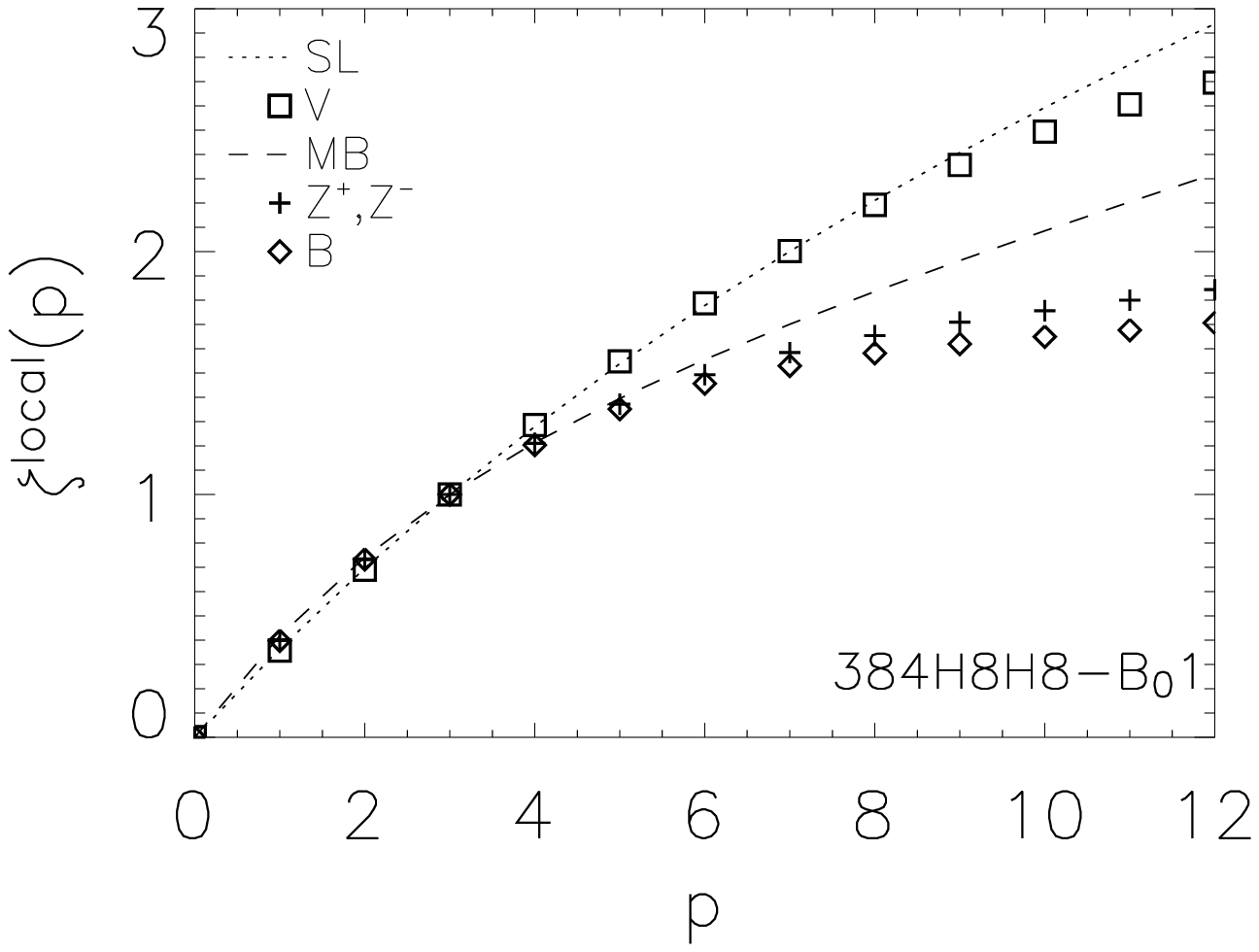}
\hfill
  \includegraphics[width=0.3\textwidth]{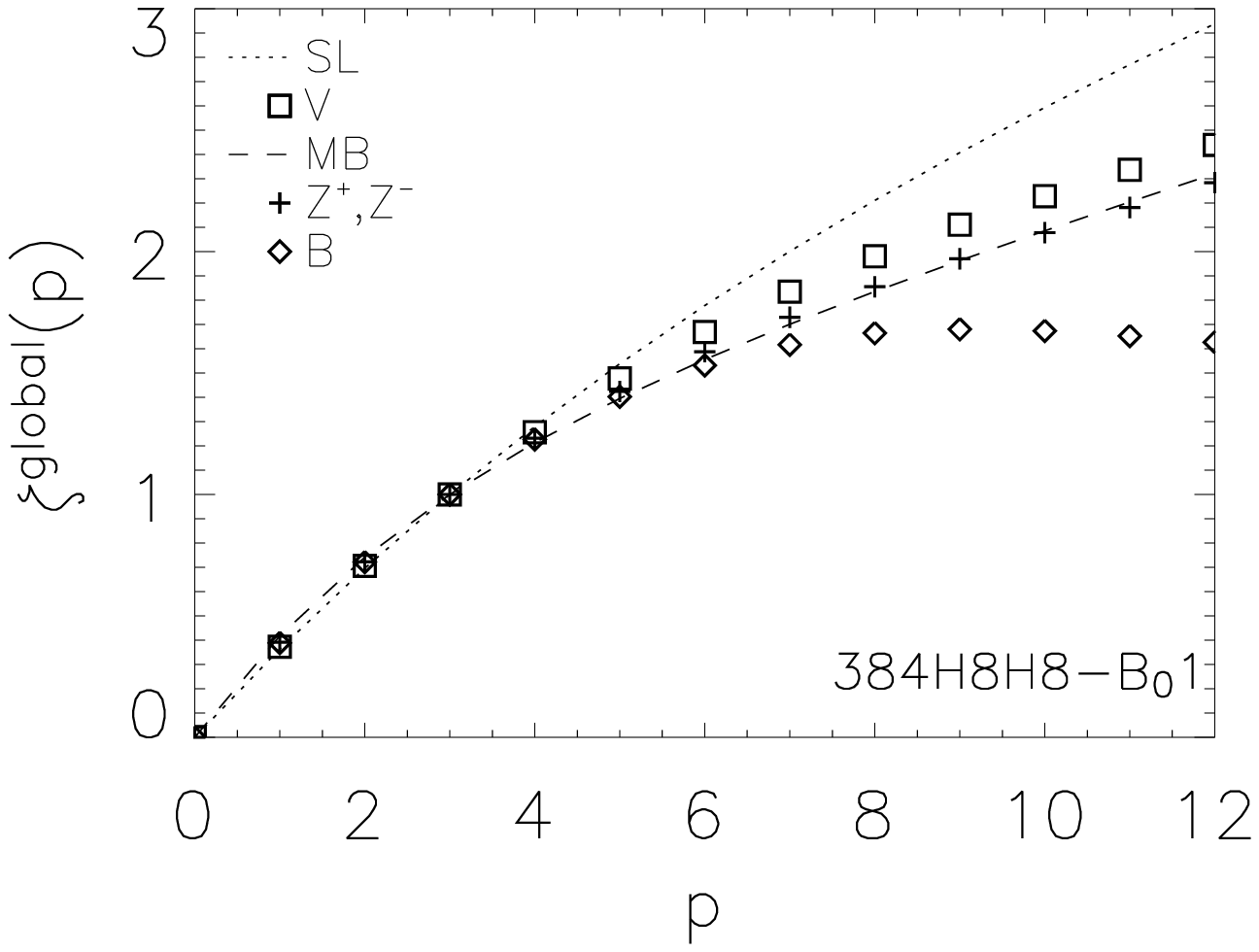}
\hfill
  \includegraphics[width=0.3\textwidth]{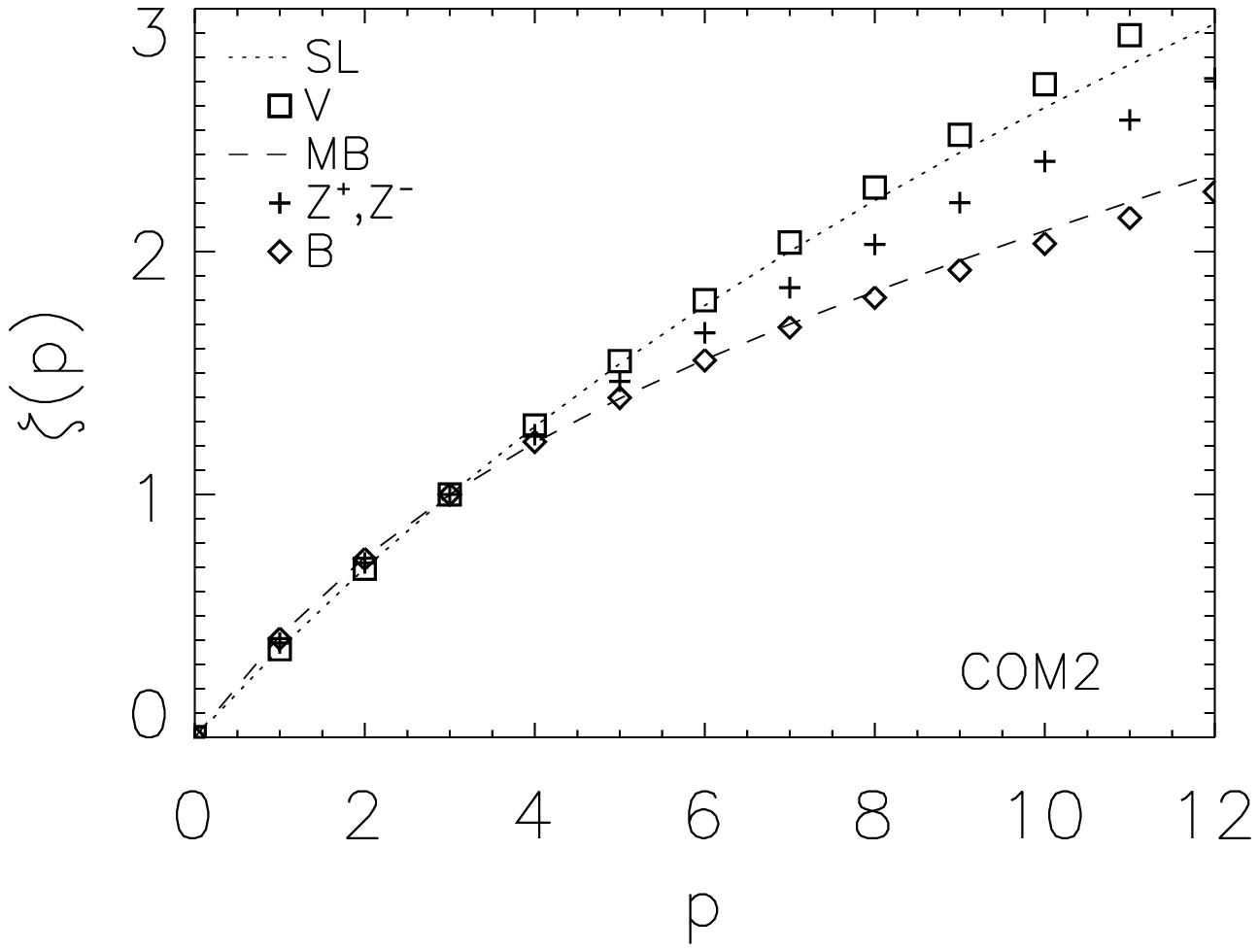}
  \caption{ {\it left panel}: Intermittency  
        exponents for incompressible MHD turbulence
        in perpendicular directions 
        in the local frame.  
        The velocity exponents show a scaling similar to the 
        She-Leveque model. 
        The magnetic field shows a different scaling.
 {\it central panel}: Intermittency 
        exponents for incompressible MHD turbulence
        in the global frame. 
        Note that the result for $z^{\pm}$ is very similar to  
        the M\"{u}ller-Biskamp model \cite{Muller2000}.
{\it right panel}: Intermittency exponents for
superAlfv\'enic compressible 
turbulence in the global frame. From CLV03
}
\label{f9}
\end{figure*}

In their pioneering study
\cite{Muller2000} applied the She-Leveque model to incompressible 
MHD turbulence and 
attracted the attention of the MHD researchers to this tool. They
used Els\"asser variables and claimed  that their results are consistent 
with dissipation within 2D structures (e.g. 2D current sheets). The consequent
study \cite{CLV02a} used velocities
instead of Els\"asser variables and provided a different
answer, namely, that the dimension of dissipation structures is
the same as in incompressible hydro, i.e.
the dissipation structures are 1D. The difference between the two
results was explained in Cho, Lazarian \& Vishniac (2003, \cite{Cho2003b} henceforth
CLV03). They noted that, first of all, 
the measurements in \cite{Muller2000} were done
in the reference frame related to the {\it mean} magnetic field, while
the measurements in \cite{CLV02a} were done in
the frame related to the {\it local} magnetic field. We believe that
the latter
 is more physically motivated frame, as it is the local magnetic field
is the field that is felt by the eddies. It is also in this reference frame
that the scale-dependent anisotropy predicted in the GS95
model is seen. Computations in CLV03 confirmed that the dissipation structures
that can be identified as velocity
 vortices in the local magnetic field reference frame can also
be identified with
two dimensional sheets in terms of Els\"asser variables 
in the mean magnetic field reference frame.
This, first of all, confirms a mental picture where motions perpendicular
to magnetic field lines are similar to hydrodynamic eddies. More importantly,
it sends a warning message about the naive interpretation
of the She-Leveque scalings in the MHD turbulence.

\subsection{Intermittency of compressible turbulence}

\begin{figure*}[t]
 \includegraphics[width=0.33\textwidth]{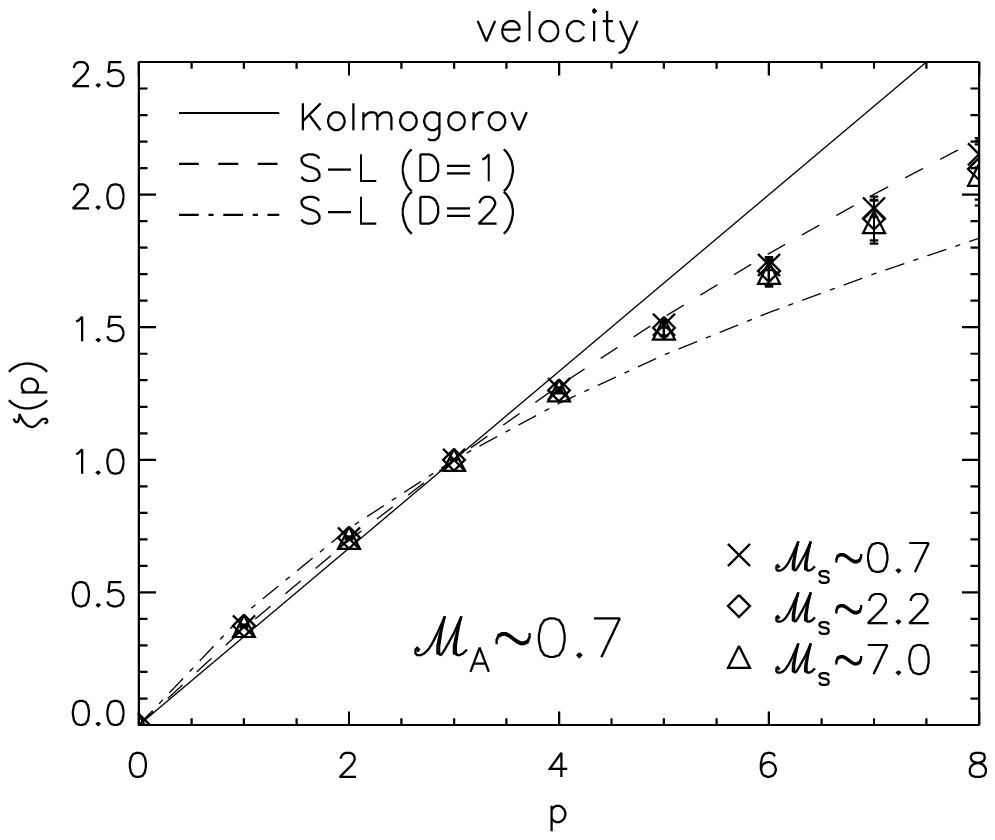}
 \includegraphics[width=0.33\textwidth]{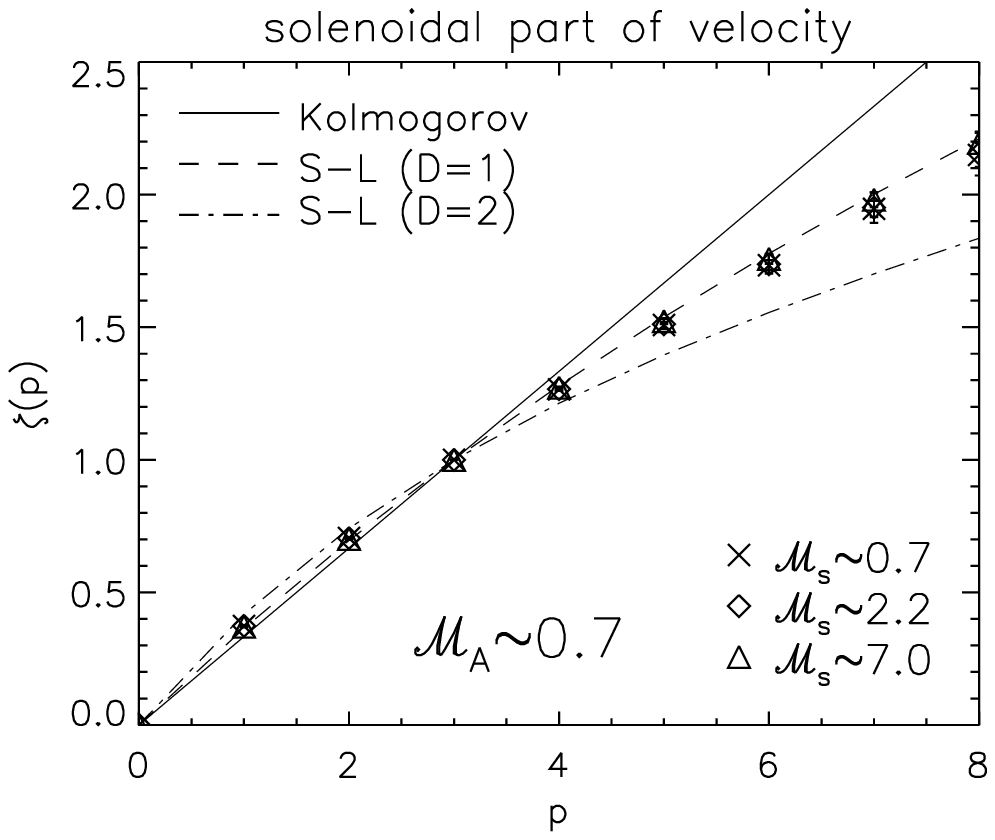}
 \includegraphics[width=0.33\textwidth]{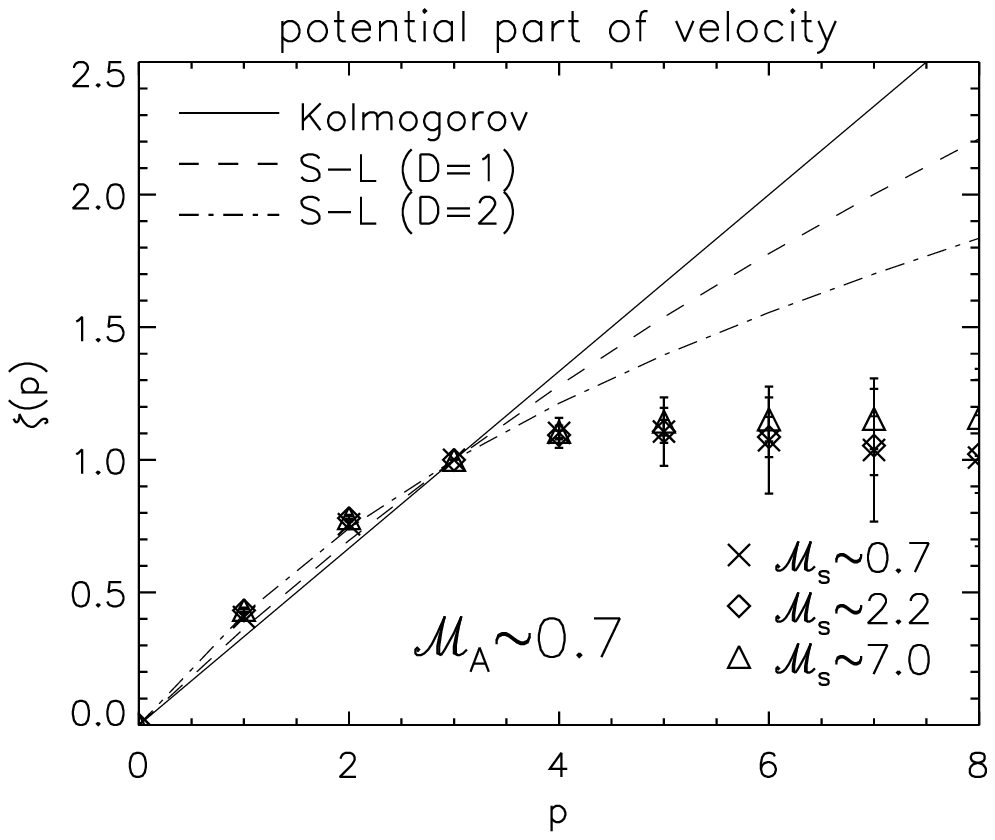}
 \includegraphics[width=0.33\textwidth]{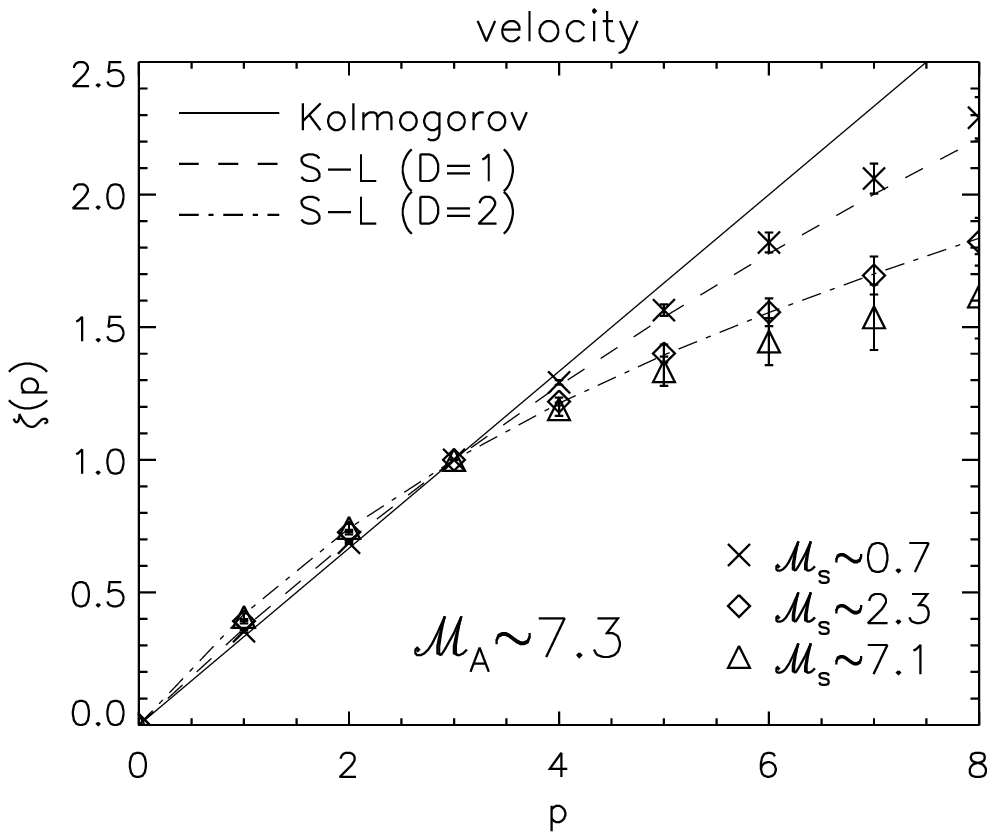}
 \includegraphics[width=0.33\textwidth]{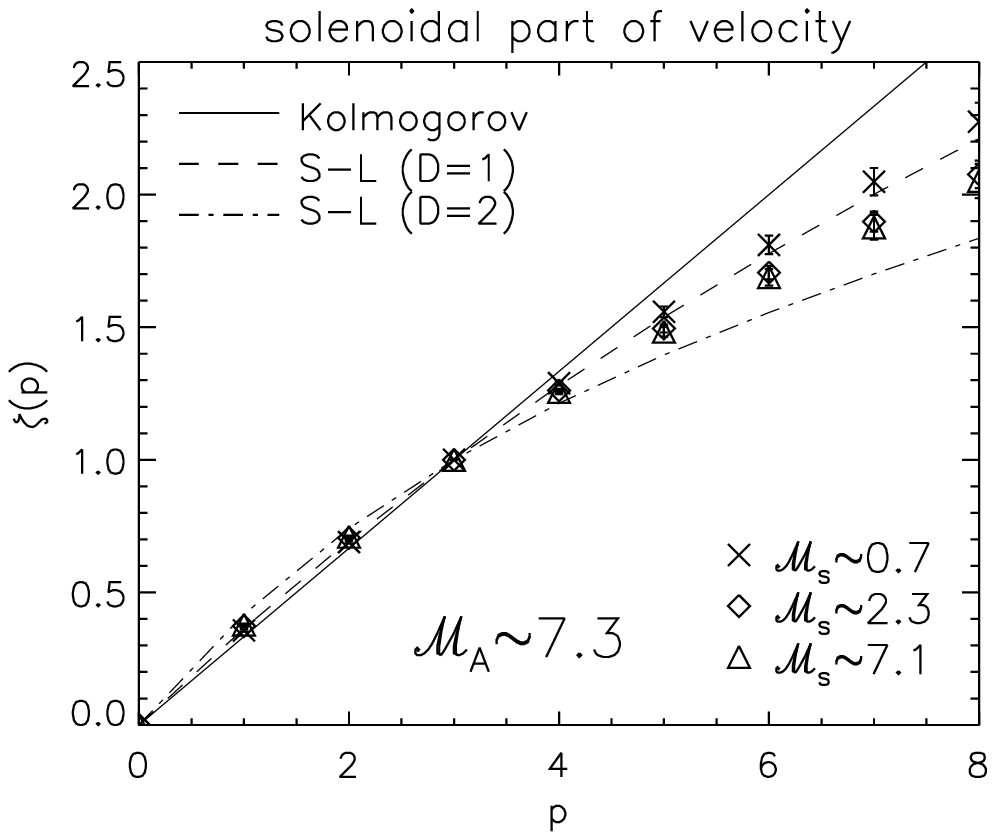}
 \includegraphics[width=0.33\textwidth]{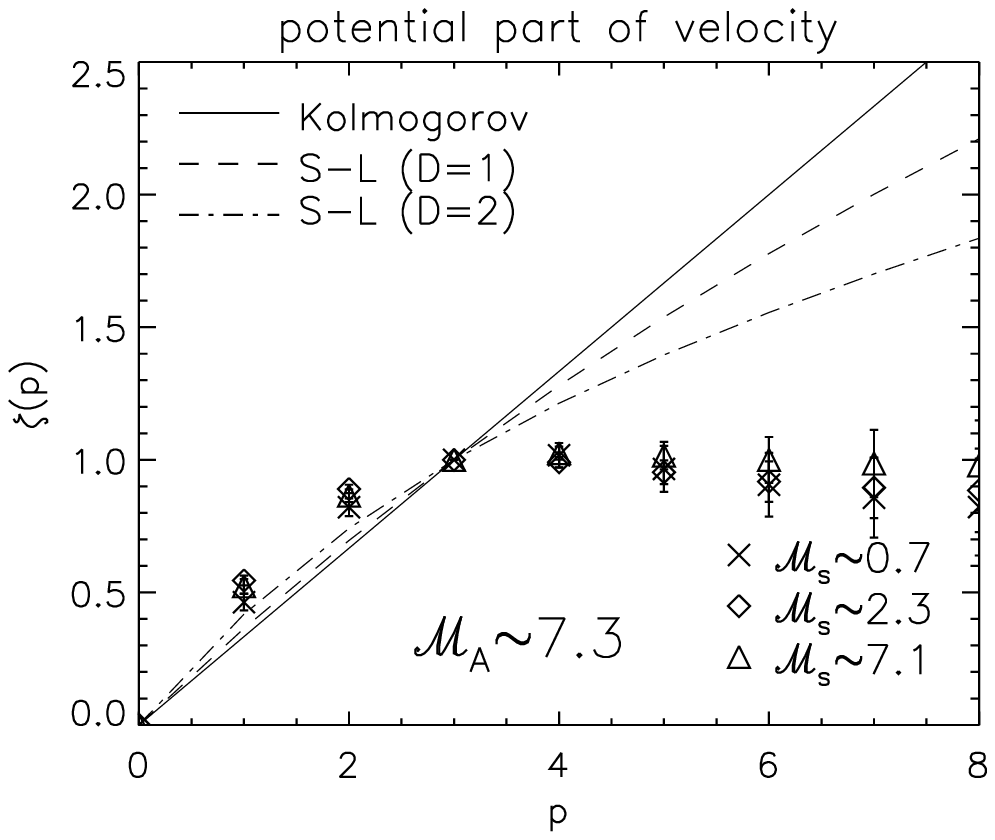}
 \caption{Scaling exponents of the velocity (left column) and its incompressible and compressible parts (middle and right columns, respectively)
 for experiments with different sonic Mach numbers in two regimes: subAlfv\'{e}nic (upper row) and superAlfv\'{e}nic (lower row). From KL10.
\label{f10}}
\end{figure*}

Intermittency in compressible MHD turbulence was discussed in Boldyrev (2002 \cite{Boldyrev2002})
who assumed that the dissipation there happens in shocks\footnote{The cited paper
introduces the model of compressible turbulence which it calls Kolmogorov-Burgers model. Within this
model turbulence goes first along the Kolmogorov scaling and then, at small scales forms shocks. The
model was motivated by the numerical measurements of the turbulence spectrum that indicated the
index of supersonic turbulence close to -5/3. This however was shown to be an artifact of numerical
simulations with lower resolution. Simulations in \cite{Kritsuk2007} showed that the slope with -5/3
is the result of the numerical bottleneck and the actual slope of the highly compressible turbulence
is -2, as was expected earlier.}  and therefore
the dimension of the dissipation structures is 2. The idea of the dominance of
shock dissipation does not agree well with the numerical simulations in
CL02, CL03, where the dominance of the vortical motions
in {\it subAlfv\'enic} turbulence (i.e. magnetic pressure is larger than the
gaseous one)  was reported. 
Nevertheless, numerical simulations
in \cite{Padoan2004} showed that for {\it superAlfv\'enic} turbulence 
(i.e. magnetic
pressure is less than the gas pressure) the dimension of the
dissipation structures was gradually changing from one to somewhat
higher than two as the
Mach number was increasing from 0.4 to 9.5. The very 
fact that the superAlfv\'enic
turbulence, which for most of the inertial scale resolvable by simulations
does not have a dynamically important magnetic field is different from
subAlfv\'enic is not surprising. The difference between the results
in \cite{Padoan2004} at low Mach number and the incompressible runs in
\cite{Muller2000} deserves a discussion, however. First of all,
the results in \cite{Padoan2004} are obtained for the velocity, while
the results in \cite{Muller2000} are obtained for the Els\"asser.
CLV03 has shown that the magnetic field and velocity have different
intermittencies. Indeed, it is clear from Fig.~1 that 
$\zeta^{\rm magnetic}<\zeta^{\rm velocity}$  which means that magnetic
field is more intermittent than velocity. An interesting feature of
superAlfv\'enic simulations in Fig. \ref{f9} is that the velocity follows the
She-Leveque  hydro scaling with vortical
dissipation, while magnetic field exhibits a pronounced dissipation
in current sheets. Both features are expected if magnetic field is
not dynamically important and the turbulence stays essentially hydrodynamic.
We also see that the dynamically important magnetic field does changes
the intermittency. The flattening of magnetic field scaling is pronounced
in Fig.~\ref{f9}.

A more recent study of intermittency of the velocity field of compressible turbulence was performed in KL10.
In Figure~\ref{f10}  we show scaling
exponents for the velocity and all its parts and waves calculated in the global
reference frame.  In the top left plot of Figure~\ref{f10} we see
that for the subAlfv\'{e}nic turbulence the scaling exponents of velocity follow
the She-L\'{e}v\^{e}que scaling with $D=1$.  Supported by the theoretical
considerations we can say that most of the dissipative structures are
one-dimensional.  Even though the scalings are not perfectly independent of the
value of ${\cal M}_s$, since we see somewhat lower values of $\zeta$ for higher
$p$, the differences between these values for models with different sonic Mach
numbers are within their error bars, thus it is relatively difficult to state
that the scalings are completely independent or only weakly dependent of the
values of ${\cal M}_s$.  Looking in the corresponding plot for models with a
weak magnetic field we clearly see that the spread of curves for different sonic
Mach numbers is much higher than in the previous case.  For subsonic model the
scaling exponents of velocity follow very well the theoretical curve defined by
the S-L scaling with parameter $D$ corresponding to one-dimensional structures.
The model with ${\cal M}_s \sim 2.3$, however, follows perfectly the S-L scaling
with $D=2$ corresponding to the two-dimensional dissipative structures.
Moreover, models with even higher values of the sonic Mach number have the
scaling exponents for $p>3$ somewhat below the S-L scaling with $D=2$.  These
observations suggest that the scaling exponents of the velocity change with the
sonic Mach number but only in the case of weak magnetic field turbulence.  The
presence of a strong magnetic field significantly reduces these changes and
preserves the generation of the dissipative structures of higher than one
dimensions.

After the decomposition of velocity into its incompressible and compressible
parts we also calculate their scaling exponents.  In the middle and right
columns of Figure~\ref{f10} we show the incompressible and
compressible parts of the velocity field, respectively.  The incompressible part
it strong.  It constitutes most of the velocity field thus it is not surprising
that its scaling exponents are very similar to those observed in velocity.  This
is true in the case of subAlfv\'enic models, because all curves in the middle
left plot in Figure~\ref{f10} are tightly covering the S-L scaling
with $D=1$.  The similarity between the velocity and its solenoidal part is also
confirmed in the case of superAlfv\'{e}nic models but only for subsonic case,
when the role of shocks is strongly diminished.  Two supersonic models show
exponents following a scaling more closer to the S-L one with $D=1$, yet still
with lower values for $p>3$.  

\begin{figure*}[t]
\includegraphics[width=0.44\textwidth]{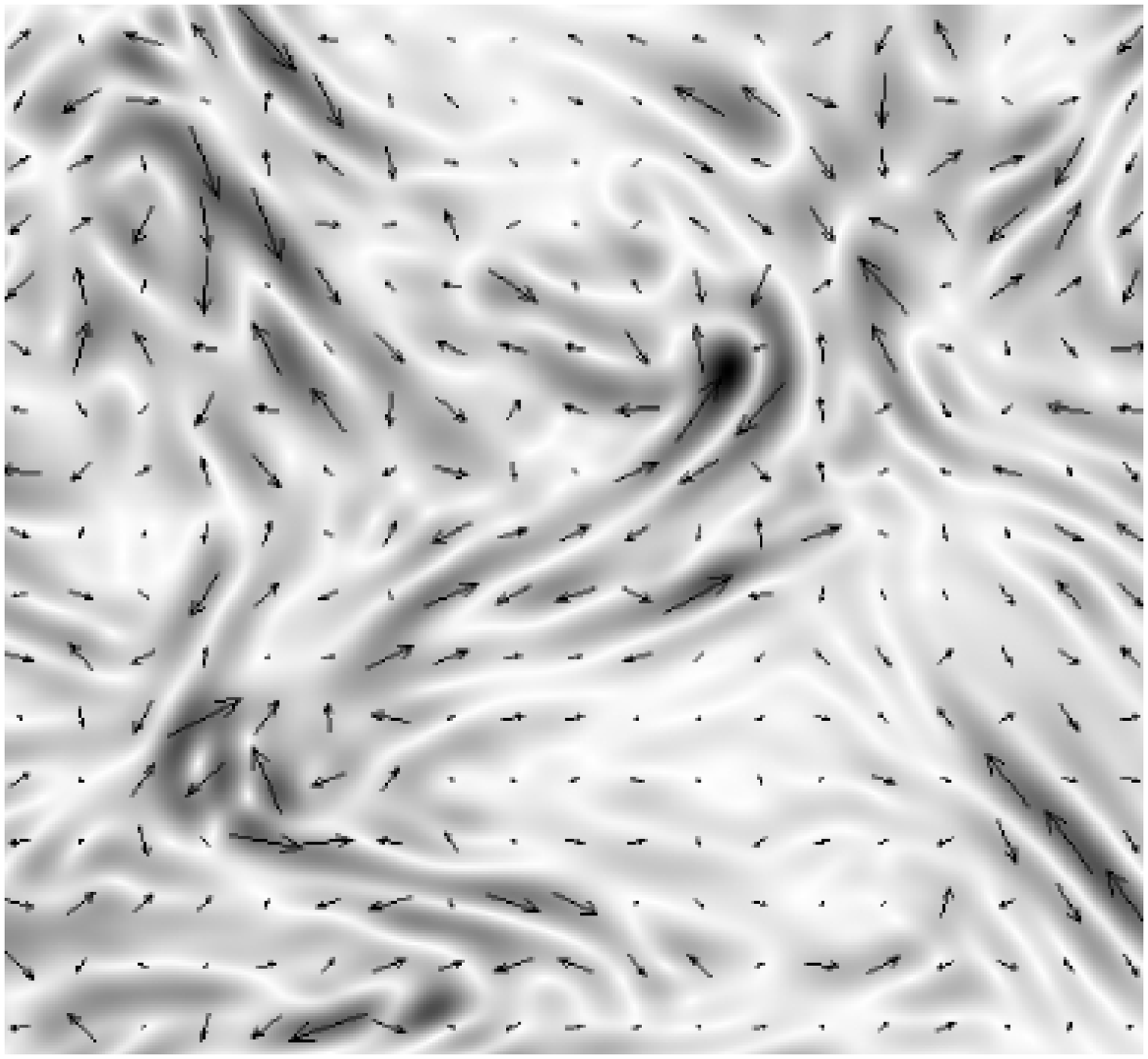}
\hfill
  \includegraphics[width=0.50\textwidth]{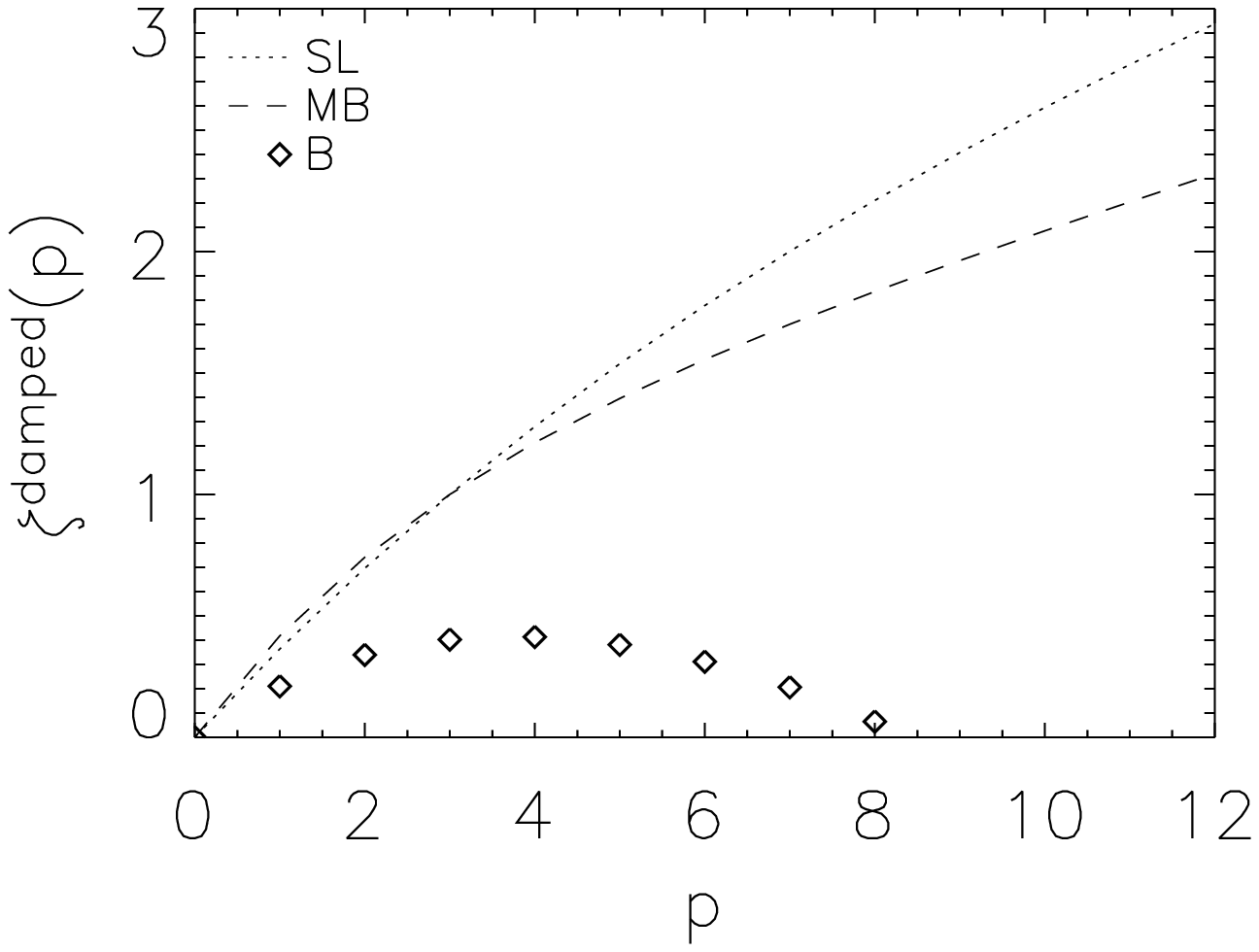}
  \caption{
The incompressible viscously damped simulations:
{\it Left}: Magnetic reversals (in the plane $\bot$ to mean $\langle{\bf B}\rangle$) that create current layers and makes turbulence highly intermittent.
Darker regions correspond to higher magnetic field.
{\it Right}: Intermittency indexes.
}
\label{f11}
\end{figure*}

\subsection{Intermittency of viscosity-damped turbulence}

For the extreme intermittency of the magnetic field suggested in LVC04
the higher moments of structure functions 
$S_p\sim \hat{b_l}^p \phi_l$ which means that $S_p\sim l^{1-p/2}$. 
The concentration of magnetic field in thin filaments gives rise to 
resistive loses that should eventually make  $\xi_p=0$ for sufficiently
large $p$. In Fig.~\ref{f11} we see this general tendency for high $p$.
For the absence of the more precise correspondence we may blame
(a) our crude model for estimating $\xi$, (b) numerical effects, and (c)
LVC04 model itself. Addressing the issue (b), we would say that the 
compelling arguments in the model provide $k^{-1}$ spectrum and this
would provide $\xi(2)=0$ in accordance with the intermittency model
above. However, due to numerical effects identified in LVC04
the spectrum of magnetic fluctuations is slightly steeper.

\section{Selected implications of MHD turbulence and turbulent dynamo}

Astrophysical fluids are turbulent and therefore one must take into account properties of turbulence while
describing astrophysical processes. We have discussed various implications in recent reviews,
e.g. in \cite{Lazarian2012}. Below is provided a brief summary of selected applications of the scalings
obtained.

\subsection{Magnetic reconnection in the presence of MHD turbulence}

\begin{figure*}[t]
  \includegraphics[width=0.45\textwidth]{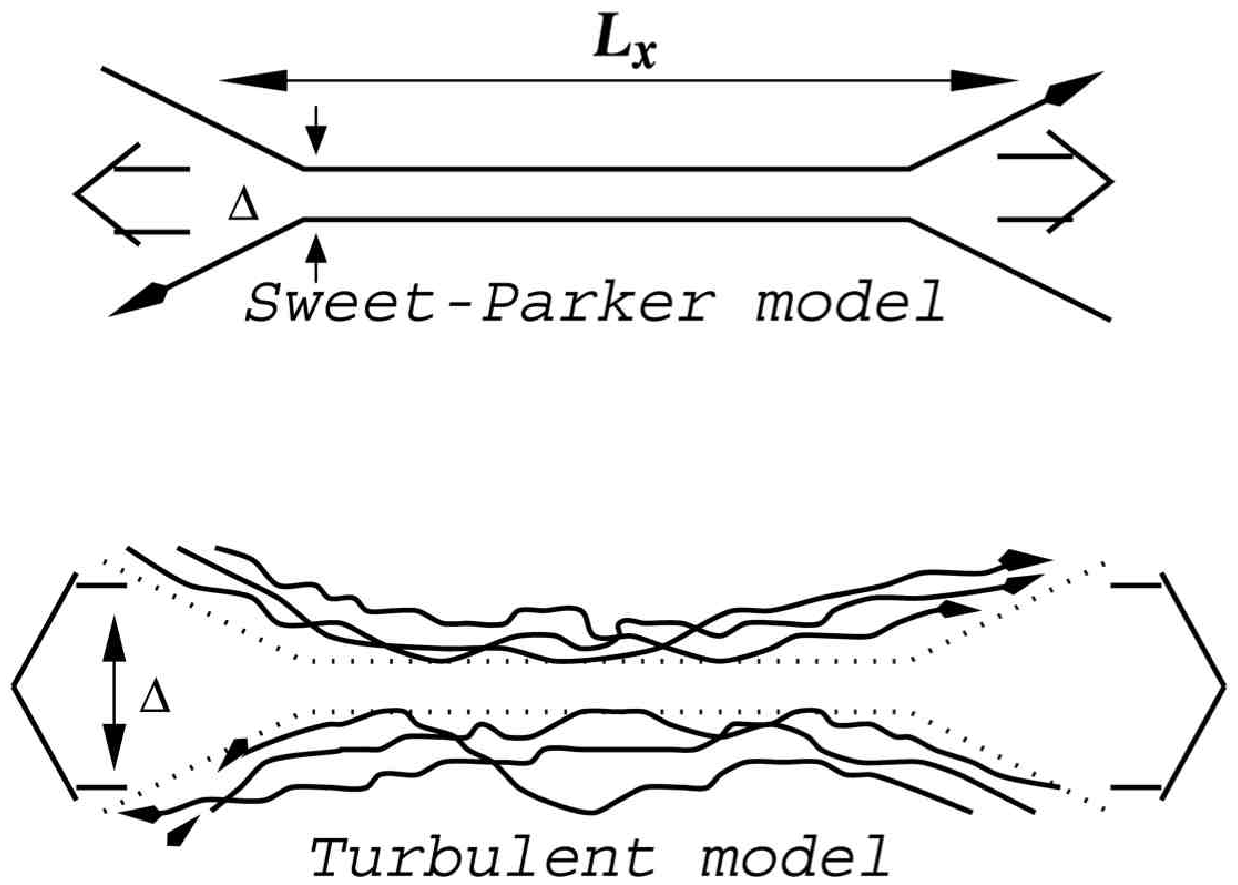}
\hfill
  \includegraphics[width=0.45\textwidth]{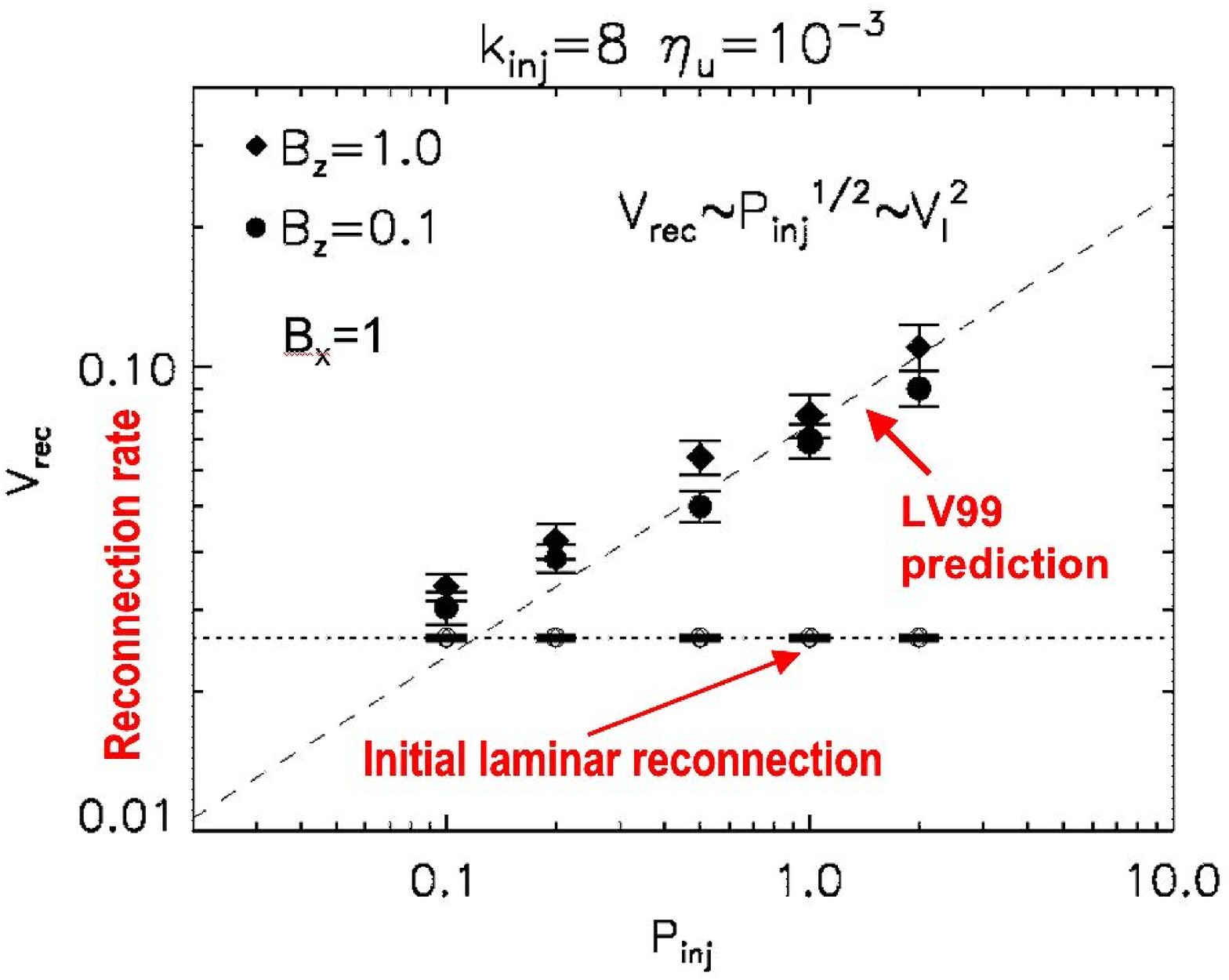}
  \caption{{\it Left}: {\it Upper panel}. Sweet-Parker reconnection. $\Delta$ is limited by resistivity and is small.
 {\it Lower panel}: reconnection according to the LV99 model. $\Delta$ is determined by turbulent field wandering and can be large.  From \cite{Lazarian2004}.
 {\it Right}: Testing of LV99 model with numerical simulations in \cite{Kowal2009}. The dependence on the power of
 turbulent driving is shown.
}
\label{f12}
\end{figure*}

Magnetic reconnection is a long standing problem. Is it associated with the fundamental ability of magnetic flux tubes to change their topology, while
being submerged within conducting fluids \cite{Biskamp2000,Priest2000}. Lazarian and Vishniac (\cite{Lazarian1999} henceforth LV99) considered
turbulence as the agent that makes magnetic reconnection fast (see Fig.~\ref{f12}). The scheme proposed in LV99 there differs appreciably from the earlier attempts to 
enhance reconnection via turbulence, \cite{Speiser1970,Jacobson1984,Matthaeus1985,Matthaeus1986}, see \cite{Eyink2011b} for a detailed comparison.
The LV99 model relies on opening of 
reconnection region via magnetic field wondering, and for the Alfv\'enic turbulence model that was discussed in this review provide the scaling that are
of reconnection velocity $V_{rec}$ being proportional to square root of the injection power $P_{inj}$. Fig.~\ref{f12} shows a good relation of the numerical
testing in \cite{Kowal2009, Kowal2012} and the predictions. See also Lazarian et al, this volume, for a review of turbulent reconnection.

A study in Eyink, Lazarian \& Vishniac (2011, \cite{Eyink2011b} henceforth ELV11) reveals a very deep relation between MHD turbulence
and magnetic reconnection. In fact, it was shown back in LV99 that the predicted reconnection rates are necessary to make the GS95 model self-consistent,
i.e. to resolve magnetic knots that emerge as eddy-like motions perpendicular to the direction of local magnetic field twist magnetic field lines. ELV11 demonstrates
that the Lagrangian properties of MHD turbulence require the violation\footnote{The violation of frozen in condition in turbulence is implicit in LV99. It was stated explicitly
in \cite{Vishniac1999} and discussed in terms of star formation in \cite{Lazarian2005}. The first formal quantitative study was performed in \cite{Eyink2011a}.}
of magnetic flux being frozen in and re-derives the predictions of LV99 model from this established properties.
As magnetic flux frozenness is a corner stone of major astrophysical theories its violation in turbulent fluids has deep consequences.
For instance, the change of our understanding of diffusion out of star forming clouds in the presence of turbulence was recently
discussed in \cite{Lazarian2012b, Lazarian2013}. The numerical confirmation of the violation of magnetic field flux freezing in turbulent fluids
was reported in \cite{eyink13}.

\subsection{Turbulence and particle acceleration}

MHD turbulence plays an important role in accelerating energetic particles. First of all, the second order Fermi acceleration can arise directly from
the scattering of particles by turbulence, see, e.g., \cite{Melrose1980}. Properties of MHD turbulence that we discussed above are essential to understanding
this process. If turbulence is injected at large scales, the anisotropy of Alfv\'enic modes at small scales makes them inefficient for scattering and
acceleration of cosmic rays \cite{Chandran2000,yan2002}. In this situation, fast modes were identified in \cite{yan2002} as the major
scattering and acceleration agent for cosmic rays and energetic particles in interstellar medium (see also \cite{Yan2004a,Yan2008a}). This conclusion was extended for solar environments in  Petrosian, Yan \& Lazarian (2006) and intracluster medium in \cite{Brunetti2007}. 

Turbulent magnetic field in the pre-shock and post-shock environment are important for the first order Fermi acceleration associated with shocks \cite{Schlickeiser2002}.
In particular, magnetic field enhancement compared to its typical interstellar values is important in the pre-shock region for the acceleration of high energy
particles. Turbulent dynamo that we discussed in Section~2 can provide a way of generating magnetic field in the precursor of the shock. In \cite{BJL09} it was shown
that the interactions of the density inhomogeneities pre-existing in the interstellar medium with the precursor generate strong magnetic fields in the shock precursor,
which allows particle acceleration up to the energy of $10^{16}$ eV.

In addition, fast magnetic reconnection of turbulent magnetic field can itself induce the first order Fermi acceleration \cite{DeGouveia2005,Lazarian2005}. Recent numerical simulations
in \cite{Kowal2012b},  demonstrate the efficiency of this process.

\subsection{Thin structures in the Interstellar Medium}

The viscosity-dominated regime of turbulence can be responsible for
the formation of structures in interstellar medium and other astrophysical 
environments. The magnetic pressure
compresses the gas as demonstrated in Fig.~\ref{f8}. More importantly,
extended current sheets that naturally emerge as magnetic field
fluctuates in the plane perpendicular to the mean magnetic field. It was speculated in \cite{Lazarian2007} that these current sheets can
account for the origin of the small ionized and neutral structures (SINS) on AU spatial scales 
\cite{Heiles1997,Stanimirovic2004}.

Goldreich and Sridhar \cite{Goldreich2006} appealed
to the generation of the magnetic field in the high $Pt$ turbulent plasma \cite{Schekochihin2004} to account for the high amplitude, but small scale fluctuations of plasma density
observed in the direction of the Galactic center. They argued that the plasma viscosity parallel to magnetic 
field can act in the same way as the normal viscosity of unmagnetized fluids. \cite{Lazarian2009} argued that the regime of dynamo
in \cite{Schekochihin2004} and the turbulence in \cite{Lazarian2004} have similarities in terms of the density enhancement that are created. Although
in the case of magnetic turbulence with sufficiently strong
mean magnetic field, global reversals, that \cite{Goldreich2006} appeal to in compressing plasma, do not happen, the reversals of the magnetic
field direction occur in the direction perpendicular to the mean magnetic field. As the mean magnetic field goes to zero, the two regimes get indistinguishable
as far as the density enhancements are concerned. Thus high intensity fluctuations of plasma density towards the Galactic center may also be the
result of viscosity-damped turbulence. 

\subsection{Intermittent turbulent heating of interstellar gas}

E. Falgarone and her collaborators \cite{Falgarone1995,Joulain1998,Godard2009,Godard2014}
attracted the attention of the interstellar community to the potential important implications of
intermittency, see also Falgarone et al, this volume, for a review.
A small and transient volume with high temperatures or violent
turbulence can have significant effects on the net rates of processes
within the ISM. For instance, many interstellar chemical reactions
(e.g., the strongly endothermic formation of CH$^+$) might take place
within very intense intermittent vortices. 

\begin{figure}
 \includegraphics[width=0.45\textwidth]{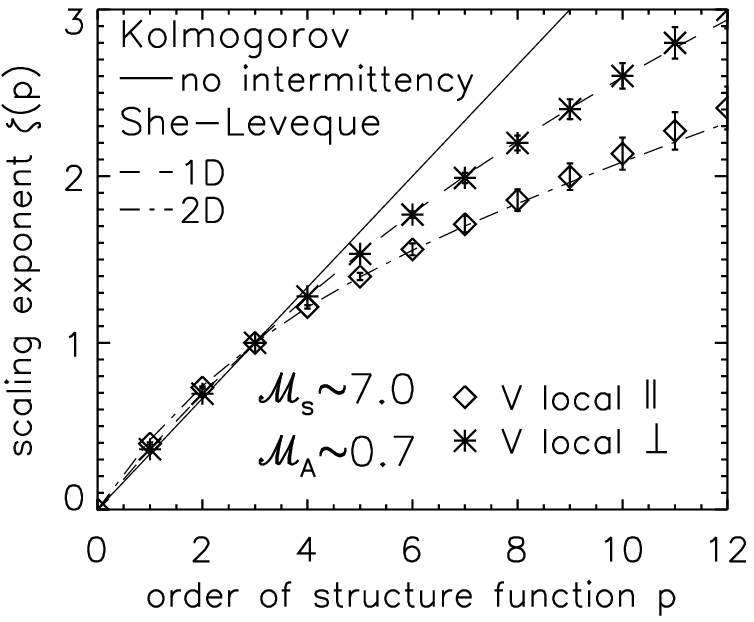}
\hfill
  \includegraphics[width=0.45\textwidth]{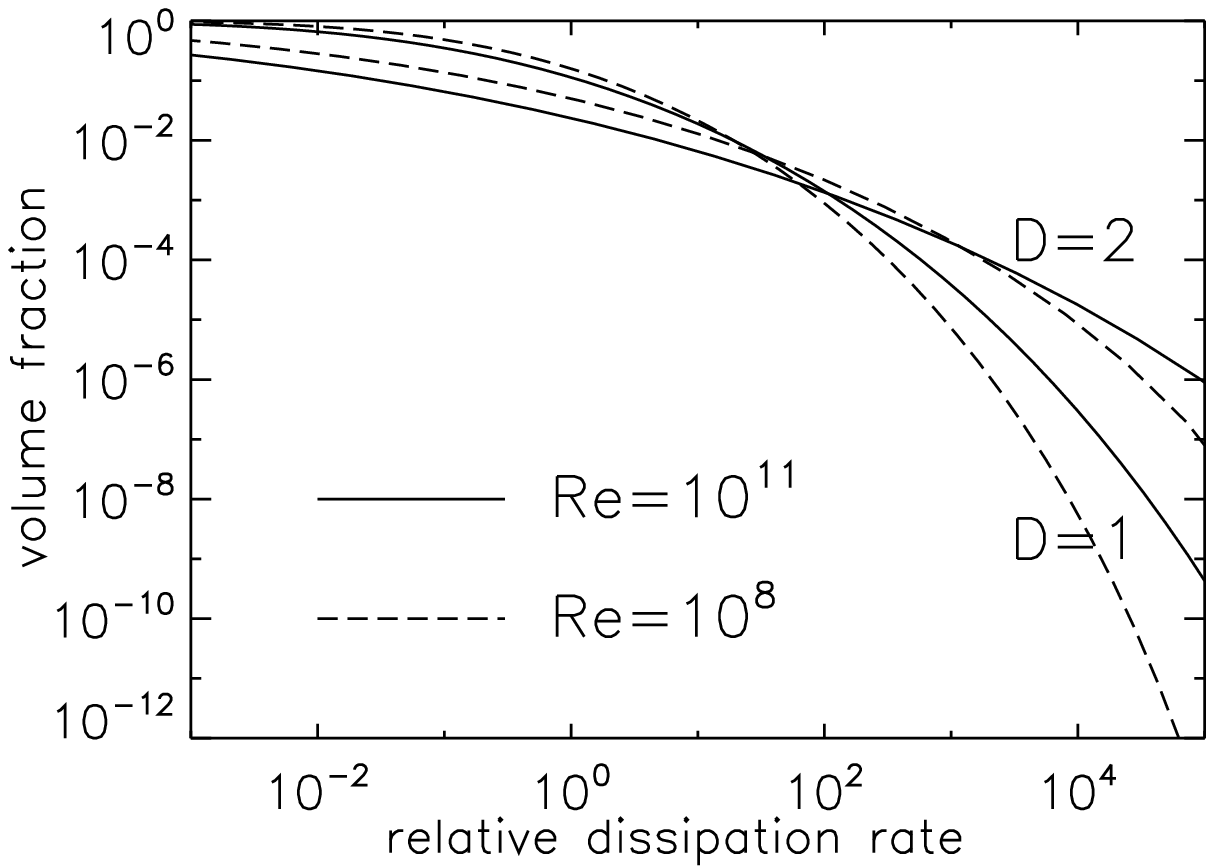}
\caption{\small {\it Left Panel}: The
intermittencies of velocities in the subAlfv\'enic, $M_A=0.7$ supersonic
$M_s=7$ MHD simulations. From \cite{Kowal2007}. {\it Right Panel}: Volume fraction with the
dissipation rate is higher than the mean rate for the She-Leveque
model of intermittency with $D=1$ and $2$.}
\label{f13}
\end{figure}

The bottom part of Fig.~\ref{f13} shows our calculations for the volume fractions of various dissipation rates (i.e., heating rates), based on the distribution function
suggested by She, Leveque and Dubrulle. While the temperatures achieved will depend upon the cooling
functions, some important conclusions are available from the analysis of Fig.~\ref{f13}.  This figure demonstrates that
even for Re$\sim 10^{11}$ the bulk of the energy dissipates within volume with dissipation rate around a factor of a few of the mean value,
provided that the She-Leveque model is valid. Furthermore, only 0.1\% of the volume has the dissipation rate higher than the factor of 100 of the mean value.
This provides stringent constraints on what chemistry we could expect to be induced by intermittent turbulent heating. 

Interestingly enough, the case of intermittency studies supports our point of the futility of
the "brute force" numerical approach. For instance, for a
typical ISM injection scale of 50 pc, the Reynolds number can be as high as ${\rm Re}=10^{11}$.
In comparison, numerical simulations can only reach ${\rm Re}\leq 10^{5}$ for the present record resolution of $4096^3$.

\subsection{Suppression of instabilities by Alfvenic turbulence}

Alfvenic turbulence can suppress instabilities, in particular, streaming instability that arises as energetic particles stream 
in one direction along magnetic field lines \cite{Yan2004,Farmer2004,BL08b}. The effect is based on
cascading of slab waves induced as a result of the instability development by the ambient turbulence. Thus the effect
is not limited by suppressing of streaming instability. For instance, in \cite{LB06} and \cite{Yan2011}
the suppression of gyroresonance instability by turbulence was considered. 

\begin{figure*}
  \includegraphics[width=0.45\textwidth]{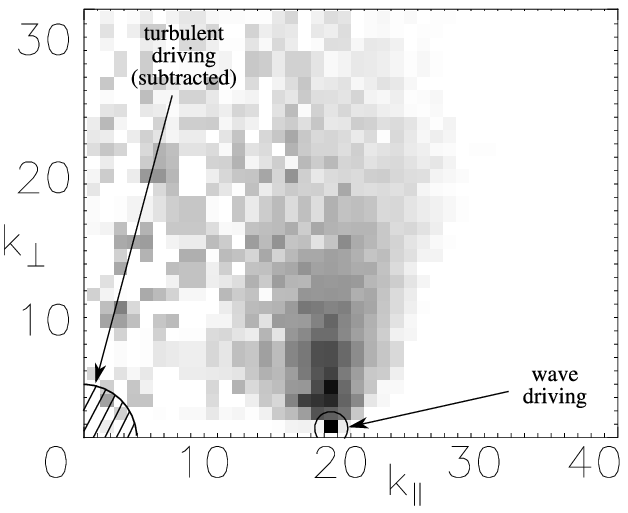}
\hfill
  \includegraphics[width=0.52\textwidth]{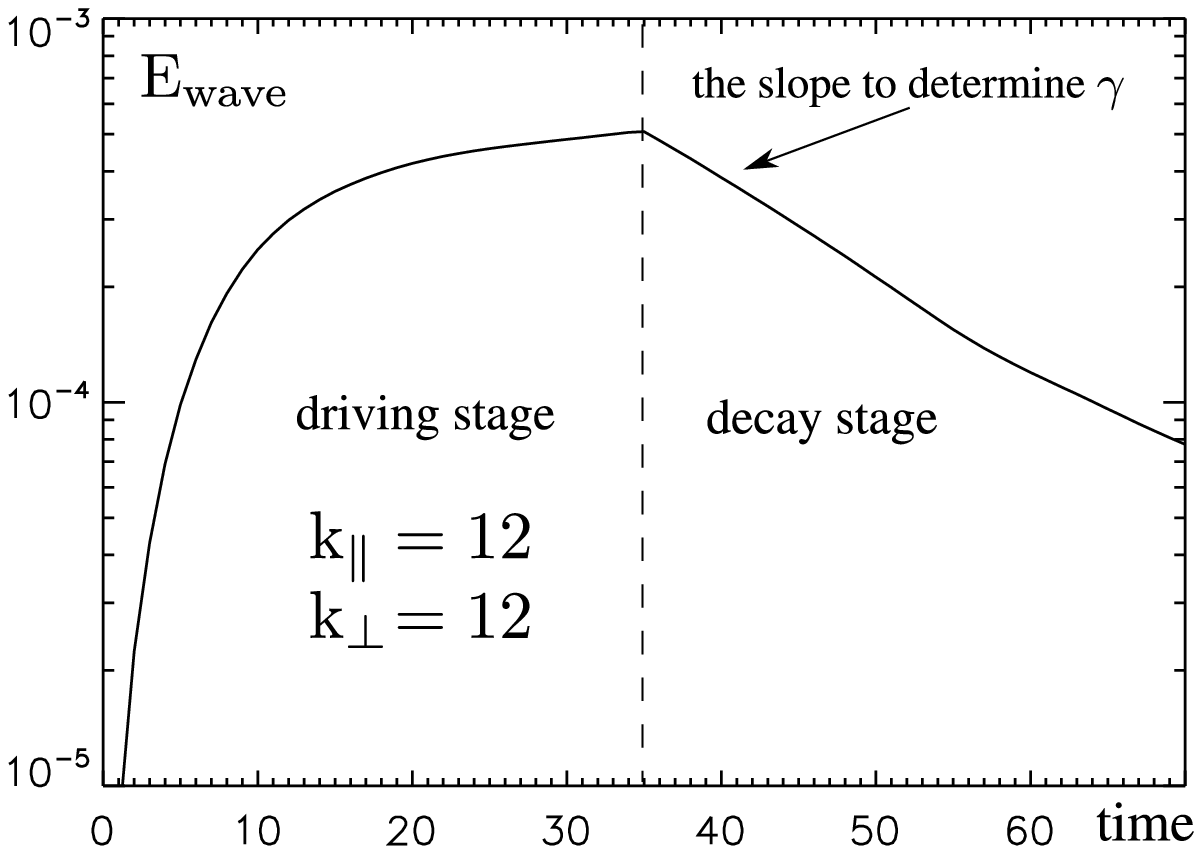}
  \caption{ The wave perturbation excited on top of existing turbulence exhibit exponential decay, as one expects a linear decay for a test perturbation in such
perturbation simulation, despite the turbulence itself is strongly nonlinear. This linear decay rate is independent of amplitude, unlike nonlinear damping, so
turbulence does not result in saturation of an instability but rather damps it completely or not.
 From \cite{BL08b}.}
\label{wave_decay}
\end{figure*}

The thorough numerical study of the suppression of the slab waves by Alfvenic turbulence was performed in \cite{BL08b}, see Fig.~\ref{wave_decay},
not only for the slab waves moving parallel to magnetic field, but also for waves moving at arbitrary angles to the mean magnetic 
field. The numerical simulations confirmed the theoretical expectations in the paper.

\subsection{Turbulent dynamo  and high redshift physics}

Magnetic turbulent dynamo that we discussed in \S~\ref{sec:1} is essential for understanding magnetic field in the early Universe.
Indeed, as long as the Universe becomes ionized and highly conductive, the viscosity is also greatly reduced due to ions scattering
in the magnetic fields \cite{Schekochihin2006,Schekochihin2008}. This creates high-Re environment which naturally produces turbulence. Despite a lot of discussion
of early dynamo is concentrated on the initial field generation mechanisms, such as Biermann battery and its modifications \cite{Lazarian1992},
we now understand that in the limit of very high Re the level of the initial field is not very important. Instead, around 5\% of the energy of the
turbulent cascade is deposited into magnetic energy due to high-Re small-scale dynamo \cite{B12a}. The magnetic field, therefore, could get dynamically
important at high redshifts. This will change the nature of many processes, for example we expect that magnetic field should be important
in the process of formation of the first stars. These possibility is discussed
in \cite{shober12, shober13}, but we believe that the relative role of non-linear dynamo
is even more important than it was presented in the latter study \cite{Xu14}.

{\bf Acknowledgements} AB was supported by Humboldt Fellowship. AL acknowledges the support of the NSF grant AST-1212096, the Vilas Associate Award as well as the support of the NSF Center for Magnetic Self- Organization.
In addition, AL thanks the International Institute of Physics (Natal, Brazil) and the Observatoire de Nice for its hospitality during writing this review.
Discussions with Ethan Vishniac and Greg Eyink are acknowledged. We thank Siyao Xu for reading the manuscript.

\def\apj{{\rm Astrophys. J.}}           
\def\apjl{{\rm ApJ }}          
\def\apjs{{\rm ApJ }}          
\def\grl{{\rm GRL }}
\def\aap{{\rm A\&A } }
\def\araa{{\rm Ann. Rep. A\&A } }
\def\mnras{{\rm MNRAS } }
\def\physrep{{\rm Phys. Rep. } }               
\def\prl{{\rm Phys. Rev. Lett.}} 
\def\pre{{\rm Phys. Rev. E}} 
\def\prd{{\rm Phys. Rev. D}} 
\def\pra{{\rm Phys. Rev. A}} 
\def\ssr{{\rm SSR}}
\def\planss{{\rm Plan. Space Sciences}}
\def\apss{{\rm Astrophysics and Space Science}}
\def\nat{{\rm Nature}}
\bibliographystyle{spphys.bst}
\bibliography{all}

\end{document}